\documentclass[prd,nofootinbib,floats,aps,twocolumn,tightenlines,superscriptaddress]
{revtex4-2}

\usepackage{natbib}
\usepackage{cancel}
\usepackage[skins,breakable]{tcolorbox}
\usepackage{float} 
\usepackage{graphicx}
\usepackage{mathtools}
\usepackage[normalem]{ulem}
\usepackage{hyperref}
\hypersetup{
	colorlinks=true,
	linkcolor=blue,
	filecolor=magenta,      
	urlcolor=cyan,
}

\usepackage{commath}
\usepackage{bm}
\usepackage{amsfonts,amsmath,amssymb,amsthm}
\usepackage[usenames,dvipsnames]{xcolor}
\usepackage{dsfont}
\usepackage{enumitem}
\usepackage{color}
\usepackage{tikz}
\usepackage{physics}
\usepackage{verbatim}
\tcbuselibrary{breakable}

\newcommand{\ii}{\mathrm{i}}

\renewcommand*\d[2][]{%
	\mathrm{d}%
	\ifx\relax#1\relax\else
	\rule{-0.02em}{1.5ex}^{#1}\rule{0.08em}{0ex}\!
	\fi
	#2\,
}

\makeatletter 

\renewcommand\onecolumngrid{% <<<<<<
	\do@columngrid{one}{\@ne}%
	\def\set@footnotewidth{\onecolumngrid}% <<<<<<<<<<<<<<<<
	\def\footnoterule{\kern-6pt\hrule width 1.5in\kern6pt}%
}

\renewcommand\twocolumngrid{% <<<<<<
	\def\footnoterule{% restore rule
		\dimen@\skip\footins\divide\dimen@\thr@@
		\kern-\dimen@\hrule width.5in\kern\dimen@}
	\do@columngrid{mlt}{\tw@}
}%
\makeatother
\newcommand\restr[2]{{% we make the whole thing an ordinary symbol
		\left.\kern-\nulldelimiterspace % automatically resize the bar with \right
		#1 % the function
		\vphantom{\normal|} % pretend it's a little taller at normal size
		\right|_{#2} % this is the delimiter
}}
\definecolor{ForestGreen}{RGB}{35,120,35}

\newtheorem{definition}{Definition}
\newtheorem{theorem}{Theorem}
\newtheorem{proposition}{Proposition}
\newtheorem{lemma}{Lemma}
\newtheorem*{claim*}{Claim}

\theoremstyle{definition}

\theoremstyle{remark}

\begin{document}
	
\title{Correlation and Entanglement partners in Gaussian systems}

\author{Ivan Agullo}
\email{agullo@lsu.edu}
\affiliation{Department of Physics and Astronomy, Louisiana State University, Baton Rouge, LA 70803, USA}
\author{Eduardo Mart\'{i}n-Mart\'{i}nez}
\email{emartinmartinez@uwaterloo.ca}
\affiliation{Department of Applied Mathematics, University of Waterloo, Waterloo, Ontario, N2L 3G1, Canada}
\affiliation{Institute for Quantum Computing, University of Waterloo, Waterloo, Ontario, N2L 3G1, Canada}
\affiliation{Perimeter Institute for Theoretical Physics, Waterloo, Ontario, N2L 2Y5, Canada}

\author{Sergi Nadal-Gisbert}
\email{sergi.nadalgisbert@helsinki.fi}
\affiliation{QTF Centre of Excellence, Department of Physics,
University of Helsinki, FI-00014 Helsinki, Finland}

\author{Patricia Ribes-Metidieri}
\email{patricia.ribesmetidieri@york.ac.uk}
\affiliation{Department of Mathematics, University of York, Heslington, York YO10 5DD, UK}

\author{Koji Yamaguchi}
%\email{yamaguchi.koji.848@m.kyushu-u.ac.jp}
\email{koji.yamaguchi@uwaterloo.ca}
\affiliation{Department of Informatics, Faculty of Information Science and Electrical Engineering,
Kyushu University, 744 Motooka, Nishi-ku, Fukuoka, 819-0395, Japan}

\begin{abstract}

We introduce a framework to identify where the total correlations and entanglement  with a chosen degree of freedom reside within the rest of a system, in the context of bosonic many-body Gaussian quantum systems. Our results are organized into two main propositions.  First, for pure Gaussian states, we show that every correlated mode possesses a unique single–degree-of-freedom partner that fully captures its correlations (consisting of entanglement), and we provide an explicit construction of this partner from the complex structure of the system’s state.  Second, for mixed Gaussian states, we constructively demonstrate that the notion of a partner subsystem splits into two: a \emph{correlation partner}, which contains all classical and quantum correlations and need not correspond to a single degree of freedom, and an \emph{entanglement partner}, which is always at most single-mode.  Finally, we extend the construction of partners to multi-mode subsystems. Together, these results provide conceptual practical tools to study how bipartite correlations and entanglement are structured and where they can be found in complex Gaussian many-body systems.
\end{abstract}

\maketitle

\section{Introduction}

Entanglement and correlations are central resources in quantum science, underpinning both foundational questions and practical applications in quantum technologies. However, characterizing how these correlations are distributed across subsystems of many-body quantum states remains a notoriously difficult problem. In particular, for mixed states, even the basic task of deciding whether a bipartite state is entangled is NP-hard in general~\cite{entanglementNPHard}. This motivates the search for structured settings in which correlations can be analyzed more transparently. Gaussian states provide one such setting.

Gaussian states play a distinguished role in continuous-variable quantum information, quantum optics, and quantum field theory~\cite{GQMRev,Serafini2017,Adesso2014}. They can be fully characterized by their first and second statistical moments, which makes them amenable to powerful analytical and numerical methods. Moreover, many physically relevant states---including ground and thermal states of quadratic Hamiltonians, coherent and squeezed states, and states generated in typical quantum optics laboratory scenarios are Gaussian. Because of this, Gaussian states offer a natural ground to address questions of entanglement structure in high-dimensional quantum systems. 

The study of entanglement structure in Gaussian systems has its roots in early work on covariance-matrix separability criteria for continuous variables~\cite{Simon2000,Duan:2000awi,Werner:2000pxg}. The modewise structure of bipartite entanglement in multimode pure Gaussian states was established in~\cite{Botero:2003wov}, where it was shown that, under local symplectic transformations, any pure Gaussian state decomposes into a product of two-mode squeezed states and local vacua. A related extension to more general bipartite Gaussian correlations was later provided in~\cite{Wolf:2008wad}.

In recent years, there have been further developments to understand the entanglement structure of Gaussian states. In particular, purification partners for the vacuum~\cite{HottaPartner} and more general Gaussian pure states~\cite{Trevison2019} in quantum field theory (QFT) have been identified. Such characterization  have stimulated further investigations, including studies on entanglement structures~\cite{tomitsuka_partner_2020,nambu_entanglement_2023,de_s_l_torres_entanglement_2023,osawa_final_2024,montes-armenteros_quantum_2025,Agullo:2023fnp,Ribes-Metidieri:2024vjn,Ribes-Metidieri:2025nfw} and entanglement harvesting~\cite{trevison_pure_2018,hackl_minimal_2019,osawa_entanglement_2025}, as well as some theoretical developments such as extensions to fermionic cases~\cite{hackl_minimal_2019,hackl_bosonic_2021,Jonsson:2021lko} and scenarios involving multiple modes~\cite{yamaguchi_quantum_2020,KojiCapacity,yamaguchi_quantum_2021,Agullo:2024har}.
Also,  in the context of QFT in the lattice, Refs.~\cite{Natalie1,NatalieUVIR,Natalie2,Natalie3} introduced a procedure to identify the most entangled pair of modes localized in two non-overlapping spatial regions. This framework was subsequently implemented in lattice field theories across different spacetime dimensions making use of Gaussian quantum mechanics and phase space analysis.

Inspired by these results and building on them, in this work we introduce a framework to identify and isolate the ``correlation partners'' in Gaussian systems. Specifically, we ask: given a particular mode, which other degrees of freedom encode all its correlations and entanglement? Addressing this question is important for both conceptual and practical reasons. Conceptually, it sheds light on the structure of correlations in continuous-variable systems, clarifying how and when complex many-body states can be reduced to effective two-mode problems. Practically, it provides constructive tools to locate and extract entanglement resources, with potential applications in quantum communication, simulation, and the study of correlations in quantum fields.

We base our analysis on an extension of the phase-space methods commonly used for continuous-variable systems. Specifically, we shift the focus from the covariance matrix and the real phase space to the so-called \emph{restricted complex structure} and a complex phase space; see \cite{hackl_minimal_2019,hackl_bosonic_2021,Jonsson:2021lko} for previous work emphasizing some of these tools. Combined with techniques from symplectic geometry on complex vector spaces, we show that this framework provides a powerful, geometric, and basis-independent approach to unraveling the correlation structure of arbitrary Gaussian states.

In addition to further developing and promoting the complex phase-space formalism for Gaussian systems, this article employs these techniques to formalize and extend previous results for pure states \cite{HottaPartner,Trevison2019,NatalieUVIR,Natalie2}. In particular, we establish the existence and uniqueness of partner modes for generic Gaussian states. Furthermore, our framework naturally generalizes the concept of partner modes to mixed Gaussian states and multimode subsystems.

Our results are organized into four main parts. First, we provide a quick review of quantum mechanics in phase space, with an emphasis on Gaussian systems. This serves both as an introduction to the basic concepts used in the rest of this article and to set up  the notation. Second, we show that for pure Gaussian states, every correlated mode has a unique single-mode partner that fully captures its correlations with the rest of the system. 
Third, we prove that in mixed Gaussian states the partner concept naturally splits into two: a correlation partner, containing all classical and quantum correlations, and an entanglement partner, which is always at most single-mode and exists precisely when the state is non-PPT. Finally, we extend the construction of partners in both pure and mixed Gaussian states to multi-mode subsystems.

\section{Quantum linear systems, Gaussian states and the classical phase space\label{sec:defs}}

\subsection{Some basic elements of the phase space for finite dimensional bosonic linear systems}

Consider a quantum mechanical system with $N$ bosonic degrees of freedom. (Our discussion applies equally to distinguishable degrees of freedom.)
 The classical phase space $\Gamma$ is a $2N$-dimensional manifold equipped with a symplectic two-form $\Omega_{ab}$. This form is anti-symmetric and invertible; we denote the twice contravariant tensor that is its contraction-inverse by $\Omega^{ab}$, such that $\Omega^{ac}\Omega_{cb} = \Omega_{bc}\Omega^{ca} = \delta^a_b$.\footnote{For index contraction, we use the conventions $\Omega^{ac}\Omega_{cb} =\delta^a_b$ and $\gamma^a = \Omega^{ab} \gamma_b$. Other contractions can be derived from these ones using the anti-symmetry of $\Omega_{ab}$.
 }

The symplectic form induces a Poisson bracket between pairs of dynamical functions. For any two functions $f$ and $g$ on $\Gamma$:
\begin{equation}
\{f, g\} \coloneq \Omega^{ab} \partial_a f \partial_b g .
\end{equation}

Consider a set of 2N real functions $(r^1,  \ldots, r^{2N})= (x^1, p^1, \ldots, x^N, p^N)$ that satisfy canonical Poisson brackets\footnote{Note the distinction: the letters $a, b, c, \ldots$ denote abstract tensorial indices, while $i, j, k, \ldots$ label  components in the canonical coordinates $r^i$.}:
\begin{equation} 
\{r^i, r^j\} = \Omega^{ij},
\end{equation}
where $\Omega^{ij}$ represent the components of $\Omega^{ab}$ in these coordinates. This set is called a Darboux set of functions, and it constitutes a chart that assigns \textit{canonical} coordinates to points in an open subset of $\Gamma$. 

The matrix formed with the components $\Omega^{ij}$  is block-diagonal
\begin{equation}
    (\Omega^{ij})=\bigoplus_N\begin{pmatrix} 0 & 1 \\ -1 & 0 \end{pmatrix}
\end{equation}

We will focus on linear systems. For these systems, Darboux coordinates exist globally in $\Gamma$, giving $\Gamma$ the structure of a $2N$-dimensional vector space. Elements of $\Gamma$ can be represented as $2N$-component vectors, which we denote as $\gamma^a$.

The vector space structure allows us to identify $\Gamma$ with its tangent space at any given point, enabling us to ``pull down'' $\Omega_{ab}$ from the tangent space of $\Gamma$ to $\Gamma$ itself and define the symplectic product between elements in $\Gamma$: 
\begin{equation}  \label{sympprod} \Omega(\gamma,\gamma') = \sum_{I=1}^N \left(p_I x'_I - x_I p'_I\right),
\end{equation} 
where $(x_I,p_I)$ and $(x'_I,p'_I)$ denote the components of $\gamma^a$ and $\gamma'^a$ in the canonical basis. 

To transition to quantum theory, we complexify the phase space, transforming it into a $2N$-complex-dimensional phase space $\Gamma_{\mathbb{C}}$ by allowing all possible linear combinations of elements in $\Gamma$ with complex coefficients. The symplectic structure $\Omega_{ab}$ and its inverse $\Omega^{ab}$ extend to $\Gamma_{\mathbb{C}}$ by linearity.

Finally, in $\Gamma_{\mathbb{C}}$, it will be handy to use a complexified version of the symplectic product, defined as
\begin{equation}
\langle \gamma, \gamma' \rangle \coloneqq -\frac{\ii}{\hbar} \, \Omega(\gamma^{*}, \gamma'),
\end{equation}
where the star denotes complex conjugation\footnote{There is no unique way of defining the operation  ``complex conjugation'' within a complex vector space. The definition requires a choice of basis. Given such a choice, the operation acts by conjugating the components of vectors. In this paper, the preferred basis is determined by the canonical coordinates used to provide $\Gamma$ with the structure of a vector space.}.

Mathematically, the quadratic form $\langle \cdot,\cdot \rangle$  defines a Hermitian pseudo-inner product in $\Gamma_{\mathbb{C}}$, referred to as the complexified symplectic product (analogous to the Klein-Gordon product in scalar field theory). The reason it is a pseudo-inner product is that, due to the antisymmetry of $\Omega$, the product is not positive semidefinite. In particular, all real $\gamma$ are orthogonal to themselves, $\langle \gamma, \gamma \rangle = 0$ for all $\gamma \in \Gamma$, or in other words,  real $\gamma$ are ``null vectors'' of the complexified symplectic product. Furthermore, for any complex $\gamma$ and $\gamma'$, it is straightforward to verify that $\langle \gamma, \gamma' \rangle^{*}=- \langle \gamma^{*}, \gamma'^{*} \rangle = \langle \gamma' ,\gamma \rangle $. It follows that any $\gamma$ is orthogonal to $\gamma^*$, and that the ``norms'' of $\gamma$ and $\gamma^*$ differ by a sign: $\langle \gamma, \gamma \rangle=- \langle \gamma^{*}, \gamma^{*} \rangle$.

The linear transformations that leave the symplectic form invariant (and therefore preserve the volume of phase space) are called the symplectic (or linear canonical) transformations and they form a group called the symplectic group. Symplectic transformations can be understood as changes of canonical coordinates, since the symplectic form (and therefore the Poisson brackets) are preserved.

\subsection{Linear operators in the quantized theory}

The quantum counterpart of $(x^1, p^1, \ldots, x^N, p^N)$ is the vector of ``Darboux operators"  $\hat{\bm r} \equiv (\hat{x}^1, \hat{p}^1, \ldots, \hat{x}^N, \hat{p}^N)$, satisfying the canonical commutation relations:
\begin{equation} \label{ccr}
[\hat{r}^i, \hat{r}^j] = \ii \hbar \, \Omega^{ij}\openone.
\end{equation} 
Of special interest are operators obtained as \emph{linear combinations} of Darboux operators: $c_i\, \hat{r}^i$, with $c_i \in \mathbb{C}$. These linear operators can be thought of as ``elementary observables'' because the full algebra of observables is generated by taking linear combinations of their products. Restricting to $c_i \in \mathbb{R}$ yields self-adjoint  operators (the ones usually called ``observables'' in quantum mechanics); however, we will also consider non-self-adjoint operators, such as creation and annihilation operators, which are relevant to our discussion. Thus, each vector $\bm{c} \in \mathbb{C}^{2N}$ naturally defines a linear operator via $c_i\, \hat{r}^i$. 

It will be beneficial to organize the linear operators in a slightly different manner. We can associate a linear operator with each vector $\gamma^a$ in the complexified classical phase space $\Gamma_{\mathbb{C}}$ as follows:
\begin{equation} \label{mapgtop} 
\gamma \in \Gamma_{\mathbb{C}} \longrightarrow \hat{O}_{\gamma} = \ii\, \langle \gamma, \hat{\bm{r}} \rangle.
\end{equation}
In components, this expression becomes 
\begin{equation}
\hat{O}_{\gamma} = \hbar^{-1}\, \Omega( \gamma^*, \hat{\bm{r}}  )=\hbar^{-1} \Omega_{ij} \gamma^{* i} \hat{r}^j,
\end{equation}
showing that $\hat{O}_{\gamma}$ is simply a linear combination of the operators in $\hat{\bm{r}}$, with $\hbar^{-1} \Omega_{ij} \gamma^{* i}$ playing the role of the complex coefficients  $c_j$.

The constant $\hbar$ is introduced for convenience, so that all linear operators $\hat{O}_{\gamma}$ are dimensionless. The complex conjugation in $\gamma$ is a convention, which makes $\hat{O}_{\gamma}$ anti-linear in $\gamma$ (i.e., $\hat{O}_{(z \gamma)} = z^* \hat{O}_{\gamma}$ for any $z \in \mathbb{C}$). This definition also implies that $\hat{O}_{\gamma}^{\dagger} = \hat{O}_{\gamma^{*}}$, so that $\hat{O}_{\gamma}$ is self-adjoint when $\gamma$ is real.

The commutation relations between any two such operators simplify to:
\begin{equation} \label{comm}
[\hat{O}_{\gamma}, \hat{O}_{\gamma'}^{\dagger}] = \langle \gamma, \gamma' \rangle.
\end{equation}

\begin{tcolorbox}[breakable,colframe=black!35,
colback=yellow!2]

\textbf{Example:}

Consider a system made of two harmonic oscillators having the same frequency $\omega$ and mass $m$. The classical phase space is two-dimensional ($N=2$, two-mode system). The following phase space vectors:
\begin{equation}
\gamma_1 = -\begin{pmatrix}  \sqrt{\frac{\hbar}{m\omega}} \\ 0 \\ 0 \\ 0 \end{pmatrix}, \quad \gamma_2 = \begin{pmatrix} 0 \\ \sqrt{\hbar m\omega} \\ 0 \\ 0 \end{pmatrix}
\end{equation}
correspond to the operators:
\begin{align}
\nonumber\hat{O}_{\gamma_1} =& \ii \langle \gamma_1, \hat{\bm{r}} \rangle\\
=& \frac{1}{\hbar}\begin{pmatrix} - \sqrt{\frac{\hbar}{m\omega}} \!\!& 0 & 0 & 0 \end{pmatrix}\!
\begin{pmatrix}
0 & -1 & 0 & 0  \\
1 & 0 & 0 & 0  \\
0 & 0 & 0 & -1 \\
0 & 0 & 1 & 0  
\end{pmatrix}\!\!
\begin{pmatrix}
\hat{x}_1 \\
\hat{p}_1 \\
\hat{x}_2 \\
\hat{p}_2 \\
\end{pmatrix}\nonumber \\
=& \sqrt{\frac{1}{\hbar \, m\omega}}\, \hat{p}_1,
\end{align}
and
\begin{equation}
\hat{O}_{\gamma_2} = \ii \langle \gamma_2, \hat{\bm{r}} \rangle = \sqrt{\frac{ m\omega}{\hbar}}\, \hat{x}_1.
\end{equation}

Their commutator is:
\begin{equation}
[\hat{O}_{\gamma_1}, \hat{O}_{\gamma_2}] = \langle \gamma_1, \gamma_2 \rangle =- \, \ii\,  
\end{equation}

\end{tcolorbox}

\subsection{Subsystems}\label{sec:subsystems}

\subsubsection{Classical Theory}\label{subsec:calssicalsubsystems}
In general, classical systems (not necessarily linear) subsystems are in one-to-one correspondence with \emph{symplectic submanifolds} of the phase-space manifold $\Gamma$.

In contrast, for linear systems---namely, systems whose classical phase space is endowed with a vector-space structure and whose Hamiltonian is a quadratic polynomial in the canonical coordinates ${\bm r}$---\emph{linear subsystems} constitute a distinguished class of subsystems, as they preserve the underlying linear structure. These linear subsystems are in one-to-one correspondence with \emph{symplectic vector subspaces} of $\Gamma$. Their canonical coordinates are obtained simply by restricting the canonical coordinates ${\bm r}$ defined on the full phase space $\Gamma$ to the corresponding symplectic subspace. This linear structure underlies the covariance-matrix description of Gaussian states and the symplectic techniques employed throughout this work.

For example, for a systems of three harmonic oscillators with equal frequencies and masses, the subspace spanned by $\gamma_1=(- \sqrt{\frac{\hbar}{m\omega}},0,0,0,0,0)$ and $\gamma_2=(0,\sqrt{\hbar m\omega},0,0,0,0)$, i.e., $\Gamma_{\textsc a}=\mathrm{span}(\gamma_1,\gamma_2)$, is a symplectic subspace and constitutes the state space  of a one-degree-of-freedom subsystem (i.e., a single-mode). In contrast, the subspace $\mathrm{span}(\gamma_1,\gamma_3)$, with $\gamma_3=(0,0,0,0,- \sqrt{\frac{\hbar}{m\omega}},0)$, is not symplectic, since the restriction of $\Omega$ to this subspace produces the zero tensor (this follows from the fact that $\gamma_1$ and $\gamma_3$ are symplectically orthogonal).

It is essential to identify subsystems with subspaces of $\Gamma$, rather than with the specific basis vectors chosen to span the subspace. For instance, the pair of vectors $\gamma_1$ and $\gamma_2$ spans the same symplectic subspace as the pair $S_{\textsc a} \cdot \gamma_1$ and $S_{\textsc a} \cdot \gamma_2$, where $S_{\textsc a}$ is any element of the symplectic group associated with the restricted symplectic structure $\Omega_{\textsc a}$. Thus, the two pairs $(\gamma_1, \gamma_2)$ and $(S_{\textsc a} \cdot \gamma_1, S_{\textsc a} \cdot \gamma_2)$ are two different bases for the same subsystem.

The elements of the symplectic group restricted to the subsystem $A$ are termed ``system-local'' symplectic transformations. Using this terminology, we say that subsystems are invariant under system-local symplectic transformations.

Two subsystems $A$ and $B$ are said to be independent if they are symplectically orthogonal, i.e., $\langle \gamma_{\textsc a}, \gamma_{\textsc b} \rangle = 0$ for all $\gamma_{\textsc a} \in \Gamma_{\textsc a}$ and $\gamma_{\textsc b} \in \Gamma_{\textsc b}$. We will denote symplectic orthogonality among subsystems as $A\perp B$.

\subsubsection{Quantum theory} 
For quantized systems, an $N_{\textsc a}$-mode subsystem, with $N_{\textsc a} < N$, is defined in direct analogy with the classical case. Quantum subsystems are in one-to-one correspondence with (tensor product) Hilbert space {\em factors}.\footnote{Classically, $\Gamma = \Gamma_{\textsc a} \oplus \Gamma_{\bar{{\textsc a}}}$, while quantum mechanically, $\mathcal{H} = \mathcal{H}_{\textsc a} \otimes \mathcal{H}_{\bar{{\textsc a}}}$, where $\bar{A}$ is the $(N - N_{\textsc a})$-dimensional symplectic complement of $A$. Thus, $\mathcal{H}_{{\textsc a}}$ is a \emph{factor} of $\mathcal{H}$. \\
Note that, for an $N$-mode system, there are no canonical or preferred subsystems. In particular, there are infinitely many inequivalent ways to factor out the Hilbert space into single-mode subsystems:  $\mathcal{H} = \mathcal{H}_{{\textsc a}_1} \otimes \cdots \otimes \mathcal{H}_{{\textsc a}_N}.$}
In the algebraic approach to quantum theory, where the primary objects are algebras of observables rather than Hilbert spaces, subsystems are defined via \emph{subalgebras} of observables (see, e.g., \cite{Fewster2019AQFT}). 

There is a direct, one-to-one correspondence between classical subsystems and their quantum counterparts. The subalgebra of observables defining the quantum subsystem $A$ is the algebra generated by the linear operators $\hat{O}_{\gamma}$, with $\gamma$ restricted to $\Gamma_{\textsc a}$. It automatically follows that quantum subsystems are also invariant under local symplectic transformations. A Hilbert space $\mathcal{H}_{\textsc a}$ for a subsystem is obtained by representing the subsystem's quantum subalgebra.

In quantum mechanics, two subsystems $A$ and $B$ are independent if their subalgebras commute. Using \eqref{comm},  we observe that this  is guaranteed if the classical subsystems are independent, i.e., 
symplectically orthogonal, $A \perp B$.

\subsection{Gaussian states}\label{GaussianCov}

Gaussian states constitute a distinguished family of quantum states that admit a natural representation as a probability distribution in phase space. Namely, their Wigner functions are Gaussian distributions, which implies that the state is completely specified by the first and second statistical moments of this distribution (see, e.g., \cite{serafini2017quantum}). The description of Gaussian states, therefore, scales only linearly with the number of degrees of freedom, in contrast to the exponential complexity of generic quantum states. Although for this reason, they are arguably “classical-like” states, they are sufficiently rich to encompass many physically relevant scenarios. In particular, Gaussian states include ground and thermal states of quadratic Hamiltonians, as well as coherent and squeezed states which can display quantum entanglement.\footnote{In fact, a two mode squeezed state of two harmonic oscillators is the most entangled state that one can get at a given energy in the joint system, since the partial states are thermal and therefore the von Neumann entropy of the partial states is maximal at that constant energy.}

Mathematically, Gaussian states are particularly simple. In particular, all calculations involving Gaussian states---whether pure or mixed---can be reduced to straightforward operations involving matrix and vector multiplication in the classical phase space.  This subsection introduces a minimal review of the description of Gaussian states using this phase space formalism that can be found in literature~\cite{serafini2017quantum}, with some extra emphasis on the extension to the complexified phase space.

Given any quantum state---Gaussian or not, pure or mixed---one can associate a covector $\mu_{a}$ in the classical phase space $\Gamma$, defined as follows. Let $\hat{\rho}$ denote the density operator representing the state. Recall that a covector is a linear map from vectors to scalars. We define the covector $\mu_{\ a}$ by specifying its action on any $\gamma \in \Gamma$ as:
\begin{equation}\label{muPhaseSpace}
\mu(\gamma) \equiv \text{Tr}[\hat \rho\, \hat O_{\gamma}],\  \ \gamma\in\Gamma.
\end{equation}
Thus, the real number $\mu(\gamma)$ is the expectation value of the operator $\hat O_{\gamma}$ associated with $\gamma$. 

It is convenient to define the \emph{centered} version (zero expectation value) of the linear operator $\hat{O}_{\gamma}$, defined as:
\begin{equation} 
\hat{\overline{O}}_{\gamma} \equiv \hat{O}_{\gamma} - \mathrm{Tr}[\hat{\rho} \hat{O}_{\gamma}]\openone.
\end{equation} 
We will adopt this notation from now on.

In the same philosophy as the covector of means $\mu$,  we can also define the following twice-covariant tensor $\sigma_{ab}$---which encodes the second moments of the state---by specifying its action on any two vectors $\gamma$ and $\gamma'$: 
\begin{equation}\label{CovDefPhaseSpace}
\sigma(\gamma,\gamma') \coloneqq \text{Tr}[\hat \rho \, \{\hat{\overline O}_{\gamma}, \hat{\overline O}_{\gamma'} \}], \ \ \gamma,\gamma'\in\Gamma,
\end{equation}
where the curly brackets denote the anti-commutator, $\{\hat{\overline O}_{\gamma}, \hat{\overline O}_{\gamma'} \} = \hat{\overline O}_{\gamma} \hat{\overline O}_{\gamma'} + \hat{\overline O}_{\gamma'} \hat{\overline O}_{\gamma}$. In other words, $\sigma(\gamma,\gamma')$ is the expectation value of the symmetrized product of the centered operators associated with $\gamma$ and $\gamma'$.

We can extend the action of $\mu$ and $\sigma$  to the complexified phase space $\Gamma_{\mathbb{C}}$ as follows:
\begin{align} \mu(\gamma) &= \text{Tr}[\hat \rho \, \hat O^{\dagger}_{\gamma}] , \ \ \gamma\in\Gamma_{\mathbb{C}} , \\ \sigma(\gamma,\gamma') &= \text{Tr}[\hat \rho \, \{\hat{\overline O}^{\dagger}_{\gamma}, \hat{\overline O}^{\dagger}_{\gamma'}\}],  \ \ \gamma,\gamma'\in\Gamma_{\mathbb{C}}. \end{align} 
With these definitions, we built a covector $\mu_{a}$ and a twice-covariant tensor $\sigma_{ab}$ in $\Gamma_{\mathbb{C}}$, respectively.\footnote{The adjoint conjugates appearing in these definitions are introduced to ensure that $\mu$ is linear in $\gamma$, and $\sigma$ is bilinear in $\gamma$ and $\gamma'$. Without the adjoint conjugates, they would be anti-linear, and thus would not define a covector and a rank-two covariant tensor, respectively.}

As mentioned before, for Gaussian states $\mu$ and $\sigma$ exhaust all the information in the state; that is, they \emph{fully characterize} it. 
In fact, the Wigner function (and therefore the full density operator) can be reconstructed. Given the components of $\mu$ and $\sigma$ in a canonical basis, the Wigner function of a Gaussian state is 
\begin{equation}
W(\boldsymbol{R})=\frac{1}{\pi^n \sqrt{\operatorname{det}({\sigma})}} \exp \left(-(\boldsymbol{R}-{\mu})_{i}(\boldsymbol{R}-{\mu})_{j} ({{\sigma}}^{-1})^{ij}\right),
\end{equation}
where $\boldsymbol{R}=( R_1,\cdots,R_{2N})$ denote coordinates in any Darboux basis and ${\mu}_i$ and $(\sigma^{-1})^{ij}$ the components of $\mu$ and the inverse of $\sigma$ in that basis.
The density operator matrix elements in the position representation can be obtained by direct Fourier transform of the Wigner function~\cite{serafini2017quantum}.

The tensor of covariances $\sigma$ has the following properties:
\begin{enumerate}
    \item $\sigma$ is symmetric, i.e., $\sigma(\gamma, \gamma') = \sigma(\gamma', \gamma)$ for all $\gamma, \gamma' \in \Gamma_{\mathbb{C}}$.
    
    \item $\sigma(\gamma^*, \gamma) \pm \langle \gamma, \gamma \rangle \geq 0$ for all $\gamma \in \Gamma_{\mathbb{C}}$.

    \item When restricted to $\Gamma$, $\sigma$ is positive definite: $\sigma(\gamma, \gamma) > 0$ for all nonzero $\gamma$ in $\Gamma$.

\end{enumerate}

Properties 1 and 3 indicate that $\sigma$ defines an inner product on $\Gamma$, which implies that any quantum state defines a metric in the classical phase space. For this reason, we refer to $\sigma$ as the {\em covariance metric}. A more common term is the covariance matrix, which we reserve for the components $\sigma_{ij}$ of $\sigma$ in a given basis.  Property 2 follows from the positivity of $\hat \rho$, i.e. $\hat \rho\geq 0$---which leads to the uncertainty principle---and is usually stated using the components of $\sigma$ and $\Omega$ as $\sigma +\ii \Omega^{-1}\geq 0$. The proof of these properties, as well as many examples of application can be found in the literature (see, for example, \cite{notesQT,adesso2007entanglementgaussianstates,Serafini2017,demarie2012}), but for completion we include them in Appendix~\ref{Proof1}.

\subsection{Gaussian states and restricted complex structures}

This subsection introduces an object closely related to $\sigma$, namely a restricted complex structure $J$. The relationship between $J$ and $\hat{\rho}$ will be instrumental for the rest of this article. Complex structures have played an important role in quantum field theory in curved spacetime \cite{Ashtekar:1975zn}, and their relevance to Gaussian quantum information has been nicely highlighted in \cite{hackl_bosonic_2021}. Our presentation here follows this trend.

The object of interest, $J$, is obtained by raising one index of $\sigma_{ab}$ with the symplectic structure:
\begin{equation}
J^a_{\ b} \coloneqq -\hbar \, \Omega^{ac} \sigma_{cb} \, ,
\end{equation} 
or, equivalently,  $J^a_{\ b} \coloneqq \hbar^{-1} \sigma^{ac} \Omega_{cb}$ (the indices of $\sigma_{ab}$ and other tensors are raised and lowered using $\Omega_{ab}$ and its inverse). Since $\Omega$ is fixed, $J$ and $\sigma$ can be considered as containing the same information. While $\sigma$ is a twice-covariant tensor, $J$ is a linear map in $\Gamma_{\mathbb{C}}$, meaning its action on a vector produces another vector. This makes it possible to consider the eigenvalues and eigenvectors of $J$. The linear map $J$ is real, in the sense that $(J \gamma)^{*} = J \gamma^{*}$ for all $\gamma \in \Gamma_{\mathbb{C}}$. This follows from the reality of $\sigma$ and $\Omega$. The restricted complex structure has the following properties:

\begin{enumerate}

\item The relation between $J$ and $\sigma$ can be inverted, yielding: $\sigma(\cdot,\cdot) =-  \hbar^{-1} \Omega(\cdot,J \cdot)$. 

\item $\Omega(\cdot,J\cdot)=-\Omega(J\cdot,\cdot)$.

This allows us to write the relation between $\sigma$, $\Omega$ and $J$ as $\sigma(\cdot,\cdot)=\hbar^{-1}\Omega(J\cdot, \cdot)$.

\item $J$ is diagonalizable, with purely imaginary eigenvalues of the form $\pm \ii \, \nu_I$, with $\nu_I \in \mathbb{R}_+$, \mbox{$I \in\{ 1, \dots, N\}$}. The $N$ real numbers $\nu_I$ are commonly referred to as the ``symplectic eigenvalues of $\sigma$''.

\item $\nu_I \geq 1$ for all $I$, which is equivalent to saying that $J^2 \leq -\mathbb{I}$.

\item The eigenvectors of $J$ appear in pairs of complex conjugate vectors $e_I, e_I^{*}$, for $I = 1, \dots, N$. (The real and imaginary parts of the eigenvectors $e_I$ are sometimes referred to as the ``normal'' or ``Williamson'' modes of $\sigma$.) All together, and after normalization, the set of eigenvectors of $J$ forms a symplectic-orthonormal basis in $\Gamma_{\mathbb{C}}$. Specifically: 
\begin{equation} \langle e_I|e_J\rangle =\delta_{IJ}\, , \ \ \ \langle e^*_I|e^*_J\rangle =-\delta_{IJ}\, , \ \ \ \langle e_I|e^*_J\rangle =0\, \end{equation} 
for $I,J=1,\cdots N$.
\end{enumerate}

The proof of these properties can be found in Appendix~\ref{ProofJ}.

The purity of a Gaussian state $\hat \rho$, defined as $P[\hat \rho]={\rm Tr}[\hat \rho^2]$, is equal to \begin{equation} \label{purity} P[\hat \rho]=\frac{1}{ \sqrt{\det J}}=\prod_{I=1}^N \frac{1}{\nu_I}\, .\end{equation} 
A Gaussian state $\hat \rho$ is pure if and only if $\nu_I=1$ for all $I$. Equivalently,  $\hat \rho$ is pure if and only if $J^2=-\openone$.

Linear maps whose square is equal to minus the identity are called complex structures. Hence, pure Gaussian states define a complex structure in the classical phase. When $J$ is such that $J^2<-\mathbb{I}$, we call it, following \cite{hackl_bosonic_2021}, a restricted complex structure. Therefore, while pure states define complex structures,  mixed states define restricted complex structures.

One can further check that, if $\hat \rho$ is pure:

\begin{enumerate}
\item $J$ defines an orthogonal transformation: $\sigma(J\cdot,J\cdot)=\sigma(\cdot,\cdot)$.
\item $J$ is a symplectic transformation: $\Omega(J\cdot,J\cdot)=\Omega(\cdot,\cdot)$.
\item $\sigma^{ab}$, obtained by raising the two indices of $\sigma_{ab}$ using $\Omega^{ab}$, coincides with the inverse of $\sigma_{ab}$, i.e., $\sigma^{ac}\sigma_{cb}=\delta^{a}_{b}.$

\end{enumerate}

\subsection{Reduced States and Correlations}

Consider a bipartite linear system, i.e., a system composed of two subsystems, $A$ and $B$. The classical phase space has the form $\Gamma_{\mathbb{C}} = \Gamma_{\textsc a} \oplus \Gamma_{\textsc b}$ (from now on, $\Gamma_I$, $I = A, B, \ldots$, denote complex vector spaces, although we omit the symbol $\mathbb{C}$ to lighten the notation). Let $N_{\textsc a}$ and $N_{\textsc b}$ be the number of modes within each subsystem.  The Hilbert space of the system is  $\mathcal{H} = \mathcal{H}_{\textsc a} \otimes \mathcal{H}_{\textsc b}$. 

A Darboux set of operators can be obtained by combining the corresponding sets within each subsystem, yielding $\hat{\bm r} = (\hat{x}^1_{\textsc a}, \hat{p}^1_{\textsc a}, \cdots, \hat{x}^{N_{\textsc a}}_{\textsc a}, \hat{p}^{N_{\textsc a}}_{\textsc a}, \hat{x}^1_{\textsc b}, \hat{p}^1_{\textsc b}, \cdots, \hat{x}^{N_{\textsc b}}_{\textsc b}, \hat{p}^{N_{\textsc b}}_{\textsc b})$.

Let $\mu$ and $\sigma$ represent the mean and covariance metric of a Gaussian state $\hat \rho$.
The reduced state describing subsystem $A$, $\hat \rho_{\textsc a} ={\rm Tr}_{\textsc b} \hat \rho$, is a Gaussian state with mean and covariance metric defined by the restrictions of $\mu$ and $\sigma$ to $\Gamma_{\textsc a}$. We denote these restrictions as $\mu_{\textsc a}$ and $\sigma_{\textsc a}$, respectively.  Thus, for a bipartite system, $\mu_{\textsc a}$ takes the form $\mu = \mu_{\textsc a} \oplus \mu_{\textsc b}$. On the contrary, $\sigma$ is not, in general, equal to $\sigma_{\textsc a} \oplus \sigma_{\textsc b}$. The difference
\begin{equation}
C^{\rm corr} \equiv \sigma - \sigma_{\textsc a} \oplus \sigma_{\textsc b},
\end{equation}
is a tensor encoding the correlations present in the state $\hat \rho$ between the two subsystems. If $C^{\rm corr} = 0$, the subsystems are uncorrelated, and $\hat \rho$ is a product state of the form $\hat \rho = \hat \rho_{\textsc a} \otimes \hat \rho_{\textsc b}$.

The union of Darboux bases in $\Gamma_{\textsc a}$ and $\Gamma_{\textsc b}$, constitutes a Darboux basis in $\Gamma_{\textsc a} \oplus \Gamma_{\textsc b}$. The matrix representation of $\sigma$ in any such basis takes the form:
\begin{equation}
\sigma = \begin{pmatrix} \sigma_{\textsc a} & C \\ C^{\top} & \sigma_{\textsc b} \end{pmatrix}, 
\end{equation}
where $\sigma_{\textsc a}$ and $\sigma_{\textsc b}$ are $2N_{\textsc a} \times 2N_{\textsc a}$ and $2N_{\textsc b} \times 2N_{\textsc b}$ matrices, respectively, containing the components of the reduced covariance metrics. $C$ is a $2N_{\textsc a} \times 2N_{\textsc b}$ correlation matrix  which are the only possibly non-zero components of $C^{\rm corr}$.

\section{Partner modes for finite-dimensional  Gaussian systems}

Consider an $N$-mode system prepared in an arbitrary Gaussian state $\hat \rho$, and let $A$  be a single-mode subsystem. The subsystem $A$ is generally correlated and entangled with the remaining $N-1$ modes in the system. In this section, we aim to address the following question: is it possible to find a single-mode subsystem, independent of $A$, that encodes all correlations and entanglement with $A$? We will call such a hypothetical mode {\em the partner of $A$}, and denote it by $A_p$. This partner mode would be composed of a specific combination of the $N-1$ modes of the system that are independent of $A$, with each mode contributing with some weight. These weights would inform us about the distribution of entanglement and correlations with $A$ within the system.

In subsection \ref{subsec:partnerpure}, we will construct the partner mode from the complex structure of the system for a pure Gaussian state. Existence and uniqueness are proven in appendices \ref{app:ExistenceProof_PureState} and \ref{app:UniquenessProof_PureState}, respectively. Subsection \ref{subsec:partnermixed} shows that, if $\hat \rho$ is mixed and Gaussian, the entanglement with $A$ can always be encoded in a unique single-mode subsystem---the ``entanglement partner'' of $A$. On the other hand, the correlations with $A$ cannot always be encoded in a single mode  when $\hat \rho$ is mixed---the ``correlation partner'' of $A$ exists and is unique but is generally composed of more than one mode. Existence and uniqueness  of the correlation and entanglement partner are proven in appendices \ref{app:ExistenceUniquenessProof_CorrelationPartner}, \ref{app:proof_symplectic_subspace_ep} and \ref{app:UniquenessProof_EntanglementPartner}.

Before proceeding, we clarify the scope of these results. Throughout this article, we restrict our attention to subsystems that respect the linear symplectic structure introduced in Sec.~\ref{sec:subsystems}. In the quantum theory, such subsystems are characterized by the operator subalgebras generated, through linear combinations and products, by the linear observables associated with symplectic subspaces of the phase space (see the beginning of subsection \ref{subsec:calssicalsubsystems} for the definition of classical linear subsytems). This restriction reflects the distinguished role of the linear structure in the bosonic systems considered here: it underlies the canonical commutation relations, covariance matrices, Gaussianity, and quadratic Hamiltonians. From this perspective, linear symplectic subsystems constitute the physically natural class of subsystems, and all existence and uniqueness results presented in this work are to be understood within this framework. If one were instead to consider more general subsystems---for example, those generated by nonlinear combinations of the canonical operators $\hat{\bm r}$---the partner need no longer be unique.\footnote{The authors are grateful to Justin Wilson for pointing out this possibility.}

The rest of this section proves these assertions and provides an algorithm to construct entanglement partners and correlation partners for arbitrary Gaussian states.

We begin defining, for a generic Gaussian state, the concept of uncorrelated subsystem.\\ 

 \begin{definition}[Uncorrelated subsystem]
Let $A$ be an $N'$-mode subsystem  of a  system containing $N$ modes, with $N' < N$. We say that $A$ is uncorrelated if
\begin{equation} \label{uncorr}
\mathrm{Tr}\left[\hat \rho\, \hat{\overline O}_{\gamma}\hat{\overline O}_{\gamma'}\right] = 0 ,
\quad \forall \gamma \in \Gamma_A,\ \forall \gamma' \in \Gamma_{\bar A},
\end{equation}
where $\bar A$ denotes the symplectic orthogonal complement of $A$ in $\Gamma_{\mathbb{C}}$
(see Appendix~\ref{app:projectors} for the definition of symplectic orthogonal complement). 

\end{definition}

Note that \eqref{uncorr} is equivalent to the condition that the expectation value of product of operators equals the product of the expectation values of each operator---i.e., the familiar definition of correlations:
\begin{equation}
\mathrm{Tr}[\hat \rho\, \hat O_{\gamma} \hat O_{\gamma'}] = \mathrm{Tr}[\hat \rho\, \hat O_{\gamma}] \cdot \mathrm{Tr}[\hat \rho\, \hat O_{\gamma'}].
\end{equation}

If $A$ is uncorrelated, then the state $\hat \rho$ factorizes as a product:
\begin{equation}
\hat \rho = \hat \rho_{\textsc a} \otimes \hat \rho_{\bar {\textsc a}}.
\end{equation}

$A$ is said to be correlated  when it is not uncorrelated. \\ 

\begin{tcolorbox}[breakable,colframe=black!35,
colback=blue!2]

 \begin{proposition}[Uncorrelated subsystems]
A subsystem $A$ is uncorrelated in the Gaussian state $\hat \rho$ if and only if the (possibly restricted) complex structure $J$ leaves $\Gamma_{\textsc a}$ invariant:
\begin{equation}
\label{eq:uncorrelated}
J \Gamma_{\textsc A} = \Gamma_{\textsc A} .
\end{equation}
Equivalently,
\begin{equation}
[\Pi_{\textsc A}, J] = 0,
\end{equation}
where $\Pi_{\textsc A}$ denotes the symplectic projector onto $\Gamma_{\textsc A}$.
\end{proposition}

\end{tcolorbox}
Here, a symplectic projector denotes a projection operator onto a symplectic subspace; see Appendix~\ref{app:projectors} for its mathematical definition.

\begin{proof}
The proof is straightforward: Let $\gamma \in \Gamma_{\textsc a}$ and $\gamma' \in \Gamma_{\bar{\textsc{a}}}$. Then:
\begin{equation}\label{TPF_decomp_sym_antisym}
\mathrm{Tr}\left[\hat \rho\, \hat{\overline{O}}_{\gamma} \hat{\overline{O}}_{\gamma'}\right] 
= \frac{1}{2} \left( \sigma(\gamma^{*}, \gamma'^{*}) + \langle \gamma, \gamma'^{*} \rangle \right)
= \frac{1}{2} \sigma(\gamma^{*}, \gamma'^{*}),
\end{equation}
because the product $\langle \gamma, \gamma'^{*} \rangle = 0$ due to symplectic orthogonality between $\Gamma_{\textsc a}$ and $\Gamma_{\bar{\textsc{a}}}$. Thus, $A$ is uncorrelated if and only if
\begin{equation}
\sigma(\gamma^*, \gamma'^*) = 0 \quad \forall \gamma \in \Gamma_{\textsc a},\, \gamma' \in \Gamma_{\bar{\textsc{a}}},
\end{equation}
which implies that the covariance matrix $\sigma$ decomposes as a direct sum:
\begin{equation}
\sigma = \sigma_{\textsc a} \oplus \sigma_{\bar {\textsc a}}.
\end{equation}
This in turn implies that the complex structure $J$ also decomposes:
\begin{equation}
J = J_{\textsc a} \oplus J_{\bar {\textsc a}},
\end{equation}
proving the direct implication.

For the converse, if $J \Gamma_{\textsc a} = \Gamma_{\textsc a}$, then $\Omega(J\gamma, \gamma'^*) = 0 \quad , \forall \, \gamma \in \Gamma_{\textsc a}$, $\gamma' \in \Gamma_{\bar {\textsc a}}$---because $J\gamma\in \Gamma_{\textsc a}$ is symplectic orthogonal to all vectors $\gamma'$ in $\Gamma_{\bar {\textsc a}}$. Since   $\hbar^{-1}\Omega(J\gamma^*, \gamma'^*) =\sigma(\gamma^*,\gamma'^*)$,  it follows $\mathrm{Tr}\left[\hat \rho\, \hat{\overline{O}}_{\gamma} \hat{\overline{O}}_{\gamma'}\right] =\sigma(\gamma^*,{\gamma'}^{*})=0$ and
    $A$ is uncorrelated. 
\end{proof}

\subsection{Pure Gaussian states}\label{subsec:partnerpure}

Let $\hat \rho$ be a pure Gaussian state, so $J^2 = -\mathbb{I}$. If a subsystem $A$ is uncorrelated ---so that $J = J_{\textsc a} \oplus J_{\bar {\textsc a}}$--- it automatically follows that $J_{\textsc a}^2 = -\mathbb{I}_{\textsc a}$, with $\mathbb{I}_{\textsc a}$ the identity map on $\Gamma_{\textsc a}$. This is equivalent to saying that, if $\hat \rho = \hat \rho_{\textsc a} \otimes \hat \rho_{\bar {\textsc a}}$ and $\hat \rho$ is pure, then $\hat \rho_{\textsc a}$ must also be pure.\\

\begin{tcolorbox}[breakable,colframe=black!35,
colback=blue!2]

\begin{proposition} [Partner subsystem]\label{prop:partner}
Let $A$ be a single-mode subsystem. If $A$ is correlated when the system is prepared in the pure Gaussian state $\hat{\rho}$, its partner subsystem $A_p$ is defined by
\begin{equation}\label{singl_corr_partner_formula}
\Gamma_{\textsc{a}_p} = \Pi^\perp_{\textsc a} (J \Gamma_{\textsc a})\, ,
\end{equation} 
where $\Pi^\perp_{\textsc a}\equiv(\mathbb{I}-\Pi_{\textsc a})$.  When $A$ is correlated, the partner mode exists and is unique.
\end{proposition}

\end{tcolorbox}

\begin{proof}
 To prove that $A_p$ is the partner of $A$, we will show that the subsystem $A \oplus A_p$ is uncorrelated — this implies that $A_p$ encodes all correlations with $A$ and that the reduced state $\hat{\rho}_{\textsc{a}\textsc{a}_p}$ is pure (so $A_p$ purifies $A$).  
Following the definition of an uncorrelated subsystem, we need to prove that $J$ leaves $\Gamma_{\textsc a} \oplus \Gamma_{{\textsc a}_p}$ invariant. 

Note first that the sum of subspaces $ \Gamma_{\textsc a} \oplus \Gamma_{{\textsc a}_p} $ can be rewritten as
\begin{equation} \label{abc}
\Gamma_{\textsc a} \oplus \Gamma_{{\textsc a}_p} = \Gamma_{\textsc a} \oplus \Pi^\perp_{\textsc a} J \Gamma_{\textsc a} = \Gamma_{\textsc a} + J \Gamma_{\textsc a},
\end{equation}
where the last equality follows from the fact that the symplectic projector $ \Pi^\perp_{\textsc a} $ in the second term is innocuous: it merely ensures that the two terms are disjoint. In other words, $ \Gamma_{\textsc a} \oplus \Pi^\perp_{\textsc a} J \Gamma_{\textsc a} $ is a direct sum, whereas $ \Gamma_{\textsc a} + J \Gamma_{\textsc a} $ denotes the same subspace without enforcing disjointness.

With this:
\begin{equation}
J(\Gamma_{\textsc a} \oplus \Gamma_{A_p}) = J(\Gamma_{\textsc a} + J \Gamma_{\textsc a}) = J \Gamma_{\textsc a} + \Gamma_{\textsc a} =\Gamma_{{\textsc a}_p} \oplus \Gamma_{{\textsc a}},\label{eq:uncorrelated_pure}
\end{equation} 
where we have used $J^2 = -\mathbb{I}$ in the second equality, and used \eqref{abc} in the last equality. 

Proving the existence of the partner amounts to showing that
$\Gamma_{{\textsc a}_p}$ is a symplectic subspace, thereby defining a
single-mode subsystem. This is proved in Appendix
\ref{app:ExistenceProof_PureState}.

To prove uniqueness, it is sufficient to show that, if $A$ is correlated and there exist two
single-mode subsystems, $A_{{\textsc a}_p}$ and $A_{{\textsc a}_p'}$, such
that both
\begin{equation}
\Gamma_{\textsc a}\oplus \Gamma_{{\textsc a}_p}
\qquad\text{and}\qquad
\Gamma_{\textsc a}\oplus \Gamma_{{\textsc a}_p'}
\end{equation}
are uncorrelated, then necessarily
\begin{equation}
\Gamma_{{\textsc a}_p}=\Gamma_{{\textsc a}_p'}.
\end{equation}
This statement is proved in Appendix
\ref{app:UniquenessProof_PureState}.

\end{proof}

The definition of partner mode does not require specifying any basis. However, in application, it is useful to characterize $A_p$ using a basis. This can be done as follows: Let $\gamma_{\textsc a}^{(1)},\gamma_{\textsc a}^{(2)}$ be any basis in $\Gamma_{\textsc a}$. Then
\begin{equation}
  N\, \Pi_{\textsc a}^{\perp}(J\gamma_{\textsc a}^{(1)})\, , \quad N\, \Pi_{\textsc a}^{\perp}(J\gamma_{\textsc a}^{(2)})\, \end{equation}
is a basis for the partner subsystem, where $N=1/\sqrt{{\rm det} J_{\textsc a}-1}$ is a normalization factor.

The ``partner formula'' for pure states in Proposition~\hyperlink{prop:partner}{2} is equivalent to an expression previously reported in \cite{HottaPartner,Trevison2019}. Thus, this part of the paper essentially gives a formal characterization of the partner formula. Nevertheless, Eqn.~\eqref{singl_corr_partner_formula} provides a more concise and mathematically transparent expression (see Appendix~\ref{app:QIC_complex_structure} for more details on the relation between the operator-based results and the phase-space approach),  which has allowed us to prove the existence and uniqueness of the partner (see Appendices \ref{app:ExistenceProof_PureState} and \ref{app:UniquenessProof_PureState}). Importantly,  expression \eqref{singl_corr_partner_formula} serves as the foundation for extending the notion of partners to mixed Gaussian states---a task we tackle in the next subsection. The usefulness of the phase-space approach in the context of purification partners for pure Gaussian states was also pointed out in \cite{hackl_minimal_2019}, which presented a partner formula for bosonic and fermionic systems within a unified framework. In that work, the standard form of the covariance matrix was employed to derive the formula. In contrast, we use a manifestly basis-independent formalism based on the complex structure of the quantum state and symplectic projectors, without requiring  specification of a basis or the use of the standard form of the covariance matrix. We also remark that Proposition~\hyperlink{prop:partner}{2} can be extended to fermionic systems (see Appendix~\ref{app:fermion}). In the main body of this article we restrict the discussion to bosonic systems.

\begin{tcolorbox}[breakable,colframe=black!35,
colback=yellow!2]
{\bf Example:}

Consider a three-mode system in a pure Gaussian state with complex structure $J$, and let $\{ e_1, e_1^*, e_2, e_2^*, e_3, e_3^* \}$ denote the eigenvectors of $J$. (These eigenvectors uniquely characterize $J$.) Since the state is pure, the eigenvalues are $(\ii, -\ii,\ii,-\ii,\ii,-\ii)$, respectively. 

Consider the vector $\gamma_{\textsc a} = 2e_1 - \sqrt{3}e_3^*$. Then, $\Gamma_{A} = \mathrm{span}\left[\gamma_{\textsc a}, \gamma_{\textsc a}^*\right]$ is a symplectic subspace of $\Gamma$ and defines a correlated single-mode subsystem. 

A basis of the partner mode of $A$ is made of $(1 - \Pi_{\textsc a})J\gamma_{\textsc a}$ and its conjugate. 

The action of $J$ on $\gamma_{\textsc a}$ is 
\begin{equation}
J\gamma_{\textsc a} = \ii (2e_1 + \sqrt{3}e_3^*).
\end{equation}

On the other hand, the symplectic projector $\Pi_{\textsc a}$ onto $A$ can be written as
\begin{equation}
\Pi_{\textsc a} = \gamma_{\textsc a} \langle \gamma_{\textsc a}, \cdot \rangle - \gamma_{\textsc a}^* \langle \gamma_{\textsc a}^*, \cdot \rangle.
\end{equation}

With this:
\begin{equation}
\Pi^\perp_{\textsc a} (J\gamma_{\textsc a}) = \ii (12e_1 + 2\sqrt{3}e_3^*).
\end{equation}

Normalizing this vector, we conclude
\begin{equation}
\Gamma_{{\textsc a}_p} = \mathrm{span}\left[\sqrt{3}e_1 + 2e_3^*,\, \sqrt{3}e_1^* + 2e_3\right].
\end{equation}

One can check that $\Gamma_{{\textsc a}_p}$ actually defines the partner of $A$ by verifying that the restriction of $J$ to $\Gamma_{\textsc a} \oplus \Gamma_{{\textsc a}_p}$, given by $\Pi_{{\textsc a} \oplus {\textsc a}_p} J \Pi_{{\textsc a} \oplus {\textsc a}_p}$, is a complex structure — i.e., it has eigenvalues $\pm \ii$, thus representing a pure state.
\end{tcolorbox}

\subsection{Mixed Gaussian states: correlation partner and entanglement partner\label{subsec:partnermixed}}

The concept of partner modes becomes substantially richer when the system is prepared in a mixed Gaussian state $\hat{\rho}$. In this section, we show that, for any correlated single-mode subsystem $A$, although one can always identify a partner subsystem that captures all correlations with $A$, this partner subsystem may comprise more than one mode when $\hat{\rho}$ is mixed. This refers to {\em total} correlations, encompassing both classical correlations and quantum entanglement.

In contrast, if we restrict our attention to entanglement, it remains true that all entanglement with a single-mode subsystem $A$ can always be captured by another {\em single-mode} subsystem, even when $\hat{\rho}$ is mixed.

Consequently, for a mixed state $\hat{\rho}$, we must distinguish between the {\em correlation partner} and the {\em entanglement partner} subsystems, which we will denote as $A_{cp}$ and $A_{ep}$, respectively. The entanglement partner $A_{ep}$ consists of at most one mode, while the correlation partner $A_{cp}$ is generally a multi-mode subsystem. Moreover, $A_{ep}$ is always a subsystem of $A_{cp}$, since entanglement is one form of correlation.

When $\hat{\rho}$ is pure, the distinction disappears: $A_{cp}$ and $A_{ep}$ coincide. This is expected, since in pure states all possible correlations necessarily include entanglement. Another key difference in the mixed-state case is that neither $A_{cp}$ nor $A_{ep}$ necessarily purify $A$.

The remainder of this section is devoted to proving these results.

\subsubsection{Correlation partner of a single-mode subsystem}

Given a single-mode subsystem $ A $ and a Gaussian state $ \hat{\rho} $, we define the \emph{correlation partner} of $ A $ as the smallest subsystem---distinct from and independent of $ A $---such that the combined system $ A \oplus A_{cp} $ is uncorrelated. That is, $ A_{cp} $ contains all degrees of freedom in the system that are correlated with those in $ A $. Existence and uniqueness are proven in Appendix \ref{app:ExistenceUniquenessProof_CorrelationPartner}.

Let $J$ be the restricted complex structure associated with $\hat \rho$. Let $\pm\ii\, \nu_I$ be its eigenvalues, each with degeneracy $d_I$.
We denote by $\Pi^+_I$ and $\Pi^-_I$ the symplectic projectors onto the eigenspaces associated with the eigenvalues $\ii\, \nu_I$ and $-\ii\, \nu_I$, respectively.\footnote{Given a basis of eigenvectors $e^{(I)}_{\mu}$ and $e^{(I)*}_{\mu}$,  with $\mu=1,\ldots,d_I$, these projectors can be written as
\begin{equation}
\Pi^+_I = \sum_{\mu=1}^{d_I} e^{(I)}_{\mu} \langle e^{(I)}_{\mu}, \cdot \rangle, \quad \Pi^-_I = -\sum_{\mu=1}^{d_I} e^{(I)*}_{\mu} \langle e^{(I)*}_{\mu}, \cdot \rangle.\label{eq:projectors_basis_decomposition}
\end{equation}}

\begin{tcolorbox}[breakable,colframe=black!35,
colback=blue!2]

\begin{proposition} [Correlation partner subsystem]\label{prop:CPpartner}

The correlation partner of $A$ contains a number of modes equal to $\sum_I{\rm dim}(\Pi_I^+\Gamma_{\textsc a})-1$, and is defined by the symplectic subspace
\begin{equation}
\Gamma_{{\textsc a}_{cp}} = \Pi^\perp_{\textsc a}\left[ \bigoplus_I \left(\Pi^+_I\Gamma_{\textsc a} \oplus \Pi^-_I \Gamma_{\textsc a}\right)\right],
\end{equation}
where $\Pi^+=\sum_I \Pi^+_I$.
\end{proposition} 
\end{tcolorbox}

\begin{proof}
The proof follows from three observations:

\noindent 1. The subsystem $B$, with
\begin{equation}
\Gamma_{\textsc b} = \bigoplus_I \left(\Pi^+_I\Gamma_{\textsc a} \oplus \Pi^-_I \Gamma_{\textsc a}\right),
\end{equation}
is uncorrelated. This holds because $J$ leaves invariant both $\Pi^+_I\Gamma_{\textsc a}$ and $\Pi^-_I\Gamma_{\textsc a}$, for all $I$, as $\Pi^\pm_I$ project onto eigenspaces of $J$.

\noindent 2. $B$ contains $A$.\footnote{A simple analogy helps visualizing this: let $V$ be a one dimensional subspace of $\mathbb{R}^2$ and let  $\Pi_x$ and $\Pi_y$ be projectors onto the  $x$ and $y$ axes, respectively. Then,   $V$  is always a subset of $\Pi_xV \oplus \Pi_yV$.} This follows from the fact that the sum $\sum_I (\Pi_I^+ + \Pi_I^-)$ equals the identity map on $\Gamma$.

\noindent 3. $B$ is the smallest subsystem that satisfies points 1 and 2. That $\Gamma_B$ is the minimal subspace $J$-invariant also guarantees that it is unique (see appendix \ref{app:ExistenceUniquenessProof_CorrelationPartner}).

To determine the dimension of $\Gamma_{A_{cp}}$ we note that, because  $\Gamma_{\textsc a}$ is a symplectic subspace,  it can be written as $\Gamma_{\textsc a} = \mathrm{span}[\gamma_{\textsc a}, \gamma_{\textsc a}^*]$ for some $\gamma_{\textsc a}$ with $\langle \gamma_{\textsc a}, \gamma_{\textsc a} \rangle \neq 0$. Then for each $I$, $\Pi^+_I\Gamma_{\textsc a}$ and $\Pi^-_I\Gamma_{\textsc a}$ are complex conjugates of each other. Hence,
\begin{equation}
\mathrm{dim}(\Pi^+_I\Gamma_{\textsc a}) = \mathrm{dim}(\Pi^-_I\Gamma_{\textsc a}),
\end{equation}
which implies
\begin{equation}
\mathrm{dim}(\Gamma_{\textsc b}) = 2\,\sum_I \mathrm{dim}(\Pi_I^+\Gamma_{\textsc a}).
\end{equation} 
Therefore, because $B$ contains both $A_{cp}$ and $A$, and $\mathrm{dim}\Gamma_{\textsc a}=2$, we conclude
\begin{equation}
\mathrm{dim}(\Gamma_{{\textsc a}_p}) = \mathrm{dim}(\Gamma_{\textsc b}) -2= 2\,\sum_I \mathrm{dim}(\Pi_I^+\Gamma_{\textsc a}) - 2,
\end{equation}
or equivalently, $\Gamma_{{\textsc a}_p}$ describes a subsystem containing $\,\sum_I \mathrm{dim}(\Pi_I^+\Gamma_{\textsc a})-1$ modes.

\end{proof}

Although the definition of the correlation partner is basis-independent, in practice it is convenient to work with a basis. If $\gamma_{\textsc a}^{(1)}$ and $\gamma_{\textsc a}^{(2)}$ form a basis of $\Gamma_{\textsc a}$, then the vectors
\begin{equation}\nonumber 
\Pi_{\textsc a}^{\perp}(\Pi^+_I \gamma_{\textsc a}^{(1)}),\, \,  \Pi_{\textsc a}^{\perp}(\Pi^+_I \gamma_{\textsc a}^{(2)}),\, \, \Pi_{\textsc a}^{\perp}(\Pi^-_I \gamma_{\textsc a}^{(1)}),\, \, \Pi_{\textsc a}^{\perp}(\Pi^-_I \gamma_{\textsc a}^{(2)}),
\end{equation}
for all $I$, span $\Gamma_{A_p}$. Some care is needed in defining a basis, as these vectors may not all be linearly independent.

An illustrative example of the calculation of correlation partners can be found in Appendix~\ref{AppExampleCP}. 

\subsubsection{Entanglement partner of a single-mode subsystem}

As we saw above, when $\hat \rho$ is pure, the correlation partner and the entanglement partner are the same, and they both can be derived from $J$. For mixed $\hat \rho$, this is no longer true, and in order to find the entanglement partner, it is insufficient to just look at $J$ (or, equivalently, $\sigma$). We need entanglement-specific tools. 

 The key idea for identifying the entanglement partner of a single-mode subsystem in arbitrary Gaussian states is contained in \cite{Natalie2}. This reference introduced a clever strategy for localizing bound entanglement within a given $N$-mode subsystem of a larger Gaussian system, where $N$ is finite but otherwise arbitrary. The approach, based on the Positivity of the Partial Transpose (PPT) criterion, was proposed in \cite{Natalie2} without a formal proof, specifically in the context of quantum field theory in Minkowski spacetime with the field prepared in the vacuum state. 

In this section, we adopt some of the ideas put forward in \cite{Natalie2}, provide a formal proof, and identify their effectiveness in identifying entanglement partners of single-mode subsystems in arbitrary Gaussian many-body systems.

The key result from Appendix \ref{Negativity}---which reviews the definitions of the partial transpose and  negativity for Gaussian states---needed for this subsection is the following: Let $J$ be the restricted complex structure of a Gaussian state $\hat \rho$, and let $J^{T_{\textsc a}}$ denote its partial transposed with restrict to a single-mode subsystem $A$. (See Appendix \ref{Negativity} for the definition of partial transposed.) 
We will say $J$ is non-PPT$_{\textsc a}$, if any of the  symplectic eigenvalues of $J^{T_{\textsc a}}$ is smaller than one.\\

\begin{tcolorbox}[breakable,colframe=black!35,
colback=blue!2]

 \begin{proposition}[Entanglement partner subsystem]\label{prop:EPpartner}
If $J$ is non-PPT$_{\textsc a}$, there exists a unique single-mode subsystem, different and independent from $A$,  encoding all entanglement between $A$ and the rest of the system. This mode, called the entanglement partner of $A$,  and denoted $A_{ep}$, is obtained as 
\begin{equation}\Gamma_{{\textsc a}_{ep}}={\rm span}\left[\Pi_{\textsc a}^{\perp}e_1^{T_{\textsc a}},(\Pi_{\textsc a}^{\perp}e_1^{T_{\textsc a}})^*\right],\label{eq_entanglement_parnter}
\end{equation}
where $e_1^{T_{\textsc a}}$ is the only eigenvector of $J^{T_{\textsc a}}$ with symplectic eigenvalue smaller than one, $\nu^{T_{\textsc a}}_1<1$. 
\end{proposition}
\end{tcolorbox}

We will prove this statement by showing a) that the partner mode exists and that entanglement between $ A $ and its complement is equal to the entanglement between $ A $ and its entanglement partner $ A_{ep}$, b) that the entanglement partner of $A_{ep}$ is $A$, and c) the partner is unique. In doing so, we will demonstrate that $ A_{\text{ep}} $ captures all the entanglement between $ A $ and the rest of the system. \\

\begin{proof}[Proof of statement a)]
%\textbf{Proof of statement a)}\\

As argued in Appendix \ref{Negativity}, $J$ is non-PPT$_{\textsc a}$ if and only if the quadratic form 
\begin{equation} \label{pap}
\langle \,\cdot,(-\mathbb{I}-\ii\, J^{T_{\textsc a}})\, \cdot\, \rangle
\end{equation}
fails to be positive semi-definite  in $\Gamma_{\mathbb{C}}$.

First, $J^{T_{\textsc a}}$ has, at most, one symplectic eigenvalue smaller than one. The proof can be found in \cite{serafini2017quantum} (in a Little Lemma of section 7.4). For completeness, we review this proof using the complexified phase space and restricted complex structures in Appendix \ref{numbdim}.

Let $e^{T_{\textsc a}}_1$ denote the eigenvector of 
$J^{T_{\textsc a}}$ associated with such symplectic eigenvalue. Therefore, $e^{T_{\textsc a}}_1$ spans the one-dimensional vector subspace in which \eqref{pap} is negative definite.  

$e^{T_{\textsc a}}_1 $ cannot lie entirely within $ \Gamma_{\textsc a} $, since that would imply that one of the symplectic eigenvalues of $ J_{\textsc a} $ (the restriction of $ J $ to $ A $) is smaller than one ---which is not possible, because the reduced density operator $ \hat{\rho}_{\textsc a} $ is a bona fide state.

Because $e^{T_{\textsc a}}_1$ cannot lie entirely within $\Gamma_{\textsc a}$,  we conclude that the  subspace 
\begin{equation} \Gamma_{{\textsc a}_{ep}}={\rm span}\left[\Pi_{\textsc a}^{\perp}e_1^{T_{\textsc a}},(\Pi_{\textsc a}^{\perp}e_1^{T_{\textsc a}})^*\right],\end{equation}
cannot be empty whenever $J$ is non-PPT$_{\textsc a}$. 

Furthermore, in Appendix \ref{app:proof_symplectic_subspace_ep}, we prove that $\Gamma_{{\textsc a}_{ep}}$ is indeed a symplectic subspace and therefore defines a single-mode subsystem. This completes the proof of the existence of the entanglement partner.

It remains to be proven that the entanglement between $ A $ and the rest of the system is identical to the entanglement between $ A $ and its entanglement partner $ A_{ep} $. To show this, recall that the entanglement between $ A $ and its complement increases monotonically with $ |\nu^{T_{\textsc a}}_1| $ (see Appendix~\ref{Negativity}). Therefore, it suffices to show that the value of $|\nu^{T_{\textsc a}}_1|$ computed from the full complex structure $ J $, and the value obtained from the restriction of $ J $ to $ \Gamma_{\textsc a} \oplus \Gamma_{A_{ep}} $, are identical. 

The restriction of $ J $ to $ \Gamma_{\textsc a} \oplus \Gamma_{{\textsc a}_{ep}} $ is given by
\begin{equation}\label{J_AAep}
\Pi_{{\textsc a}{\textsc a}_{ep}}\, J\, \Pi_{{\textsc a}{\textsc a}_{ep}},
\end{equation}
where $ \Pi_{{\textsc a}{\textsc a}_{ep}} $ denotes the symplectic projector onto $ \Gamma_{\textsc a} \oplus \Gamma_{{\textsc a}_{ep}} $.

To compute $ \nu^{T_{\textsc a}}_1 $ from \eqref{J_AAep}, we apply the partial transposition $ T_{\textsc a} $:
\begin{equation}
\left(\Pi_{{\textsc a}{\textsc a}_{ep}}\, J\, \Pi_{{\textsc a}{\textsc a}_{ep}}\right)^{T_{\textsc a}} = \Pi_{{\textsc a}{\textsc a}_{ep}}\, J^{T_{\textsc a}}\, \Pi_{{\textsc a}{\textsc a}_{ep}},\label{JT_AAep}
\end{equation}
where we used that the operations $ T_{\textsc a} $ and projection onto $ \Gamma_{\textsc a} \oplus \Gamma_{{\textsc a}_{ep}} $ commute.  

Recall that $ \Gamma_{\textsc a} \oplus \Gamma_{{\textsc a}_{ep}} $ contains the symplectic subspace
\begin{equation}
\Gamma_1 \equiv  \mathrm{span}\left\{ e^{T_{\textsc a}}_1, (e^{T_{\textsc a}}_1)^* \right\}.
\end{equation}
Consequently, we can decompose the projector $ \Pi_{AA_{ep}} $ as sum of a projector onto $\Gamma_1$ plus a projector onto the symplectically orthogonal complement to $\Gamma_1$ within $\Gamma_{\textsc a} \oplus \Gamma_{A_{ep}}$:
\begin{equation}\label{PiAAep_decomp_Pi_1}
\Pi_{{\textsc a}{\textsc a}_{ep}} = \Pi_1 + \Pi_{{\textsc a}{\textsc a}_{ep}} \Pi_1^\perp,
\end{equation}
where $ \Pi_1 $ projects onto $ \Gamma_1 $, and $ \Pi_1^\perp $ denotes the projector onto the symplectically orthogonal complement to $ \Gamma_1 $ in the full phase space ---the presence of   $ \Pi_{{\textsc a}{\textsc a}_{ep}} $ multiplying $ \Pi_1^\perp $  ensures that the image lies within $ \Gamma_{\textsc a} \oplus \Gamma_{{\textsc a}_{ep}} $.

Applying \eqref{PiAAep_decomp_Pi_1}:
\begin{align}\label{secondt}
\nonumber &\Pi_{{\textsc a}{\textsc a}_{ep}}\, J^{T_{\textsc a}}\, \Pi_{{\textsc a}{\textsc a}_{ep}}\\ &= \Pi_1\, J^{T_{\textsc a}}\, \Pi_1 \oplus \left( \Pi_{{\textsc a}{\textsc a}_{ep}} \Pi_1^\perp \right)\, J^{T_{\textsc a}}\, \left( \Pi_{{\textsc a}{\textsc a}_{ep}} \Pi_1^\perp \right). 
\end{align}

We know from the discussion above that  $e^{T_{\textsc a}}_1$ is an eigenvector of $J^{T_{\textsc a}}$. Thus,  $ \Pi_1\, J^{T_{\textsc a}}\, \Pi_1 $ has eigenvectors $ e^{T_{\textsc a}}_1 $ and $ (e^{T_{\textsc a}}_1)^* $, with symplectic eigenvalue $ \nu_1^{T_{\textsc a}} < 1 $. 

On the other hand, the symplectic eigenvalue of the second term in  \eqref{secondt} must be greater than or equal to 1. Otherwise, $ \langle \cdot, (-\mathbb{I} - \ii J^{T_{\textsc a}})\, \cdot \rangle $ would not be positive semi-definite on the corresponding eigenspace. This would contradict the fact that $ \langle \cdot, (-\mathbb{I} - \ii J^{T_{\textsc a}})\, \cdot \rangle $ is positive semi-definite on the symplectic orthogonal complement of $\Gamma_1$. 
We conclude that the only symplectic eigenvalue of $ \Pi_{{\textsc a}{\textsc a}_{ep}}\, J^{T_{\textsc a}}\, \Pi_{{\textsc a}{\textsc a}_{ep}} $ smaller than 1 is precisely $ \nu_1^{T_{\textsc a}} $. This completes the proof.
\end{proof}

\begin{proof}[Proof of statement b)]

Let $J$ be the restricted complex structure of an $N$-mode Gaussian state with covariance metric $\sigma$ and $A$ a single-mode subsystem.  
Recall that with our convention \mbox{$J = -\hbar\,\Omega\,\sigma$}, the partially transposed complex structure with respect to $A$ is
\begin{equation}
J^{T_{\textsc a}}
= -\,\hbar\,\Omega\,T_{\textsc a}\,\sigma\,T_{\textsc a},
\end{equation}
where $T_{\textsc a}$ is the momentum flip on $A$ (see Appendix~\ref{Negativity}).

Let us assume the state is non-PPT with respect to $A$, i.e.\ the smallest symplectic eigenvalue of $J^{T_{\textsc a}}$ satisfies $\nu<1$. 
Equivalently, $J^{T_{\textsc a}}$ has a conjugate eigenpair
\begin{equation}\label{Eiegneq}
J^{T_{\textsc a}} e = \ii\nu e, \qquad
J^{T_{\textsc a}} e^\ast = -\ii\nu e^\ast,
\quad\text{with } \nu\in(0,1).
\end{equation}
Note that for a single-mode $A$, the subunity symplectic eigenvalue under partial transpose is unique.
For clarity, in this proof we simplify the notation by writing
$
e:=e^{T_{\textsc a}}_1
$ and
$
\nu :=\nu^{T_{\textsc a}}_1
$.

We have defined the partner of $A$ by
\begin{equation}
    \Gamma_{\textsc b} \;=\; \Gamma_{{\textsc a}_{\mathrm{ep}}}\;=\; \mathrm{span}\!\left\{\,\Pi_{\textsc a}^{\perp} e,\ \big(\Pi_{\textsc a}^{\perp} e\big)^\ast \right\},
\end{equation}
so that $e,e^\ast\in \Gamma_{\textsc a}\oplus\Gamma_{\textsc b}$.

\medskip

On $\Gamma_{{\textsc a} {\textsc b}}:=\Gamma_{\textsc a}\oplus\Gamma_{\textsc b}$, the restricted symplectic form is 
$
\Omega_{{\textsc a} {\textsc b}}=\Omega_{\textsc a}\oplus\Omega_{\textsc b}
$,
and $T_{\textsc a},T_{\textsc b}$ preserve $\Gamma_{{\textsc a} {\textsc b}}$.
One checks
\begin{equation}
T_{\textsc a}\,\Omega_{{\textsc a} {\textsc b}}\,T_{\textsc a} = (-\Omega_{\textsc a})\oplus \Omega_{\textsc b},
\quad
T_{\textsc b}\,\Omega_{{\textsc a} {\textsc b}}\,T_{\textsc b} =  \Omega_{\textsc a}  \oplus(-\Omega_{\textsc b}),
\end{equation}
hence
\begin{equation}
(T_{\textsc a} T_{\textsc b})\,\Omega_{{\textsc a} {\textsc b}}\,(T_{\textsc a} T_{\textsc b}) = -\,\Omega_{{\textsc a}{\textsc b}}.
\label{eq:signflip}
\end{equation}

\begin{lemma}[Restricted sign flip identity]\label{lemma:signflip}
For any symplectically orthogonal single-mode subsystems $A,B$, the following operator identity (restricted to $\Gamma_{{\textsc a}{\textsc b}}$) holds:
\begin{equation}
\label{eq:conj}
T_{\textsc a} T_{\textsc b} \, J^{T_{\textsc b}}\, T_{\textsc a} T_{\textsc b}
= -\,J^{T_{\textsc a}}.
\end{equation}
\end{lemma}
\begin{proof}
Using $J^{T_S} = -\,\hbar\,\Omega\,T_S\,\sigma\,T_S$, the commutativity of $T_{\textsc a}$ and $T_{\textsc b}$, and $T_S^2=\mathbb{I}_S$, we obtain
\begin{equation}
T_{\textsc a} T_{\textsc b} \, J^{T_{\textsc b}}\, T_{\textsc a} T_{\textsc b}
= -\,\hbar\,T_{\textsc a}\,(T_{\textsc b}\,\Omega_{{\textsc a}{\textsc b}}\,T_{\textsc b})\,\sigma\,T_{\textsc a}.
\end{equation}
Using Eq. \eqref{eq:signflip} on $\Gamma_{{\textsc a}{\textsc b}}$, this becomes
\begin{equation}
= \hbar\,\Omega_{{\textsc a}{\textsc b}}\,T_{\textsc a}\sigma T_{\textsc a}
= -\,J^{T_{\textsc a}},
\end{equation}
which proves \eqref{eq:conj}.
\end{proof}

\medskip
We can now apply the lemma to the PT eigenvector:
From the eigen-equation \eqref{Eiegneq} we have
$J^{T_{\textsc a}} e^\ast = -\ii\nu\, e^\ast$.
Now Identity \eqref{eq:conj} acting on $v=e^\ast$ yields
\begin{align}
&J^{T_{\textsc b}}\bigl(T_{\textsc a} T_{\textsc b} e^\ast\bigr)\\
\nonumber &\quad= -\, T_{\textsc a} T_{\textsc b}\,J^{T_{\textsc a}} e^\ast
= -\, T_{\textsc a} T_{\textsc b}(-\ii\nu e^\ast)
= \ii\nu\, \bigl(T_{\textsc a} T_{\textsc b} e^\ast\bigr).
\end{align}
Thus $e_{\textsc b} := T_{\textsc a} T_{\textsc b} e^\ast$ is an eigenvector of $J^{T_{\textsc b}}$ with eigenvalue $\ii\nu$ where $\nu<1$ and
$e_{\textsc b},e_{\textsc b}^\ast\in \Gamma_{\textsc a}\oplus\Gamma_{\textsc b}$, with $\Pi_{\textsc a} e_{\textsc b}\neq 0$ and $\Pi_{\textsc b} e_{\textsc b}\neq 0$.

By the same definition, the partner of $B$ is
\begin{equation}
\Gamma_{{\textsc b}_{\mathrm{ep}}}
\;=\;
\mathrm{span}\!\left\{\, \Pi_{\textsc b}^\perp e_{\textsc b},\ \big(\Pi_{\textsc b}^\perp e_{\textsc b}\big)^\ast \right\}.
\end{equation}
On $\Gamma_{{\textsc a} {\textsc b}}$, $\Pi_{\textsc b}^\perp$ coincides with $\Pi_{\textsc a}$, hence
$\Gamma_{{\textsc b}_{\mathrm{ep}}}=\mathrm{span}\{\Pi_{\textsc a} e_{\textsc b},(\Pi_{\textsc a} e_{\textsc b})^\ast\}=\Gamma_{\textsc a}$,
because $A$ is a single mode and $\Pi_{\textsc a} e_{\textsc b}\neq 0$.
Therefore $B_{\mathrm{ep}} = A$. This concludes the proof.\end{proof}

The proof of statement c) (uniqueness) is contained in Appendix \ref{app:UniquenessProof_EntanglementPartner}.

\begin{tcolorbox}[breakable,colframe=black!35,
colback=yellow!2]
{\bf Example:}

Let $ J = \mathrm{diag}(2\ii,-2\ii,2\ii,-2\ii,3\ii,-3\ii) $ be the matrix representation of the restricted complex structure $ J $ in the basis 
\begin{equation}
b = \{e_1, e_1^*, e_2, e_2^*, e_3, e_3^*\}.
\end{equation}

Consider the subsystem $ A $ defined by 
\begin{equation}
\Gamma_{\textsc a} = \mathrm{span} \left[\gamma_{\textsc a}, \gamma_{\textsc a}^*\right],
\quad \text{with} \quad \gamma_{\textsc a} = 2e_1 - \sqrt{3}\, e_3^*.
\end{equation}

A short calculation confirms that $ J^{T_{\textsc a}} $ has a single symplectic eigenvalue smaller than one, given by
\begin{equation}
\nu_1^{T_{\textsc a}} = \frac{1}{2} \left(-35 + \sqrt{1201}\right).
\end{equation}

The associated symplectic eigenvector is
\begin{align}
\nonumber e_1^{T_{\textsc a}} &= \sqrt{2 + \frac{2}{\sqrt{1201}}}\, e_1 + \sqrt{2 - \frac{2}{\sqrt{1201}}}\, e_3\\ 
&- \sqrt{\frac{3}{2} - \frac{3}{2 \sqrt{1201}}}\, e_1^* 
- \sqrt{\frac{3}{2} + \frac{3}{2 \sqrt{1201}}}\, e_3^*.
\end{align}

From this, the entanglement partner of $ A $ is the single-mode subsystem defined by:
\begin{equation*}
\begin{split}
\Gamma_{{\textsc a}_{ep}} &= \mathrm{span} \left[ \Pi^{\perp}_{\textsc a} e_1^{T_{\textsc a}},\ \Pi^{\perp}_{\textsc a} e_1^{T_{\textsc a}\, *} \right] \\
&= \mathrm{span} \left[ \sqrt{3}\, e_1^* - 2 e_3,\ \sqrt{3}\, e_1 - 2 e_3^* \right].
\end{split}
\end{equation*}
Similarly, if one begins with $\Gamma_{{\textsc a}_{ep}}$ and computes its entanglement partner,  the result is $\Gamma_{\textsc a}$.       
\end{tcolorbox}

It is worth mentioning that, since the entanglement partner has the same dimension as the original mode, logarithmic negativity is a faithful entanglement measure.

An important remark is in order. Although the entanglement partner $A_{ep}$ encodes all the entanglement between $A$ and its complement, and the entanglement partner of $A_{ep}$ is $A$ itself, these facts do not imply that $A$ is unentangled with the remaining modes. In particular, the joint subsystem $A \oplus A_{ep}$ is not, in general, separable from its complement.

Mathematically, this is possible because, although the logarithmic negativity provides a faithful measure of entanglement for $1$-vs-$N$ bipartitions (see Appendix~\ref{Negativity}), no monogamy relation is known for this measure in continuous-variable systems.

Consequently, one can construct mixed Gaussian states in which the entanglement between $A$ and its complement is entirely accounted for by its entanglement with $A_{ep}$, while $A$ remains simultaneously entangled with modes belonging to the complement of $A_{ep}$. An explicit example illustrating this situation is presented below.

\begin{tcolorbox}[breakable,colframe=black!35,
colback=yellow!2]
{\bf Example:} \\
Let
\begin{align}
J=
\begin{pmatrix}
 0 & -\frac{5}{3} & 0 & -1 & 0 & \frac{2}{3} \\
 \frac{4}{3} & 0 & -\frac{4}{3} & 0 & \frac{1}{2} & 0 \\
 0 & -1 & 0 & -\frac{4}{3} & 0 & \frac{2}{3} \\
 -\frac{4}{3} & 0 & 4 & 0 & -1 & 0 \\
 0 & \frac{2}{3} & 0 & \frac{2}{3} & 0 & -\frac{4}{3} \\
 \frac{1}{2} & 0 & -1 & 0 & 2 & 0 \\
\end{pmatrix}
\end{align}
be the matrix representation of the restricted complex structure $J$ in the basis
\begin{equation}
    b = \{e_{\textsc a}, e_{\textsc a}^*, e_{\textsc b}, e_{\textsc b}^*, e_{\textsc c}, e_{\textsc c}^*\}.
\end{equation}
$J^{T_{\textsc a}}$ has only one symplectic eigenvalue smaller than one, given by
\begin{equation}
    \nu_1^{T_{\textsc a}}= \frac23
\end{equation}
The associated symplectic eigenvector is given by
\begin{equation}
   e_1^{T_{\textsc a}} = \sqrt{2} e_{\textsc a} + \frac{3}{2\sqrt{2}} e_{\textsc b} +\frac{1}{2\sqrt{2}} e_{\textsc b}^*
\end{equation}
and by \eqref{eq_entanglement_parnter}, the entanglement partner of $A$ in this example is given by mode $B$.
\begin{equation}
    \Gamma_{{\textsc a}_{ep}} = \Gamma_{\textsc b}
\end{equation}

Nevertheless, the reduced state of the $A C$ system is also entangled. A straightforward calculation shows that $$J_{\textsc a c}^{T_{\textsc a}}= \Pi_{\textsc b }^{\perp} J^{T_{\textsc a}}\Pi_{\textsc b }^{\perp}$$
has a symplectic eigenvalue smaller than one, namely
\begin{equation}
    \tilde \nu_1^{T_{\textsc a}}=\frac13 \sqrt{25 - \sqrt{277}} < 1
\end{equation}
Thus, although the entanglement between $A$ and the rest of the system is fully captured by the entanglement partner (mode $B$), the subsystem $A$ remains entangled with its orthogonal complement (mode $C$).
\end{tcolorbox}

\section{Partners of multi-mode systems}\label{sec:Multimode}

Next, we extend the partner construction to multi-mode subsystems, which becomes straightforward within the geometric framework developed in this work. Consider an $N$-mode system with phase space $\Gamma$ and a subsystem $A$ made of $N_{\textsc a} < N$ modes. 
 
 We will show that, for pure states, the partner of $A$ exists and its dimension can be determined from the symplectic spectrum of the restriction $J_{\textsc a}=\Pi_{\textsc a} J \Pi_{\textsc a}$. For mixed states, we will discuss \emph{correlation partners}, which contain all correlations (classical and quantum), and  \emph{entanglement partners}, which encode the distillable entanglement. While the construction of the correlation partner is identical to the single-mode case discussed in Sec.~\ref{subsec:partnermixed},  the generalization of the entanglement partner involves additional subtleties. 

 \subsection{Pure Gaussian states}
\begin{tcolorbox}[breakable,colframe=black!35,colback=blue!2]
\begin{proposition}
Let $\hat\rho$ be a pure Gaussian state with complex structure $J$. 
For an $N_{\textsc a}$-mode subsystem $\Gamma_{\textsc a}\subset \Gamma$, the partner subsystem is
\begin{equation}\label{eq:partner_pure_multimode}
\Gamma_{{\textsc a}_p} = \Pi_{\textsc a}^\perp(J\Gamma_{\textsc a}),
\end{equation}
and contains $M$ modes, where $M$ is the number of symplectic eigenvalues of $J_{\textsc a}=\Pi_{\textsc a} J \Pi_{\textsc a}$ larger than one.
\end{proposition}
\end{tcolorbox}
\begin{proof}
The proof follows the single-mode case. The operator $J$ leaves the subspace 
$\Gamma_{\textsc a} + J\Gamma_{\textsc a}$ invariant; if $A$ is correlated, then 
$J\Gamma_{\textsc a} \not\subset \Gamma_{\textsc a}$, yielding a non-empty partner.

The eigenspaces of $J_{\textsc a}$ within $\Gamma_{\textsc a}$ associated with symplectic eigenvalues 
$\nu = 1$ define the uncorrelated modes contained in  $\Gamma_{\textsc a}$. In contrast, the single-mode subsystems of $A$ with $\nu > 1$ correspond to  subsystems in $\Gamma_{\textsc a}$ that are correlated with modes outside $\Gamma_{\textsc a}$. Therefore, the number of partner modes  agrees with the number of such subsystems.

It remains to prove that the dimensionality of the partner is $M$.

Let $\Gamma_{\textsc a} = \text{span}\{e_I, e_I^* \}_{I=1}^{N_A}$ be a symplectic eigenbasis of the restricted complex structure $J_{\textsc a}= \Pi_{\textsc a} J\Pi_{\textsc a}$ with $J_A \, e_I = \ii \nu_I \,e_I$ and normalized $\langle e_I , e_J \rangle= \delta_{IJ}$. 

For each eigenvector $e_I$, we can decompose the action of the full complex structure into components parallel and orthogonal to $A$, 
\begin{equation}
    J e_I = J_{\textsc a} e_I + \Pi_{\textsc a}^{\perp} (J e_I) 
\end{equation}
Since $\Gamma_{\textsc a}$ and $\Gamma_{\textsc a}^{\perp}$ are orthonormal, its contributions to the symplectic product add independently
\begin{align}
    \langle J e_I, J e_I \rangle= \langle J_{\textsc a} e_I, J_{\textsc a} e_I \rangle + \langle \Pi_{\textsc a}^{\perp} (J e_I), \Pi_{\textsc a}^{\perp} (J e_I) \rangle
\end{align}
Using  $ J_{\textsc a} e_I = \ii \nu_I e_I $,  the orthonormality of the vectors $e_I$, and the property $\langle J e_I, J e_I\rangle=\langle e_I, e_I \rangle$ (valid for pure states), we find
\begin{align}
\langle \Pi_{\textsc a}^{\perp} (J e_I), \Pi_{\textsc a}^{\perp} (J e_I) \rangle = (1-\nu_I^2)
\end{align}
For each eigenmode with $\nu_I > 1$, the vector 
\begin{equation}
    \eta_I:=\Pi_{\textsc a}^{\perp} (J e_I)
\end{equation}
has non-vanishing symplectic norm and the pair $ \{\eta_I, \eta_I^* \}$ spans a two‑dimensional symplectic subspace, i.e. a single partner mode. These subspaces are mutually symplectically orthogonal for different $I$, because the $e_I$ are.

To prove that the partner system has precisely $M$ modes, it remains to show that the modes with $\nu_I=1$ do not contribute the partner subsystem.

From $Je_I=\ii \nu_I e_I+\eta_I$ and $J^2=-I$, we obtain
\begin{align}
    -e_I =J^2e_I=J(\ii\nu_Ie_I+\eta_I)=-\nu_I^2e_I+\ii\nu_I\eta_I+J\eta_I.
\end{align}
When $\nu_I=1$, we get 
\begin{align}
    J\eta_I=-\ii\nu_I\eta_I.
\end{align}
Therefore,
\begin{align}
    \sigma(\eta_I^*,\eta_I)=-\hbar^{-1}\Omega(\eta_I^*,J\eta_I)=0.
\end{align}
Since $\sigma(\cdot^*,\cdot)$ is non-degenerate, this implies $\eta_I=0$.

Hence, the number of partner modes is exactly equal to the number of symplectic eigenvalues of $J_{\textsc A}$ that are strictly larger than unity.

\end{proof}

As a consistency check, when $\Gamma_{\textsc a}$ is single-mode this reduces to the result of Sec.~\ref{subsec:partnerpure}. It is also worth mentioning that  Proposition~5 is in agreement with the results in \cite{KojiCapacity} (see also Appendix~\ref{app:QIC_complex_structure}).

Finally, a basis of $\Gamma_{{\textsc a}_p}$ can be obtained by applying Eq.~\eqref{eq:partner_pure_multimode} to a basis of $\Gamma_{\textsc a}$ and symplectically orthonormalizing the resulting vectors, e.g.~via a symplectic Gram–Schmidt algorithm (see Appendix A of Ref.~\cite{Natalie2}).

\begin{tcolorbox}[breakable,colframe=black!35,colback=yellow!2]
\textbf{Example.} 
Consider a four-mode system in a pure Gaussian state with complex structure $J$ and corresponding eigenbasis $\{e_i,e_i^*\}_{i=1}^4$, and let $\Gamma_{\textsc a}$  be the two-mode system spanned by $2e_1-\sqrt{3}e_3^*,\,2e_2-\sqrt{3}e_4^*$, and their complex conjugates.   
A simple calculation shows that the restricted complex structure $J_{\textsc a}$ has two identical symplectic eigenvalues $\nu=7$, so the partner system $\Gamma_{{\textsc a}_p}$ is a two-mode subsystem. Moreover, applying Eq.~\eqref{eq:entanglement_partner_multimode} on the basis of $\Gamma_{\textsc a}$ given above and symplectically orthonormalizing the resulting vectors, one finds that $\Gamma_{{\textsc a}_p}$ is spanned by $\sqrt{3}e_1+2e_3^*,\,\sqrt{3}e_2+2e_4^*$ and their complex conjugates. 
\end{tcolorbox}

\begin{tcolorbox}[breakable,colframe=black!35,colback=yellow!2]
\textbf{Example.} 
Consider a four-mode system in a pure Gaussian state, and denote by $\{e_i,e_i^*\}_{i=1,...,4}$ the eigenbasis of the associated complex structure $J$. Let $\Gamma_{\textsc a}$ be the two-mode subsystem spanned by $\gamma_{{\textsc a}_1} = 2e_1-\sqrt{3}e_3^*$, $\gamma_{{\textsc a}_2}=\tfrac{1}{\sqrt{5}}(-2\sqrt{3}e_1^*+e_2+4e_3),$ and their complex conjugates. The reduced complex structure, $J_{\textsc a}$, has symplectic eigenvalues $\nu_1=7$ and $\nu_2=1$, so there is only one correlated mode within $A$ and the partner subsystem must is a made of  single-mode. Applying Eq.~\eqref{eq:entanglement_partner_multimode} to $\gamma_{{\textsc a}_1}$, $\gamma_{{\textsc a}_2}$, and their conjugates indeed yields a single-mode partner system spanned by $\sqrt{3}e_1+2e_3^*$ and its complex conjugate.  
\end{tcolorbox}

\subsection{Mixed Gaussian states}
For mixed states, two different notions of partner systems exist: correlation partners and entanglement partners. This section summarizes how their construction in Sec.~\ref{subsec:partnermixed} can be extended to multi-mode systems. 

\begin{tcolorbox}[breakable,colframe=black!35,colback=blue!2]
\begin{proposition}
The correlation partner of $A$ is
\begin{equation}
\Gamma_{{\textsc a}_{cp}}=\Pi_{\textsc a}^\perp\!\left[\bigoplus_I(\Pi_I^+\Gamma_{\textsc a}\oplus \Pi_I^-\Gamma_{\textsc a})\right],
\end{equation}
and contains $\sum_I{\rm dim}(\Pi_I^+\Gamma_{\textsc a})-N_{\textsc a}$ modes.
\end{proposition}
\end{tcolorbox}
The proof of this statement is identical to the proof of the single-mode case (using that the dimension of $A$ is given by $N_{\textsc a}$ instead of one).

The construction of the entanglement partner $A_{ep}$  for multi-mode systems follows the single-mode case, with three differences: i) in general, $A_{ep}$ encodes only the distillable part of the entanglement (instead of all the entanglement), since the PPT criterion underlying the construction is sufficient and necessary only for $1$ vs $N$ modes \cite{Werner:2000pxg}; ii) $A_{ep}$ may itself be multi-mode, with $\mathrm{dim} A_{ep} \leq N_{\textsc a}$; and iii) the partner of $A_{ep}$ is generally a subset of $A$ (equal to $A$ only when $\mathrm{dim} \Gamma_{A_{ep}} = \Gamma_{\mathrm{dim} A}$).

\begin{tcolorbox}[breakable,colframe=black!35,colback=blue!2]
\begin{proposition}\label{prop:entanglement_partner_general}
Let $\hat \rho$ be a mixed Gaussian state of an $N$-mode system, and let $A$ be an $N_{\textsc a}$-mode subsystem, $N_{\textsc a}<N$. 
If $A$ is non-PPT$_{\textsc a}$, its entanglement partner is
\begin{equation}\label{eq:entanglement_partner_multimode}
\Gamma_{{\textsc a}_{ep}}=\Pi_{\textsc a}^\perp E^{T_{\textsc a}},
\end{equation}
where $E^{T_{\textsc a}}$ is the symplectic subspace spanned by eigenvectors of $J^{T_{\textsc a}}$ with symplectic eigenvalues smaller than one. The dimension of $\Gamma_{A_{ep}}$ is at most $2N_{\textsc a}$.
\end{proposition}
\end{tcolorbox}
\begin{proof}
The proof of this statement is as follows.  Since at most $N_{\textsc a}$ symplectic eigenvalues of $J^{T_{\textsc a}}$ can be smaller than one (see the Little Lemma in Ref.~\cite{serafini2017quantum}), $E^{T_{\textsc a}}$ spans a subsystem of dimension $\leq N_{\textsc a}$. Moreover, $E^{T_{\textsc a}} \not \subset \Gamma_{\textsc a}$ and $E^{T_{\textsc a}} \not \subset \Gamma_{\bar{\textsc a}}$; otherwise, one would find subspaces of $A$ or its complement with symplectic eigenvalues $<1$, contradicting positivity of $J$. Hence $\Pi_{\textsc a}^\perp E^{T_{\textsc a}}$ is non-empty when $J$ is non-PPT$_{\textsc a}$. Finally, using the same argument as in the single-mode case, one reaches the conclusion that the distillable entanglement between $A$ and its complement is entirely contained in the reduced subsystem consisting of $A$ and $A_{ep}$. See Appendices~\ref{app:proof_symplectic_subspace_ep} and \ref{app:UniquenessProof_EntanglementPartner} for more details of the proof.
\end{proof}

 \begin{tcolorbox}[breakable,colframe=black!35,
colback=yellow!2]
{\bf Example:}

Let 
\begin{equation}
    J = \mathrm{diag}\{2i, -2i, 2i,-2i,3i,-3i, 4i,-4i\}
\end{equation}
denote the  matrix representation of the restricted complex structure $J$ of a mixed Gaussian state in the basis $\{e_i,e_i^*\}$ with $i=1,\dots, 4$. Consider the subsystem $A$  defined by $A = \mathrm{span}\{\gamma_{{\textsc a}_1}, \gamma_{{\textsc a}_1}^*, \gamma_{{\textsc a}_2}, \gamma_{{\textsc a}_2}^* \}$, with $\gamma_{{\textsc a}_1} = 2 e_1 - \sqrt{3}e_3^*$ and $\gamma_{{\textsc a}_{2}} = 2e_2 - \sqrt{3}e_4^*$. 

A short calculation shows that $J^{T_{\textsc a}}$ has two symplectic eigenvalues smaller than one, given by 
\begin{equation}
    \nu^{T_{\textsc a}}_1 = \sqrt{57}-7,~\text{and}~\nu_2^{T_{\textsc a}} = \frac{1}{2}(\sqrt{73} - 7).
\end{equation}
Hence, the space $E^{T_{\textsc a}}$ corresponds to a  two-mode subsystem. After projecting $E^{T_{\textsc a}}$ onto the symplectic complement of $A$, we conclude that the entanglement partner of $A$ is the two-mode subsystem spanned by 
\begin{align}
    \gamma_{{\textsc a}_{ep1}} &= -\frac{1}{4} \sqrt{9 \sqrt{57}-3} e_2 - \frac{3 \sqrt{\sqrt{57}+5}}{4} e_2^* \nonumber\\
&+ \frac{1}{2} \sqrt{3 \left(\sqrt{57}+5\right)} e_4\nonumber\\
&+ 8 \sqrt{\frac{2}{3 \sqrt{57}+1}} e_4^*,\\
\gamma_{{\textsc a}_{ep2}} &= 4 i \sqrt{\frac{6}{5 \sqrt{73}-17}} e_1 \nonumber\\ 
& -i \left(\sqrt{73}-5\right) \sqrt{\frac{3}{2 \left(5 \sqrt{73}-17\right)}} e_1^* \nonumber\\ 
&+i \left(\sqrt{73}-5\right) \sqrt{\frac{2}{5 \sqrt{73}-17}}e_3 \nonumber\\ 
&-8 i \sqrt{\frac{2}{5 \sqrt{73}-17}} e_3^*, 
\end{align}
and their complex conjugates. 
Similarly, one can show that the entanglement partner of $\Gamma_{{\textsc a}_{ep}}$ coincides with $A$, since  $E^{T_{\textsc a}}$ and $A$ have the same dimension.
\end{tcolorbox}

\section{Conclusions}

In this work, we have introduced a method to identify the degrees of freedom that carry the correlations of any chosen single mode in many-body Gaussian quantum systems. 

Our analysis is based on a complex phase-space formalism in which the primary object is a complex structure for pure Gaussian states and a restricted complex structure for mixed Gaussian states. We have shown that the distribution of correlations and entanglement in Gaussian systems is encoded in the eigenspaces of these complex structures and of their partially transposed counterparts. One of the main goals of this article is to further develop and promote these tools as a powerful geometric framework for the study of Gaussian quantum systems.

Our main findings were encapsulated in two main propositions.

First, we showed that in pure Gaussian states every correlated mode admits a unique, disjoint, single-mode partner that captures all of its correlations. We further provided a constructive procedure to obtain this partner from the eigenspaces of the complex structure.

For pure Gaussian states, our formalism yields a geometric and basis-independent characterization of the partner subsystem, and establishes its existence and uniqueness. When expressed in a particular basis, our results are consistent with previous studies~\cite{HottaPartner,Trevison2019,hackl_minimal_2019}. The main outcome of these works is that the correlation structure of an arbitrary many-body Gaussian system can be reduced to an effective two-mode description, providing an intuitive way to identify and pair the degrees of freedom responsible for the entanglement.

Second,  we have extended the analysis of partners in Gaussian systems to mixed states, demonstrating that the notion of a partner naturally splits into two distinct concepts: the \emph{correlation partner}, which collects all correlations with the reference mode (and may involve several modes), and the \emph{entanglement partner}, which is always a single mode and exists precisely when the state violates the PPT criterion. This distinction allows one to pinpoint where genuine quantum entanglement resides, and to separate it from broader, possibly classical, correlations. 

Finally, we extended the proofs of existence and uniqueness of partners for both pure and mixed Gaussian states to multi-mode subsystems. For pure Gaussian states, the construction parallels that of the partner for single-mode subsystems, with the only difference that the dimension of the partner of a multi-mode subsystem $A$ does not necessarily match the dimension of $A$.  For mixed Gaussian states, the partner concept again splits into two. The correlation partner of a multi-mode subsystem is constructed analogously to the single-mode case and retains the same interpretation, whereas the notion of entanglement partner captures only the distillable part of the entanglement.

Together, these results provide a systematic and constructive framework to track bipartite correlations and entanglement in continuous-variable systems. Beyond simplifying the analysis of complex Gaussian states, they establish a foundation for practical applications in quantum communication and for the study of correlations in quantum field theory.
%{\color{orange} in finitely-generated subsystems} in quantum field theory. 
Future directions include extending the formalism to continuum QFT (see \cite{Agullo:2025lug,Ribes-Metidieri:2025nfw} for a discussion of the subtleties that arise when extending the analysis of partner modes to infinite-dimensional systems),  and to large-scale many-body systems, where identifying the ``partners'' of specific modes could become a powerful tool for controlling and utilizing entanglement.

\begin{acknowledgments}
The content of this paper has benefited from discussions with A. Ashtekar, E. Bianchi, P. Chisvert, R. Dawkins, A. Delhom, F. Mele, B. Elizaga-Navascues, J. Wilson, and V. Vennin. 
 I.A. is supported by the NSF grants PHY-2409402 and PHY-2110273, by the RCS program of Louisiana
Boards of Regents through the grant LEQSF(2023-25)-RD-A-
04,  by the Hearne Institute for Theoretical Physics and by Perimeter Institute of Theoretical Physics through the Visitor fellow program. IA is also supported by the WOST (WithOut SpaceTime) project (https://withoutspacetime.org), led by the Center for Spacetime and the Quantum (CSTQ), and supported by Grant ID 63683 from the John Templeton Foundation (JTF).  SNG acknowledges the financial support of the Research Council of Finland through the Finnish Quantum Flagship project (358878, UH). P.R.M. acknowledges support from the Royal Commission for the Exhibition of 1851 through the 1851 Research Fellowship. 
K.Y. acknowledges support from the JSPS Overseas Research Fellowship and JSPS KAKENHI Grant No. JP24KJ0085.
 Research at Perimeter Institute is supported
in part by the Government of Canada through the Department of Innovation, Science and Industry Canada
and by the Province of Ontario through the Ministry of
Colleges and Universities.
\end{acknowledgments}

\appendix

\section{Proof of the properties of the covariance metric}\label{Proof1}
This appendix collects the proofs of several statements made in 
Section~\ref{GaussianCov} regarding the covariance metric $\sigma$. Specifically, we prove: 

\begin{enumerate}
    \item $\sigma$ is symmetric, i.e., $\sigma(\gamma, \gamma') = \sigma(\gamma', \gamma)$ for all $\gamma, \gamma' \in \Gamma_{\mathbb{C}}$.
    
    \item $\sigma(\gamma^*, \gamma) \pm \langle \gamma, \gamma \rangle \geq 0$ for all $\gamma \in \Gamma_{\mathbb{C}}$.

    \item When restricted to $\Gamma$, $\sigma$ is positive definite: $\sigma(\gamma, \gamma) > 0$ for all nonzero $\gamma$ in $\Gamma$.

\end{enumerate}

{\em {Proof:}\\}
\begin{enumerate}
    \item This property follows directly from the definition of $\sigma$.

    \item $\sigma(\gamma^*, \gamma) +  \langle \gamma, \gamma \rangle $ is equal to $2 \, {\rm Tr}[\hat{\rho} \, \hat{\overline{O}}_{\gamma} \hat{\overline{O}}^{\dagger}_{\gamma}]$. This is the expectation value of the positive semi-definite operator $\hat{\overline{O}}_{\gamma} \hat{\overline{O}}^{\dagger}_{\gamma}$. Then, the positive semi-definiteness of the state $\hat \rho$ guarantees that this quantity is $\geq 0$. 
    
    On the other hand,  $\sigma(\gamma^*, \gamma) -  \langle \gamma, \gamma \rangle $ is equal to $\sigma(\gamma, \gamma^*) +  \langle \gamma^*, \gamma^* \rangle$. This is the same as
       $2 \, {\rm Tr}[\hat{\rho} \, \hat{\overline{O}}^{\dagger}_{\gamma} \hat{\overline{O}}_{\gamma}]$. This is also the expectation value of a semi-positive definite operator, thus its expectation value is $\geq 0$ for all $\gamma$ in $\Gamma_{\mathbb{C}}$.

    \item We will prove a slightly more general statement. Let us define the sesquilinear  map $(\cdot, \cdot)$ in $\Gamma_{\mathbb{C}}$ as $(\gamma, \gamma') \equiv \sigma(\gamma^*, \gamma')$. We will show that $(\cdot, \cdot)$ is positive definite in $\Gamma_{\mathbb{C}}$, so it defines a Hermitian inner product in $\Gamma_\mathbb{C}$. Because $(\cdot, \cdot)$ and $\sigma$ agree when acting on real vectors, the positivity of $\sigma$ when restricted to $\Gamma$ follows automatically.

    First, since $(\gamma, \gamma) = {\rm Tr}[\hat{\rho} \, \{ \hat{\overline{O}}_{\gamma}, \hat{\overline{O}}^{\dagger}_{\gamma} \}]$ and the right-hand side is $\geq 0$ ---because $ \{ \hat{\overline{O}}_{\gamma}, \hat{\overline{O}}^{\dagger}_{\gamma} \}$ is a positive semi-definite operator--- we conclude that $(\cdot, \cdot) \geq 0$. 

    To prove the stronger condition $(\gamma, \gamma) > 0$ for all $\gamma \in \Gamma_{\mathbb{C}}$, we consider three different cases:
    \begin{enumerate}
        \item $\gamma$ satisfies $\langle \gamma, \gamma \rangle < 0$.\\
        Property 2 above can be written as $(\gamma, \gamma) + \langle \gamma, \gamma \rangle \geq 0$. This implies $(\gamma, \gamma) \geq -\langle \gamma, \gamma \rangle  $
           from which $(\gamma, \gamma) > 0$ follows (when $\langle \gamma, \gamma \rangle < 0$).

        \item $\gamma$ satisfies  $\langle \gamma, \gamma \rangle > 0$.\\
        This implies $\langle \gamma^*, \gamma^* \rangle < 0$. Property 2 implies $(\gamma^*, \gamma^*) + \langle \gamma^{*}, \gamma^* \rangle \geq 0$, from which $(\gamma^*, \gamma^*) > 0$ follows when $\langle \gamma^*, \gamma^* \rangle < 0$. Since $(\gamma^*, \gamma^*) = (\gamma, \gamma)$ for all $\gamma \in \Gamma_{\mathbb{C}}$, it follows that $(\gamma, \gamma) > 0$ for all $\gamma \in \Gamma_{\mathbb{C}}$.

        \item $\gamma$ satisfies  $\langle \gamma, \gamma \rangle = 0$.\\
        For this case, we will prove $(\gamma, \gamma) > 0$ by contradiction. 
        
Assume there exists a nonzero $\gamma$ for which $(\gamma, \gamma) = 0$. We can always find a vector $\gamma'$ such that $(\gamma, \gamma') \neq 0$. Consider one such $\gamma'$, and construct from it the vector $\gamma'' = a \, \gamma + \gamma'$, with $a \in \mathbb{C}$. Then
\begin{align}
&(\gamma'', \gamma'') = |a|^2 \, (\gamma, \gamma) + 2 \, {\rm Re}[a \, (\gamma', \gamma)] + (\gamma', \gamma') \, .
\end{align}
If $(\gamma, \gamma)$ were to vanish, we could always choose $a$ such that this expression is negative, contradicting $(\cdot, \cdot) \geq 0$. Thus, $(\gamma, \gamma)$ cannot vanish.
\end{enumerate}
\end{enumerate}

\section{Proof of the properties of the restricted complex structure}\label{ProofJ}

\begin{enumerate}
\item Multiplying $J^a_{\ b} = -\hbar\Omega^{ac} \sigma_{cb}$ times $\Omega_{da}$ and using $\Omega_{da} \Omega^{ac} = \delta^c_d$, we find $-\sigma_{db} = \hbar^{-1}\Omega_{da} J^a_{\ b}$. This is equivalent to $\sigma_{ab} = -\hbar^{-1}\Omega_{ac} J^c_{\ b}$ or, equivalently, $\sigma(\cdot,\cdot)=-\hbar^{-1}\Omega(\cdot,J\cdot)$.

\item Because $\Omega(\gamma_1, J\gamma_2) \propto \sigma(\gamma_1, \gamma_2)$, the bilinear form $\Omega(\cdot, J\cdot)$ is symmetric, and we have:
\begin{equation} \label{a}
\Omega(\gamma_1, J\gamma_2) = \Omega(\gamma_2, J\gamma_1) \quad \forall\, \gamma_1, \gamma_2.
\end{equation}

On the other hand, since $\omega$ is antisymmetric, it follows that:
\begin{equation} \label{b}
\Omega(\gamma_1, J\gamma_2) = -\Omega(J\gamma_2, \gamma_1) \quad \forall\, \gamma_1, \gamma_2.
\end{equation}

Combining equations \eqref{a} and \eqref{b} yields the desired result.

\item All anti-Hermitian linear maps are diagonalizable with purely imaginary eigenvalues, which come in pairs $\pm \ii\, c$, with $c \in \mathbb{R}_+$. Therefore, it suffices to show that $J$ is anti-Hermitian with respect to a suitable Hermitian inner product on $\Gamma_{\mathbb{C}}$.
Specifically, we will show that $J$ is anti-Hermitian in $\Gamma_{\mathbb{C}}$ with respect to the Hermitian inner product $(\cdot, \cdot) = \sigma(\cdot^{*}, \cdot)$.

Anti-Hermiticity means   $J^{\dagger} = -J$, where $J^{\dagger}$ is the linear operator satisfying $(J^{\dagger} \cdot, \cdot) = (\cdot, J \cdot)$. Thus, $J$ is anti-Hermitian if $(-J \cdot, \cdot) = (\cdot, J \cdot)$. This can be proven as follows:
\begin{eqnarray}(\gamma_1,J\gamma_2)&=&\sigma(\gamma_1^*,J\gamma_2)=\hbar^{-1}\Omega(J\gamma_1,J\gamma_2)=\hbar^{-1}\Omega(J^2\gamma_1,\gamma_2)\nonumber \\ &=&-\sigma(J\gamma_1^*,\gamma_2)=-(J\gamma_1,\gamma_2), \forall \gamma_1,\gamma_2 \end{eqnarray} 
where we have used $\Omega(\cdot,J\cdot)=-\Omega(J\cdot,\cdot)$ in the third equality.

\item This follows from the property $\sigma(\gamma^{*}, \gamma) - \langle \gamma, \gamma \rangle \geq 0$ for all $\gamma \in \Gamma_{\mathbb{C}}$ shown earlier.

We denote by $e_I$ (without complex conjugate) the eigenvector of $J$ with eigenvalue $+\ii \nu_I$, with $\nu_I$ a non-negative real number. We will show first that $e_I$ has positive symplectic norm, $\langle e_I, e_I \rangle > 0$.

This follows by writing $\langle e_I, J e_I \rangle$ in two ways. On the one hand, $\langle e_I, J e_I \rangle = \langle e_I, \ii \nu_I \, e_I \rangle = \ii \nu_I \, \langle e_I, e_I \rangle$. On the other hand,
\begin{eqnarray}
    \langle e_I, J e_I \rangle&=& -\ii\, \hbar^{-1}\, \Omega(e_I^*,Je_I)\nonumber \\&=&\ii\,  \sigma(e_I^*,e_I)=\ii\,  (e_I,e_I)=\ii\, \alpha\, ,\end{eqnarray} 
with $\alpha \in \mathbb{R}_+$, by the positivity of the inner product $(\cdot, \cdot)$. Comparing these two results, we conclude $\langle e_I, e_I \rangle > 0$.
 
With this, the condition $\sigma(\gamma^{*}, \gamma) - \langle \gamma, \gamma \rangle \geq 0$, when particularized to $\gamma=e_I$, can be re-written as 
 \begin{align} &\hbar^{-1}\Omega(J e_I^*,e_I)-\langle e_I, e_I \rangle \geq 0\\ &\Rightarrow \ii \langle J e_I, e_I \rangle -\langle e_I, e_I \rangle \geq 0 \nonumber \\
 & \Rightarrow \langle (-\ii J-\openone)e_I, e_I\rangle \geq 0\\ 
&\Rightarrow (\nu_I-1)\, \langle e_I, e_I\rangle \geq 0, \Rightarrow \nu_I\geq 1. \nonumber \end{align} 

\item The anti-Hermiticity of $J$ guarantees that its eigenvectors appear in pairs $e_I$ and $e^*_I$ for $I = 1, \dots, N$, forming an orthogonal basis with respect to the inner product $( \cdot, \cdot)$. Specifically, we have $( e_I, e_J) = ( e^*_I, e^*_J) = 0$ for all $I \neq J$ and $( e_I, e^*_J) = 0$ for all $I$ and $J$.

Because $( e_I, e_J) = \sigma( e^*_I, e_J) = \hbar^{-1}\Omega(J e^*_I, e_J) = -\ii \, \nu_I \,\hbar^{-1} \Omega(e^*_I, e_J) = \nu_I \, \langle e_I, e_J \rangle$, and $\nu_I \neq 0$, we have that $( e_I, e_J) = 0$ implies $\langle e_I, e_J \rangle = 0$ for $I \neq J$.

Since $\langle e^*_I, e^*_J \rangle = -\langle e_I, e_J \rangle$, it follows that $\langle e^*_I, e^*_J \rangle = 0$ for $I \neq J$.

Similarly, $( e_I, e^*_J) = 0$ for all $I$ and $J$ implies $\langle e_I, e^*_J \rangle = 0$.

On the other hand, as shown in item 4 of this list, $\langle e_I, e_I \rangle > 0$. Thus, we can normalize the eigenvectors $e_I$, and conclude that $\langle e_I, e_I \rangle = \delta_{IJ}$. This normalization also implies $\langle e^*_I, e^*_I \rangle = -\delta_{IJ}$.

\end{enumerate}

\section{Symplectic spaces and projectors}\label{app:projectors}

A symplectic space is a vector space equipped with a non-degenerate skew-symmetric bilinear form. The phase space of a bosonic linear system is naturally a symplectic space endowed with the symplectic form $\Omega$. A subsystem of a bosonic system can be specified by a symplectic subspace, and projections defined with respect to $\Omega$ play an essential role in characterizing orthogonal complements in a basis-independent manner. In what follows, we summarize the mathematical definitions and main properties of these notions.

A vector subspace $\Gamma_{\textsc a} \subset \Gamma$ of a symplectic space $\Gamma$ is said to be a symplectic subspace if the restriction of the symplectic form $\Omega$ on $\Gamma$ to $\Gamma_{\textsc a}$ is itself a \textit{bona fide} symplectic form, namely, a non-degenerate skew-symmetric bilinear form on $\Gamma_{\textsc a}$. Note that a symplectic subspace is always even-dimensional, since a non-degenerate skew-symmetric bilinear form exists only on even-dimensional vector spaces. When $\Gamma_{\textsc a}$ is a $2N_{\textsc a}$-dimensional symplectic subspace, we refer to $A$ as an $N_{\textsc a}$-mode subsystem. 

If $\Gamma_{\textsc a}$ is a symplectic subspace of $\Gamma$, then $\Gamma$ admits a direct sum decomposition
\begin{equation}
\Gamma = \Gamma_{\textsc a} \oplus \Gamma_{\bar{{\textsc a}}},
\end{equation}
where $\Gamma_{\bar{{\textsc a}}}$ denotes the symplectic orthogonal complement of $\Gamma_{\textsc a}$, defined by
\begin{equation}
\Gamma_{\bar{{\textsc a}}} := \left\{ \gamma \in \Gamma \,\middle|\, \Omega(\gamma, \gamma') = 0 \ \ \forall\, \gamma' \in \Gamma_{\textsc a} \right\}.
\end{equation}
The subspace $\Gamma_{\bar{{\textsc a}}}$ is itself a symplectic subspace of $\Gamma$, and the symplectic form decomposes as 
$\Omega = \Omega_{\textsc a} \oplus \Omega_{\bar{{\textsc a}}}$, where $\Omega_{\textsc a}$ and $\Omega_{\bar{{\textsc a}}}$ are the restrictions of $\Omega$ 
to $\Gamma_{\textsc a}$ and $\Gamma_{\bar{{\textsc a}}}$, respectively. We refer to the subsystem $\bar{A}$ associated with $\Gamma_{\bar{{\textsc a}}}$ as the symplectic complement of $A$.

Once $\Gamma$ is decomposed as the direct sum $\Gamma_{\textsc a} \oplus \Gamma_{\bar{{\textsc a}}}$, any vector $\gamma \in \Gamma$ can be uniquely written as the sum of a vector in $\Gamma_{\textsc a}$ and a vector in $\Gamma_{\bar{{\textsc a}}}$. This unique decomposition induces the definition of the symplectic orthogonal projector onto $\Gamma_{\textsc a}$, denoted by $\Pi_{\textsc a}$. The complementary projector onto $\Gamma_{\bar{{\textsc a}}}$ is then given by $\Pi_{\textsc a}^\perp \coloneqq \mathbb{I}-\Pi_{\textsc a}$.

The above definitions of subsystems and symplectic projectors apply to both classical and quantum cases. As our focus in the present paper is on quantum bosonic systems, we now provide an explicit expression for the symplectic orthogonal projector $\Pi_{\textsc a}$ using the Hermitian pseudo-inner product $\langle\cdot,\cdot\rangle$ (defined in Eq.~\eqref{sympprod}) on the complexified phase space $\Gamma_{\mathbb{C}}$. It should be emphasized that, although the projector is formulated without reference to basis vectors, it is often advantageous to represent it in terms of them.

Let $A$ be an $N_{\textsc a}$-mode symplectic subsystem. With respect to a given basis $\{\xi_i\}_{i=1}^{2N_{\textsc a}}$ of $\Gamma_{\textsc a}$, the symplectic orthogonal projector $\Pi_{\textsc a}$ is expressed as
\begin{align}
    \Pi_{\textsc a}(\cdot)=\sum_{i,j}\xi_i B_{ij}\langle \xi_j,\cdot\rangle,
\end{align}
where $B$ denotes the inverse of the Gram matrix $G$, whose entries are $G_{ij}\coloneqq \langle \xi_i,\xi_j\rangle$ for $i,j=1,\ldots, 2N_{\textsc a}$. Note that the Gram matrix is always invertible when $\Gamma_{\textsc a}$ is a symplectic subspace, since the restriction of $\langle \cdot,\cdot \rangle$ to $\Gamma_{\textsc a}$ is non-degenerate by definition.

In particular, it is always possible to choose basis vectors $\{\gamma_I,\gamma_I^*\}_{I=1}^{N_{\textsc a}}$ such that
\begin{align}
    \langle\gamma_I,\gamma_J\rangle=\delta_{IJ},\quad   \langle\gamma_I^*,\gamma_J^*\rangle=-\delta_{IJ},\quad \langle\gamma_I,\gamma_J^*\rangle=0
\end{align}
for $I,J=1,\cdots, N_{\textsc a}$. This is equivalent to 
\begin{align}
    [\hat{a}_I,\hat{a}_J^\dagger]=\delta_{IJ}\hat{\openone},\quad [\hat{a}_I,\hat{a}_J]=0  =[\hat{a}_I^{\dagger},\hat{a}_J^{\dagger}],
\end{align}
where we defined $\hat{a}_I \coloneqq \hat{O}_{\gamma_I}$ through the correspondence introduced in Eq.~\eqref{mapgtop}. This relation implies that $\{\hat{a}_I^\dag\}_{I=1}^{N_{\textsc a}}$ and $\{\hat{a}_I\}_{I=1}^{N_{\textsc a}}$ represent  creation and annihilation-like  operators of independent modes associated with $\{\gamma_I\}_{I=1}^{N_{\textsc a}}$. Given such basis vectors, the symplectic orthogonal projector $\Pi_{\textsc a}:\Gamma_{\mathbb{C}}\to\Gamma_{\textsc a}$ can be expressed as
\begin{align}
    \Pi_{\textsc a}(\cdot)=\sum_{I=1}^{N_{\textsc a}}\left(\gamma_I\langle\gamma_I,\cdot\rangle-\gamma_I^*\langle\gamma_I^*,\cdot\rangle\right),
\end{align}
where the negative sign in the coefficients of $\gamma_I^*$ reflects the fact that $\langle \gamma_I^*,\gamma_J^*\rangle= -\delta_{IJ}$. 

\section{Operator-based approach and phase-space approach}\label{app:QIC_complex_structure}
%\ky{New section}
Proposition~\hyperlink{prop:uncorrelated}{1} states that a subsystem is uncorrelated in a Gaussian state if and only if the associated subspace in phase space is invariant under the (possibly restricted) complex structure $J$. For a pure Gaussian state, Eq.~\eqref{eq:uncorrelated_pure} shows that for a given subspace $\Gamma$, the enlarged subspace $\Gamma+J\Gamma$ defines an uncorrelated subsystem. This observation underlies the purification partner formula established in Proposition~\hyperlink{prop:partner}{2}. The simplest case is when $\Gamma_{\textsc a}$ is spanned by a real vector, in which case $\Gamma_{\textsc a}+J\Gamma_{\textsc a}$ defines a two-dimensional linear subspace that characterizes a single-mode uncorrelated subsystem, reproducing the formula for a single-mode quantum information capsule (QIC)~\cite{yamaguchi_quantum_2020}. Previous studies~\cite{KojiCapacity,yamaguchi_quantum_2021} extended this idea by recursively applying the single-mode formula to identify multi-mode QICs, with the partner formula appearing as the special case of a two-mode QIC. We briefly review here the relation between these operator-based results and the phase-space approach which we adopted in the present paper. 

For a pure Gaussian state $\ket{\Psi}$, let $\hat{\overline{\bm{r}}}\coloneqq \hat{\bm{r}}-\langle\Psi|\hat{\bm{r}}|\Psi\rangle$ denote the vector of centered Darboux operators. Given a Hermitian operator $\hat{O}\coloneqq \sum_k v_k\hat{\overline{r}}^k$ with $\bm{v}\in\mathbb{R}^{2N}$, a state-dependent map $f_\Psi$ is defined~\cite{yamaguchi_quantum_2020,KojiCapacity} by
\begin{equation}
f_\Psi(\hat{O})\coloneqq -\sum_{k,l,m}\hat{\overline{r}}^k\Omega^{kl}\sigma^{lm}v_m,
\end{equation}
where $\Omega^{ij}$ and $\sigma^{ij}$ denote the matrix elements of $\Omega^{ab}$ and $\sigma^{ab}$ in the Darboux basis, and we set $\hbar=1$ for simplicity. By using this map, for a given set of Hermitian operators $\{\hat{O}_i\}_{i=1}^k$, the subalgebra generated by $\{\hat{O}_i,f_{\Psi}(\hat{O}_i)\}_{i=1}^k$ defines an uncorrelated subsystem consisting of (at most) $k$ modes~\cite{yamaguchi_quantum_2021}, referred to as a $k$-mode QIC~\cite{KojiCapacity}. 

In our phase-space formulation, linear operators are defined via the symplectic product. Since $\sum_k v_k\hat{\overline{r}}^k=\sum_{k,l}\Omega_{lk}v^l\hat{\overline{r}}^k$ with $v_k\eqqcolon\sum_l \Omega_{lk}v^l$, we obtain  
\begin{equation}
f_\Psi(\hat{O}_v) = -\sum_{k,l,m,n}\hat{\overline{r}}^k\Omega^{kl}\sigma^{lm}\Omega_{nm}v^n = \hat{O}_{Jv},
\end{equation}
where we used $\Omega^{kl}=\Omega_{lk}$ in the Darboux basis, and $J^l_{\ n} =\sum_m \sigma^{lm}\Omega_{mn}=-\sum_m \sigma^{lm}\Omega_{nm}$. Hence, the characterization of an uncorrelated subsystem as $\Gamma+J\Gamma$ is essentially equivalent to the operator-based result on QIC. 

The framework of the present paper provides a concise and geometric proof of these results in QIC. Specifically, for any given $\gamma\in \Gamma$, $\frac{1}{2}(\openone\mp\ii J)\gamma$ are eigenstates of $J$ with eigenvalue $\pm i$---consequently, the operator corresponding to $\frac{1}{2}(\openone-\ii J)\gamma$ annihilates the pure state defined by $J$---thereby demonstrating that $\mathrm{span}\{\gamma,J\gamma\}$ defines an uncorrelated single-mode system for any $\gamma\in \Gamma$. 

Despite such a connection, the phase-space approach provides distinct advantages:
\begin{itemize}
    \item[(i)] it offers a concise and mathematically transparent expression of the partner formula in pure Gaussian states,  
    \item[(ii)] it naturally extends to mixed Gaussian states, as established in the main text, and  
    \item[(iii)] it yields a unified description of partner formulas for both bosonic and fermionic Gaussian states (see Appendix~\ref{app:fermion}).
\end{itemize}

\section{Partner formula for fermionic systems in Gaussian pure states}\label{app:fermion}

Similar to the bosonic case, one can introduce a complex structure to describe
fermionic Gaussian states (see, e.g.,~\cite{hackl_bosonic_2021} for a more 
comprehensive discussion).

Consider a system consisting of $N$ fermionic modes (for instance, $N$ modes of a 
fermionic quantum field). Let $\hat a_I$ and $\hat a_I^{\dagger}$, 
$I = 1,\ldots,N$, denote the associated annihilation and creation operators. 
They satisfy the canonical anti-commutation relations
\begin{align}
    \{\hat a_I, \hat a_J\} = \{\hat a_I^{\dagger}, \hat a_J^{\dagger}\} = 0, 
    \qquad
    \{\hat a_I, \hat a_J^{\dagger}\} = \delta_{IJ}.
\end{align}

It is convenient to introduce the self-adjoint (Majorana) operators
\begin{align}
    \hat{x}_I := \left(\hat a_I + \hat a_I^{\dagger}\right), 
    \qquad
    \hat{p}_I :=-\ii
        \left(\hat a_I - \hat a_I^{\dagger}\right),
\end{align}
in terms of which the canonical anti-commutation relations become
\begin{align}\label{Cliff}
    \{\hat{\xi}^i, \hat{\xi}^j\} = 2\, g^{ij}\,\hat{\openone},
\end{align}
where we have defined the vector of canonical operators
\[
    \hat{\boldsymbol{\xi}}
    := (\hat{x}_1, \hat{p}_1, \ldots, \hat{x}_N, \hat{p}_N),
\]
and $g^{ij} = \mathrm{diag}(1,\ldots,1)$. Eqn.~\eqref{Cliff} is the familiar Clifford's algebra. 

Similar to the strategy followed for bosons, we define a $2N$-dimensional real 
vector space $\Gamma$, equipped with an Euclidean metric $g_{ab}$, and establish a 
one-to-one correspondence between vectors in $\Gamma$ and operators linear in 
$\hat{\boldsymbol{\xi}}$:
\begin{equation}
    \gamma \in \Gamma \quad \longmapsto \quad 
    \hat{O}_{\gamma} := g(\gamma, \hat{\boldsymbol{\xi}})\, .
\end{equation}
The anti-commutator of any pair of such operators is
\begin{equation}
    \{\hat{O}_{\gamma}, \hat{O}_{\gamma'}\}
    = 2\, g(\gamma, \gamma')\, .
\end{equation}

A fermionic pure Gaussian state $\ket{\psi}$ defines a two-form on $\Gamma$ by
\begin{equation}
    \Omega(\gamma, \gamma') 
    \coloneqq \ii\, \langle \psi | [\hat{O}_{\gamma}, \hat{O}_{\gamma'}] | \psi \rangle .
\end{equation}
Up to the factor of $\ii$, introduced for convenience, the action of $\Omega$ on two 
vectors equals the expectation value of the commutator of the corresponding linear 
operators. One can show that, for pure Gaussian states, the two-form $\Omega$ is non-degenerate and therefore 
defines a symplectic form on $\Gamma$.

Thus, the roles of the metric tensor and the symplectic form are interchanged 
relative to the bosonic case: for bosons, the quantum state determines the metric 
on $\Gamma$ while the symplectic form is state-independent and fixed by the 
canonical commutation relations. For fermions, the situation is reversed.

As in the bosonic case, the symplectic form and the metric can be combined to define 
a linear map $J$ by
\begin{equation}
    J^{a}{}_{b} \coloneqq -\,\Omega^{ac} g_{cb}.
\end{equation}
 For a pure Gaussian state, this linear map satisfies $J^{2} = -\mathbb{I}$ and therefore 
defines a complex structure on $\Gamma$.

A subsystem $A$ is specified by any even-dimensional subspace 
$\Gamma_{\textsc a} \subset \Gamma_{\mathbb{C}}$ satisfying that the restriction of $g_{ab}$ to $\Gamma_{\textsc a}$ is non-degenerate and  defines a bona fide metric 
on this subspace.  
In parallel with Proposition~\hyperlink{prop:uncorrelated}{1}, the subsystem is 
uncorrelated if and only if $\Gamma_{\textsc a}$ is an invariant subspace of the complex 
structure $J$.  
In contrast to the bosonic case, the projector $\Pi_{\textsc a}$ is defined with respect to 
metric $g_{ab}$, hence it is an orthogonal projector.   Using $J$ and $\Pi_{\textsc a}$, Proposition~\hyperlink{prop:partner}{2} extends to fermionic 
systems: the partner of a single-mode subsystem $A$ is given by
\begin{equation}
    \Gamma_{{\textsc a}_p} = \Pi_{\textsc a}^\perp \bigl( J \Gamma_{\textsc a} \bigr)\, .
\end{equation}

\section{Example of calculation of correlation and entanglement partners}\label{AppExampleCP}

Let $J = \mathrm{diag}(2\ii, -2\ii, 2\ii, -2\ii, 3\ii, -3\ii)$ be the matrix representation of the complex structure $J$ in the basis $b = \{e_1, e_1^{*}, e_2, e_2^{*}, e_3, e_3^{*}\}$. This $J$ has two different pairs of eigenvalues: $\pm\ii 2$ (two-fold degenerate) and $\pm\ii 3$ (non-degenerate). Accordingly, the index $I$ takes values $1$ and $2$, with the following projectors:

\begin{equation}
\Pi^{+}_1 = e_1\langle e_1, \cdot \rangle + e_2\langle e_2, \cdot \rangle, \quad 
\Pi^{-}_1 = -e^*_1\langle e^*_1, \cdot \rangle - e^*_2\langle e^*_2, \cdot \rangle,
\end{equation}
\begin{equation}
\Pi^{+}_2 = e_3\langle e_3, \cdot \rangle, \quad 
\Pi^{-}_2 = -e^*_3\langle e^*_3, \cdot \rangle.
\end{equation}
\begin{comment}
    {\color{green}\begin{equation}
\cancel{\Pi^{+}_3 = e_4\langle e_4, \cdot \rangle, \quad 
\Pi^{-}_3 = -e^*_4\langle e^*_4, \cdot \rangle.}
\end{equation}}
\end{comment}

We now consider different subsystems $A$, each defined as $\Gamma_{\textsc a} = \mathrm{span}[\gamma_{\textsc a}, \gamma_{\textsc a}^*]$, for the following choices of $\gamma_{\textsc a}$:

\begin{enumerate}
\item $\gamma_{\textsc a} = e_1 + e_2$: \\
We find that $J\gamma_{\textsc a} = \ii 2\, \gamma_{\textsc a}$ and $J\gamma_{\textsc a}^* = -\ii 2\, \gamma_{\textsc a}^*$, so $J$ leaves $\Gamma_{\textsc a}$ invariant. This implies that $\Gamma_{\textsc a}$ defines an uncorrelated subsystem. 

We compute:
\begin{align}
&\Pi^{+}_1\gamma_{\textsc a} = \gamma_{\textsc a}, \quad \Pi^{+}_1\gamma^*_{\textsc a} = 0, \quad
\Pi^{+}_2\gamma_{\textsc a} = 0, \quad \Pi^{+}_2\gamma^*_{\textsc a} = 0.
\end{align}
Thus, $\sum_I \Pi_I^+\Gamma_{\textsc a} = \mathrm{span}(\gamma_{\textsc a})$ is one-dimensional. It follows that  $\sum_I \dim(\Pi_I^+\Gamma_{\textsc a})-N_{\textsc a}=1-1=0$, and $A_{cp}$ is empty, as expected---because $\Gamma_{\textsc a}$ is uncorrelated.

\item $\gamma_{\textsc a} = e_1 + e_3$: \\
In this case, $J\gamma_{\textsc a} = \ii(2 e_1 + 3 e_3) \notin \Gamma_{\textsc a}$, so $A$ is correlated.

We compute:
\begin{align}
&\Pi^{+}_1\gamma_{\textsc a} = e_1, \quad \Pi^{+}_2\gamma_{\textsc a} = e_3, \quad
\Pi^{+}_1\gamma^*_{\textsc a} = 0, \quad  \Pi^{+}_2\gamma^*_{\textsc a} = 0.
\end{align}
Hence, $\sum_I \Pi_I^+\Gamma_{\textsc a} = \mathrm{span}(e_1, e_3)$ is two-dimensional. Therefore $\sum_I \dim( \Pi_I^+\Gamma_{\textsc a})-N_{\textsc a}=2-1=1$ so  $A_{cp}$ contains one mode.
\begin{widetext}
We have $\Gamma_{{\textsc a}_{cp}}$
\begin{equation}
\Gamma_{{\textsc a}_{cp}} = \bigoplus_{I=1}^{2}\, \mathrm{span}\big[
\Pi^{\perp}_{\textsc a}(\Pi^{+}_{I}\gamma_{\textsc a}), \,
\Pi^{\perp}_{\textsc a}(\Pi^{-}_{I}\gamma_{\textsc a}), \,
\Pi^{\perp}_{\textsc a}(\Pi^{+}_{I}\gamma^*_{\textsc a}), \,
\Pi^{\perp}_{\textsc a}(\Pi^{-}_{I}\gamma^*_{\textsc a})
\big],
\end{equation}
\end{widetext}
and a basis is given by:
\begin{equation}
\frac{1}{\sqrt{2}} (e_1 - e_3)
\end{equation}
and its complex conjugate.

\item $ \gamma_{\textsc a} = e_1 + e^*_1 + e_2$: \\
In this case $J\gamma_{\textsc a} = \ii 2 (e_1 - e_1^* + e_2) \notin \Gamma_{\textsc a}$, so $A$ is correlated. We find:
\begin{equation}
\Pi^{+}_1\gamma_{\textsc a} = e_1 + e_2, \quad \Pi^{+}_1\gamma^*_{\textsc a} = e_1, \quad
\Pi^{+}_2\gamma_{\textsc a} = 0, \quad \Pi^{+}_2\gamma^*_{\textsc a} = 0.
\end{equation}

Therefore, $\sum_I \Pi_I^+\Gamma_{\textsc a}$ is two-dimensional, and $A_{cp}$ is the single mode subsystem spanned by 
\begin{equation}
\Gamma_{{\textsc a}_{ep}} = \mathrm{span}\{ e_1^* + e_2 - e_2^*, 
e_1 + e_2^* - e_2\}.
\end{equation}

\item $\gamma_{\textsc a} = \frac{1}{\sqrt{2}}(e_1 + e^*_1 + e_2 + e_3)$: \\
This subsystem is correlated. We compute:
\begin{align}
&\Pi_1^+ \gamma_{\textsc a} = \frac{1}{\sqrt{2}}( e_1 + e_2), & \Pi_1^+ \gamma^*_{\textsc a} = \frac{1}{\sqrt{2}}e_1,\\
&\Pi_2^+ \gamma_{\textsc a} =  \frac{1}{\sqrt{2}}e_3, & \Pi_2^+ \gamma^*_{\textsc a} = 0.
\end{align}
Then,  $\sum_I \Pi_I^+\Gamma_{\textsc a}$ is three-dimensional, so $A_{cp}$ contains two modes. Explicitly, 
\begin{equation}\begin{split}
\Gamma_{{\textsc a}_{cp}} =& \mathrm{span}\{e_1+e_1^* + e_2  -e_3, e_1 -e_1^* + e_2^* -2e_3 + e_3^*, \\ 
&e_1+e_1^*+ e_2^* -e_3^* ,e_1^*  -e_1 + e_2 -2e_3^*  + e_3 \}.
\end{split}
\end{equation}
While the correlation partner of $A$ is a two-mode subsystem, its entanglement partner---defined in Prop.~\ref{prop:EPpartner}---is single-mode. 

To identify the entanglement partner of $A$, we write the covariance matrix in a symplectic orthonormal basis whose first two basis vectors are $\gamma_{\textsc a}$ and $\gamma_{\textsc a}^*$. In this basis,  the partial transposition with respect to $A$ is implemented as described in appendix~\ref{Negativity}. Applying the construction of Prop.~\ref{prop:EPpartner} yields the entanglement partner subsystem
$$\Gamma_{{\textsc a}_{ep}} = \mathrm{span}\{\gamma_{{\textsc a}_{ep}}, \gamma_{{\textsc a}_{ep}}^*\}, $$
where 
\begin{equation}\begin{split}
    \gamma_{{\textsc a}_{ep}} &= (1.0015 + 0.2342 \ii) e_1 \\ 
    &+ (0.0548 - 0.2342 \ii)e_1^* \\
    &+ (-0.5766 - 0.0113 \ii)e_2\\
    &+ (-0.5395 + 0.0484 \ii)e_2^*\\
   & + (-0.4797 + 0.0113 \ii)e_3 \\
    &+ (-0.5168 - 0.0484 \ii)e_3^*.
    \end{split}
\end{equation} 

The subsystems $\Gamma_A$ and $\Gamma_{{\textsc a}_{ep}}$ are symplectic orthogonal. Moreover, the reduced covariance matrix of the subsystem $\textsc a\oplus {\textsc a}_{ep}$, expressed on the aforementioned basis, for which 
\begin{align}
T_{\textsc a} =  \left( \begin{matrix}
    1 & 0 \\ 
    0 & -1 
\end{matrix}\right) \oplus \openone_{2},    
\end{align} 
takes the form 
\begin{align}
    \bm \sigma_{\textsc{a}\textsc{a}_{ep}} = \left(\begin{matrix}
    4.5 &  2& 1.1592& 3.0387\\ 
    2.& 4.5& 3.2377& -0.3085\\ 
    1.1592&3.2377& 3.5699& -0.4824\\
    3.0387&-0.3085& -0.4824&4.0472
\end{matrix}\right).
\end{align}
One can numerically verify that the partially transposed covariance matrix associated with $\bm \sigma_{\textsc{aa}_{ep}} $ has the same smallest symplectic eigenvalue as the partially transposed covariance matrix of the three-mode system. Consequently, $A \oplus A_{ep}$ reproduces the logarithmic negativity of the original state, as expected. 

This example illustrates the differences between the correlation and entanglement partners for mixed Gaussian states: while the correlation partner $A_{cp}$ is a two-mode subsystem, all the distillable entanglement between $A$ and its complement can be `localized' into the single-mode entanglement partner $\textsc{a}_{ep}$.

\end{enumerate}

\section{PPT and Negativity for Gaussian states}\label{Negativity}

A well-established approach to quantifying bipartite entanglement in Gaussian states is based on the Peres–Horodecki positive partial transpose (PPT) criterion for continuous variable systems~\cite{Simon2000}. For large families of Gaussian states, including one-versus-many-mode partitions~\cite{Werner:2000pxg} and bisymmetric~\cite{Serafini2005} , this criterion is both necessary and sufficient for separability. This makes computable measures such as the negativity and logarithmic negativity particularly effective tools to study entanglement in these systems~\cite{VidalNegativity,Plenio2005logneg,Audenaert2003}. 

In this appendix, we will summarize the notions of PPT and Negativity as a way to characterize entanglement in mixed states.

The \emph{negativity} of a bipartite quantum state $\hat\rho_{{\textsc a} {\textsc b}}$ is defined as
\begin{equation}\label{eq:negativity}
    \mathcal{N}(\hat{\rho}_{{\textsc a} {\textsc b}}) \coloneqq \frac{\|\hat{\rho}_{{\textsc a} {\textsc b}}^{T_{\textsc b}}\|_1 - 1}{2},
\end{equation}
where $\| \hat{O} \|_1 \equiv \Tr \sqrt{\hat{O}^\dagger \hat{O}}$ denotes the trace norm of the operator $\hat{O}$, and $\hat{\rho}_{{\textsc a} {\textsc b}}^{T_{\textsc b}}$ indicates the partial transpose of the density matrix $\hat{\rho}_{{\textsc a} {\textsc b}}$ with respect to subsystem $B$.  Since $\Tr(\hat{\rho}_{{\textsc a} {\textsc b}}^{T_{\textsc b}}) = \Tr(\hat{\rho}_{{\textsc a} {\textsc b}}) = 1$ and $\hat{\rho}_{{\textsc a} {\textsc b}}^{T_{\textsc b}}$ is self-adjoint, one can equivalently interpret~\eqref{eq:negativity} as the absolute value of the sum of the negative eigenvalues of $\hat{\rho}_{{\textsc a} {\textsc b}}^{T_{\textsc b}}$.

The intuitive reason why this quantity quantifies entanglement is that for any separable state, the partial transpose remains a valid density operator, implying that $\|\hat{\rho}_{{\textsc a} {\textsc b}}^{T_{\textsc b}}\|_1=1$ and thus $\mathcal{N}(\hat{\rho}_{{\textsc a} {\textsc b}})=0$. Consequently, a strictly positive value of $\mathcal{N}$ certifies entanglement. Such states are often called \emph{non-PPT states}, in contrast to the class of \emph{PPT states} for which $\mathcal{N}(\hat{\rho}_{{\textsc a} {\textsc b}})=0$. This provides an especially convenient entanglement test, since computing the spectrum of a partially transposed density operator is typically far easier than evaluating other mixed-state entanglement measures. Beyond being a detection tool, the negativity is also an entanglement \emph{monotone}, i.e., it cannot increase under local operations and classical communication (LOCC)~\cite{VidalNegativity}. This property supports its interpretation as a measure of the strength of entanglement for certain families of states.  

Closely related to $\mathcal{N}$ is the \emph{logarithmic negativity}, defined as
\begin{equation}
    E_{\mathcal{N}}(\hat{\rho}_{{\textsc a} {\textsc b}}) \coloneqq \log\!\big(2\mathcal{N}(\hat{\rho}_{{\textsc a} {\textsc b}})+1\big) = \log_2 \|\hat{\rho}_{{\textsc a} {\textsc b}}^{T_{\textsc b}}\|_1 \, ,
\end{equation}
which has the additional feature of bounding from above another key entanglement measure: the \emph{distillable entanglement}. The latter quantifies the number of maximally entangled pairs (EPR pairs) that can be asymptotically extracted from $\hat{\rho}_{{\textsc a} {\textsc b}}^{\otimes N}$ by LOCC in the limit $N \to \infty$.\footnote{For precise definitions and a detailed discussion, see~\cite{Audenaert2003, entanglementmeasuresreview}.} Importantly, states with vanishing negativity—commonly referred to as positive-partial-transpose (PPT) states—necessarily have zero distillable entanglement~\cite{VidalNegativity}.  

Altogether, these properties establish the negativity and logarithmic negativity as extremely practical and informative tools for the study of entanglement in mixed states.

When focusing on entanglement in Gaussian states, the simplifications are substantial. First, a general Gaussian state is completely specified by only $N(2N+3)$ real parameters, subject to the constraint $\sigma \geq \ii \Omega^{-1}$ (which ensures consistency with the canonical commutation relations).
This is in sharp contrast with general states of an infinite-dimensional Hilbert space. Second, bipartite entanglement admits a particularly transparent characterization for Gaussian bisymmetric states\footnote{These are Gaussian states that remain invariant under arbitrary mode permutations within each party of the bipartition.} and for partitions of one mode versus $M<N$ modes, where the celebrated Peres–Horodecki criterion becomes both necessary and sufficient~\cite{Simon2000, Serafini2005}. In these Gaussian scenarios, separable states coincide exactly with the class of positive-partial-transpose (PPT) states, and therefore the negativity as well as the logarithmic negativity serve as \emph{faithful} entanglement monotones.  

Even outside these cases—where negativity and logarithmic negativity are no longer strictly faithful monotones—they remain powerful tools, since they still provide relevant information about distillable entanglement~\cite{VidalNegativity, Audenaert2003, Plenio2005logneg, entanglementmeasuresreview}.  

Consider now a bipartition of the $N$ total bosonic modes into two complementary groups, labeled $A$ and $B$, consisting of $N_\textsc{a}$ and $N_\textsc{b}$ modes, respectively. If the full system is in a Gaussian state with covariance matrix $\sigma$, the covariance matrix of the partially transposed state with respect to subsystem $B$ is defined as
\begin{equation}
{\sigma}^{T_{\textsc b}} = (\openone_{\textsc a} \oplus T_\textsc{b}) \, \sigma \, (\openone_{\textsc a} \oplus T_\textsc{b}),
\end{equation}
where $T_\textsc{b}$ is (real, involutive) momentum flip operation on $A$, which in a canonical basis takes the matrix representation
\begin{equation}
T_\textsc{b} = \bigoplus_{j=1}^{N_\textsc{b}}
\begin{pmatrix}
    1 & 0 \\
    0 & -1
\end{pmatrix}.\label{T_operator_Canonical}
\end{equation}
Let $\{\tilde{\nu}_1,\ldots,\tilde{\nu}_{N}\}$ denote the symplectic eigenvalues of ${\sigma}^{T_{\textsc b}}$, i.e., the absolute values of the eigenvalues of $(J^{T_{\textsc b}})^a_{\, b}:=- \hbar \, \Omega^{ad}({\sigma}^{T_{\textsc b}})_{db}$. 
Then the logarithmic negativity is given by~\cite{VidalNegativity}
\begin{equation}\label{Eq:logneg_symplectic}
    E_{\mathcal{N}} = \sum_{j=1}^{N} \max\{0,-\log_2 \tilde{\nu}_j\}.
\end{equation}
Thus, for the cases of permutation invariant partitions, subsystems $A$ and $B$ are entangled if and only if ${\sigma}^{T_{\textsc b}}$ possesses at least one symplectic eigenvalue strictly smaller than unity, or equivalently, if ${\sigma}^{T_{\textsc b}} \ngeq \ii \Omega^{-1}$. The condition becomes only sufficient for any other partitions. In particular, relevant to this paper, for partitions of 1 versus $M<N$ modes the criterion is always necessary and sufficient and negativity and logarithmic negativity are faithful entanglement measures.

The condition of being non-PPT applied to a restricted complex structure becomes the following: a restricted complex structure  $J$ corresponds to a non-PPT state if the quadratic form $\langle  \,\cdot \, ,(-\openone-\ii J^{T_{\textsc b}}) \,  \, \,\cdot\rangle$  is not positive semi-definite, where $J^{T_{\textsc b}}\equiv  - \hbar \,\Omega\,  \sigma^{T_{\textsc b}}  =-\hbar\,\Omega\, T_{\textsc b}\,\sigma\,T_{\textsc b}$ was defined above.

\section{Derivation of the number of  symplectic eigenvalues of $J^{T_{\textsc a}}$ smaller than one}\label{numbdim}

This appendix contains a proof of the statement: 
``For a single-mode subsystem $A$, $J^{T_{\textsc a}}$ has at most one symplectic eigenvalue smaller than one,”  used in section \ref{subsec:partnermixed},  in the language used in this article.  (This proof is equivalent to the one provided in a Little Lemma in Section 7.4 in \cite{serafini2017quantum}.)

We here show a general statement that for an $N_A$-mode subsystem $A$, the number of symplectic eigenvalues of $J^{T_{\textsc a}}$ smaller than one is at most $N_A$. 

The argument becomes more transparent if we use a matrix representation of the quadratic form $\langle \,\cdot,(-\mathbb{I}-\ii J^{T_{\textsc a}})\, \cdot\, \rangle$ using a basis in $\Gamma_{\mathbb{C}}$ in which $\gamma_{x_{{\textsc a}i}}$ and $\gamma_{p_{{\textsc a}i}}$ are two basis elements for each mode, where $\{\gamma_{x_{{\textsc a}i}},\gamma_{p_{{\textsc a}i}}\}_{i=1}^{N_A}$ denote any Darboux basis in $A$. 

Using this basis, the operation  ``partial transposition with respect to the subsystem $A$'', denoted as $T_{\textsc a}$, amounts to reverse the sign of $\{\gamma_{p_{{\textsc a}i}}\}_{i=1}^{N_A}$ (see Eq. \eqref{T_operator_Canonical}). Thus, $T_{\textsc a}$ reduces to the identity in the $2N-N_A$ subspace of $\Gamma_{\mathbb{C}}$ symplectically orthogonal to the direction $\{\gamma_{p_{{\textsc a}i}}\}_{i=1}^{N_A}$. This in turn implies that the quadratic form $\langle \,\cdot,(-\mathbb{I}-\ii J^{T_{\textsc a}})\, \cdot\, \rangle$ is positive definite in a $(2N-N_A)$-dimensional subspace of $\Gamma_{\mathbb{C}}$, since $\langle \,\cdot,(-\mathbb{I}-\ii J)\, \cdot\, \rangle$ is positive definite in $\Gamma_{\mathbb{C}}$  (see property 4 in appendix \ref{ProofJ}). 

On the other hand, the linear map $(-\mathbb{I}-\ii J^{T_{\textsc a}})$ has eigenvalues $\nu_I^{T_{\textsc a}}-1$ and $-\nu_I^{T_{\textsc a}}-1$ and eigenvectors $e^{T_{\textsc a}}_I$ and $e^{T_{\textsc a}*}_I$, respectively, $I=1,\ldots,N$, where $\nu_I^{T_{\textsc a}}>0$ are the symplectic eigenvalues of $J^{T_{\textsc a}}$.  

In the eigenspaces spanned by all $e^{T_{\textsc a}*}_I$, $\langle \,\cdot,(-\mathbb{I}-\ii\, J^{T_{\textsc a}})\, \cdot\, \rangle$ is positive semi-definite, because 
\begin{equation}\langle e^{T_{\textsc a}*}_I,(-\mathbb{I}-\ii J^{T_{\textsc a}})e^{T_{\textsc a}*}_I \rangle =(-\nu_I^{T_{\textsc a}}-1)\, \langle e^{T_{\textsc a}*}_I,e^{T_{\textsc a}*}_I \rangle
\end{equation}
is larger than zero, because the two factors appearing on the right hand side are negative real numbers. On the other hand, in the eigenspaces spanned by $e^{T_{\textsc a}}_I$
\begin{equation}
\langle e^{T_{\textsc a}}_I,(-\mathbb{I}-\ii\, J^{T_{\textsc a}})e^{T_{\textsc a}}_I \rangle =(\nu_I^{T_{\textsc a}}-1)\, \langle e^{T_{\textsc a}}_I,e^{T_{\textsc a}}_I \rangle.
\end{equation}
Because $\langle e^{T_{\textsc a}}_I,e^{T_{\textsc a}}_I \rangle>0$, this quantity would be negative if $\nu_I^{T_{\textsc a}}<1$. Because the quadratic form $\langle \,\cdot,(-\mathbb{I}-\ii J^{T_{\textsc a}})\, \cdot\, \rangle$ is positive definite in a $(2N-N_A)$-dimensional subspace, $\nu_I^{T_{\textsc a}}$ can be smaller than one, at most, for $N_A$ vectors, which, without loss of generality, we choose to be $I=1,...,m$ with some $m\leq N_A$.

\section{Existence of the partner mode in the pure-state case}\label{app:ExistenceProof_PureState}
Let $A$ be a single-mode subsystem, and let $\Gamma_{\textsc a}=\mathrm{span}\{\gamma_{\textsc a},\gamma_{\textsc a}^*\}$, with $\langle \gamma_{\textsc a},\gamma_{\textsc a}\rangle=1$. 
Assume the global Gaussian state is pure, so that $J^2=-\openone$. 
The proposed partner is $\Gamma_{\textsc a_p}=\mathrm{span}\{ \eta , \eta ^*\}$, where
\begin{equation}
\eta \coloneqq \Pi_{\textsc a}^{\perp}(J\gamma_{\textsc a}) .
\end{equation}
To prove that this defines a bona fide single mode, it is enough to show that $\eta$ has non-vanishing symplectic norm,
\begin{equation}
\langle \eta,\eta\rangle \neq 0.
\end{equation}
Indeed, if $\langle \eta,\eta\rangle\neq 0$, then $\eta$ and $\eta^*$ are linearly independent and span a two-dimensional symplectic subspace, i.e.\ a single mode.

We first decompose $J\gamma_{\textsc a}$ into its components parallel and orthogonal to $\Gamma_{\textsc a}$:
\begin{equation}
J\gamma_{\textsc a}= \Pi_{\textsc a}(J\gamma_{\textsc a})+\Pi_{\textsc a}^{\perp}(J\gamma_{\textsc a})
= J_{\textsc a}\gamma_{\textsc a}+\eta ,
\end{equation}
where $J_{\textsc a}\coloneqq \Pi_{\textsc a}J\Pi_{\textsc a}$ is the restriction of $J$ to $A$.

Using that $\Gamma_{\textsc a}$ and $\Gamma_{\textsc a}^{\perp}$ are symplectically orthogonal, we find
\begin{equation}\label{eq:norm_decomp_partner_existence}
\langle J\gamma_{\textsc a},J\gamma_{\textsc a}\rangle
=
\langle J_{\textsc a}\gamma_{\textsc a},J_{\textsc a}\gamma_{\textsc a}\rangle
+
\langle \eta,\eta\rangle .
\end{equation}
Now, because the global state is pure, $J$ is a symplectic transformation, and hence it preserves the complexified symplectic product:
\begin{equation}
\langle J\gamma_{\textsc a},J\gamma_{\textsc a}\rangle
=
\langle \gamma_{\textsc a},\gamma_{\textsc a}\rangle
=1 .
\end{equation}
Therefore, it follows from \eqref{eq:norm_decomp_partner_existence}
\begin{equation}\label{eq:eta_norm_from_restriction}
\langle \eta,\eta\rangle
=
1-\langle J_{\textsc a}\gamma_{\textsc a},J_{\textsc a}\gamma_{\textsc a}\rangle .
\end{equation}

We now evaluate the second term. Since $A$ is a single mode, $J_{\textsc a}$ acts on the two-dimensional space $\Gamma_{\textsc a}=\mathrm{span}\{\gamma_{\textsc a},\gamma_{\textsc a}^*\}$. We can hence write
\begin{equation}
J_{\textsc a}\gamma_{\textsc a}=a\,\gamma_{\textsc a}+b\,\gamma_{\textsc a}^* ,
\end{equation}
for some $a,b\in\mathbb{C}$. Then, using
$
\langle \gamma_{\textsc a},\gamma_{\textsc a}\rangle=1
$,
$
\langle \gamma_{\textsc a}^*,\gamma_{\textsc a}^*\rangle=-1
$,
and
$
\langle \gamma_{\textsc a},\gamma_{\textsc a}^*\rangle=0
$,
we obtain
\begin{equation}\label{eq:innerprod_1}
\langle J_{\textsc a}\gamma_{\textsc a},J_{\textsc a}\gamma_{\textsc a}\rangle
=
|a|^2-|b|^2 .
\end{equation}
Since
\begin{equation}
J_A\gamma_A = a\,\gamma_A + b\,\gamma_A^*,
\label{eq:JA_on_gamma}
\end{equation}
and $J_A$ is real, we also have
\begin{equation}
J_A\gamma_A^* = (J_A\gamma_A)^* = b^*\,\gamma_A + a^*\,\gamma_A^*.
\label{eq:JA_on_gamma_conj}
\end{equation}
Therefore, in the basis $\{\gamma_A,\gamma_A^*\}$, the matrix representation of $J_A$ is
\begin{equation}
[J_A]_{\{\gamma_A,\gamma_A^*\}} =
\begin{pmatrix}
a & b^* \\
b & a^*
\end{pmatrix}.
\label{eq:JA_matrix}
\end{equation}
Its determinant is then
\begin{equation}
\det J_\textsc{A} = a a^* - b b^* = |a|^2 - |b|^2.
\label{eq:det_JA}
\end{equation}
Since the symplectic eigenvalues of $J_{\textsc a}$ are $\pm \ii \nu_{\textsc a}$, one has
\begin{equation}
\det J_{\textsc a}=\nu_{\textsc a}^2 .
\end{equation}
Combining this with \eqref{eq:innerprod_1} and \eqref{eq:det_JA}, we obtain
\begin{equation}
\langle J_{\textsc a}\gamma_{\textsc a},J_{\textsc a}\gamma_{\textsc a}\rangle
=\nu_{\textsc a}^2 .
\end{equation}
Substituting into \eqref{eq:eta_norm_from_restriction}, we conclude
\begin{equation}\label{eq:eta_norm_final}
\langle \eta,\eta\rangle
=
1-\nu_{\textsc a}^2 .
\end{equation}

Finally, if $A$ is correlated with its complement and the global state is pure, then the reduced state of $A$ is mixed, and therefore $\nu_{\textsc a}>1$. It follows that
\begin{equation}
\langle \eta,\eta\rangle = 1-\nu_{\textsc a}^2 \neq 0 .
\end{equation}
Therefore $\eta$ and $\eta^*$ span a two-dimensional symplectic subspace, and the partner mode exists.

Conversely, if $\nu_{\textsc a}=1$, then the reduced state of $A$ is pure. Since the global state is also pure, it follows that $A$ is uncorrelated with its complement. By Proposition~\hyperlink{prop:uncorrelated}{1}, this is equivalent to $J\Gamma_{\textsc a}=\Gamma_{\textsc a}$, and therefore
\begin{equation}
\Pi_{\textsc a}^{\perp}(J\gamma_{\textsc a})=0.
\end{equation}
Hence $\eta=0$, and no nontrivial partner exists.

\section{Uniqueness of the partner mode in the pure-state case}\label{app:UniquenessProof_PureState}
Let $A$ be a correlated single-mode subsystem of a bosonic system prepared in a pure Gaussian state, and let $J$ be the associated complex structure, so that $J^2=-\openone$. Suppose $A_p$ and $A_p'$ are two single-mode subsystems, both independent of $A$, satisfying the definition of partner. By definition, this means that both
\begin{equation}
\Gamma_{\textsc a}\oplus \Gamma_{{\textsc a}_p}
\qquad\text{and}\qquad
\Gamma_{\textsc a}\oplus \Gamma_{{\textsc a}_p'}
\end{equation}
are uncorrelated with their respective symplectic complements. By Proposition~\hyperlink{prop:uncorrelated}{1}, this is equivalent to saying that both subspaces are invariant under $J$:
\begin{equation}\label{eq:Jinv_two_partners}
J(\Gamma_{\textsc a}\oplus \Gamma_{{\textsc a}_p})
=
\Gamma_{\textsc a}\oplus \Gamma_{{\textsc a}_p},
\qquad
J(\Gamma_{\textsc a}\oplus \Gamma_{{\textsc a}_p'})
=
\Gamma_{\textsc a}\oplus \Gamma_{{\textsc a}_p'}.
\end{equation}

We will show that this implies
\begin{equation}
\Gamma_{{\textsc a}_p'}=\Gamma_{{\textsc a}_p}.
\end{equation}

Since $A$ is correlated with its complement, $J\Gamma_{\textsc a}\not\subset \Gamma_{\textsc a}$. Let $\gamma_{\textsc a}$ be a normalized such that $\Gamma_{\textsc a}=\mathrm{span}\{\gamma_{\textsc a},\gamma_{\textsc a}^*\}$ and $\langle \gamma_{\textsc a},\gamma_{\textsc a}\rangle=1$. Because $\Gamma_{\textsc a}\oplus \Gamma_{{\textsc a}_p}$ is $J$-invariant, we must have
\begin{equation}
J\gamma_{\textsc a}\in \Gamma_{\textsc a}\oplus \Gamma_{{\textsc a}_p}.
\end{equation}

Applying $\Pi_{\textsc a}^{\perp}$ (the symplectic projector  onto the complement of $A$) and using that $\Gamma_{{\textsc a}_p}\perp \Gamma_{\textsc a}$, we obtain
\begin{equation}\label{eq:proj_into_Ap}
\Pi_{\textsc a}^{\perp}(J\gamma_{\textsc a})\in \Gamma_{{\textsc a}_p}.
\end{equation}
Exactly the same argument, using the $J$-invariance of $\Gamma_{\textsc a}\oplus \Gamma_{{\textsc a}_p'}$, gives
\begin{equation}\label{eq:proj_into_App}
\Pi_{\textsc a}^{\perp}(J\gamma_{\textsc a})\in \Gamma_{{\textsc a}_p'}.
\end{equation}

From the proof in Appendix \ref{app:ExistenceProof_PureState}, for correlated $A$ the vector
\begin{equation}
\eta\coloneqq \Pi_{\textsc a}^{\perp}(J\gamma_{\textsc a})
\end{equation}
has non-vanishing symplectic norm,
\begin{equation}
\langle \eta,\eta\rangle \neq 0.
\end{equation}
Hence $\eta$ and $\eta^*$ are linearly independent and span a two-dimensional symplectic subspace:
\begin{equation}
\mathrm{span}\{\eta,\eta^*\}.
\end{equation}
But both $\Gamma_{{\textsc a}_p}$ and $\Gamma_{{\textsc a}_p'}$ are single-mode symplectic subspaces, hence both are two-dimensional. By \eqref{eq:proj_into_Ap} and \eqref{eq:proj_into_App} they both contain $\eta$, and therefore, being closed under complex conjugation, they also contain $\eta^*$. Therefore,
\begin{equation}
\Gamma_{{\textsc a}_p}
=
\mathrm{span}\{\eta,\eta^*\}
=
\Gamma_{{\textsc a}_p'}.
\end{equation}
This proves uniqueness of the partner mode.

\section{Proof of the existence and the uniqueness of the correlation partner}\label{app:ExistenceUniquenessProof_CorrelationPartner}

Let us first prove the following lemma:
\begin{lemma}\label{lem:J_inv_symplectic_subspace}
    Let $J:\Gamma_{\mathbb{C}}\to \Gamma_{\mathbb{C}}$ be a restricted complex structure such that $(J\gamma)^*= J\gamma^*$. If a subspace $V\subset \Gamma_{\mathbb{C}}$ satisfies $V^*=V$ and is $J$-invariant, i.e.,
    \begin{align}
        JV\subset V,
    \end{align}
    then $V$ is a symplectic subspace, i.e., the restriction of $\Omega$ onto $V$ is non-degenerate. 
\end{lemma}
\begin{proof}
    Suppose that $v \in V$ satisfies
    \begin{align}
        \Omega(v,u)=0,\quad \forall u\in V.
    \end{align}
    The restriction of $\Omega$ onto $V$ is non-degenerate if and only if the above condition implies $v=0$. 
    
    Since $V^*=V$, we have $v^*\in V$. Since $JV\subset V$, we also have $Jv^*\in V$. Therefore, taking $u=Jv^*$ gives
    \begin{align}
        \Omega (v,Jv^*)=0.
    \end{align}
    By using $\sigma(\cdot,\cdot)=-\hbar^{-1}\Omega(\cdot,J\cdot)$, this is equivalent to
    \begin{align}
        \sigma(v,v^*)=0.\label{eq:vanishing_metric}
    \end{align}

    Let us write $v$ as $v=x+\ii y$ with $x,y\in \Gamma$. Since $v^*=x-\ii y$, we have
    \begin{align}
        \sigma(v,v^*)&=\sigma(x,x)+\sigma(y,y)+\ii \left(\sigma(y,x)-\sigma(x,y)\right)\nonumber \\
        &=\sigma(x,x)+\sigma(y,y),
    \end{align}
    where in the last line, we used the fact that $\sigma$ is symmetric. Since $\sigma$ is positive definite on $\Gamma$, Eq.~\eqref{eq:vanishing_metric} holds if and only if $x=y=0$, or equivalently, $v=0$. Therefore, the only vector in $V$ that is $\Omega$-orthogonal to all elements of $V$ is the zero vector, implying that the restriction of $\Omega$ onto $V$ is non-degenerate.
\end{proof}

We also prove a lemma on a linear map:
\begin{lemma}\label{lem:minimal_inv_subspace}
    Let $V$ be a finite-dimensional complex vector space and $f:V\to V$ be a linear map. Assume that $f$ is diagonalizable, i.e.,
    \begin{align}
        f=\sum_{I}\lambda_I P_I,
    \end{align}
    where $\lambda_I\in\mathbb{C}$ such that $\lambda_I\neq \lambda_{I'}$ if $I\neq I'$, and $P_I$ are projectors satisfying 
    \begin{align}
        P_IP_{I'}=\delta_{II'} P_I,\qquad \sum_I P_I=\mathbb{I}. 
    \end{align}
    Then, for a subspace $W\subset V$, the minimal $f$-invariant subspace containing $W$ is uniquely given by
    \begin{align}
        \sum_I P_I(W).
    \end{align}
\end{lemma}
\begin{proof}
    Define $U\coloneqq \sum_I P_I(W)$. We first show that $U$ contains $W$. Since $\sum_I P_I=\mathbb{I}$, for any $w\in W$, we have
    \begin{align}
        w=\sum_{I}P_I(w).
    \end{align}
    Since each $P_I(w)$ belongs to $P_I(W)$, it follows that $w\in U$, implying that $U$ contains $W$. 

    Next, we show that $U$ is $f$-invariant. For any $u\in U$, there exists a vector $w_I\in W$ such that 
    \begin{align}
        u=\sum_IP_I(w_I).
    \end{align}
    Since $fP_I=\lambda_I P_I$, we obtain
    \begin{align}
        f(u)=\sum_I f(P_I(w_I))=\sum_I\lambda_I P_I(w_I).
    \end{align}
    For each $I$, we have $\lambda_IP_I(w_I) \in P_I(W)$. Hence, $f(u)\in \sum_IP_I(W)=U$. Therefore,
    \begin{align}
        f(U)\subset U,
    \end{align}
    i.e., $U$ is $f$-invariant.

    We now prove the minimality of $U$. Let $S\subset V$ be an arbitrary $f$-invariant subspace such that $W\subset S$. Since $S$ is $f$-invariant, it is also invariant under any polynomial of $f$. Because the spectral projector can be written as
    \begin{align}
        P_I=\prod_{J\neq I}\frac{f-\lambda_J\mathbb{I}}{\lambda_I-\lambda_J},
    \end{align}
    each $P_I$ is a polynomial in $f$. Therefore, 
    \begin{align}
        P_I(S)\subset S.
    \end{align}
    Since $W\subset S$, we also find $P_I(W)\subset P_I (S)$, implying that
    \begin{align}
        P_I(W)\subset S.
    \end{align}
    Since this holds for all $I$, we get $\sum_IP_I(W)\subset S$, i.e., $U\subset S$. Therefore, $U$ is the minimal $f$-invariant subspace containing $W$.

    We finally prove the uniqueness. Suppose that $U_1$ and $U_2$ are both minimal $f$-invariant subspaces containing $W$. Then, since $U_1$ is minimal, we have $U_1\subset U_2$. Similarly, since $U_2$ is minimal, $U_2\subset U_1$. Consequently, we obtain $U_1=U_2$, meaning that the minimal $f$-invariant subspace is unique, and it is given by $\sum_IP_I(W)$. 
\end{proof}

By using these lemmas, the existence and uniqueness of the correlation partner are proven as follows:
\begin{theorem}
    Let $\Gamma_{\textsc a}$ be a symplectic subspace of $\Gamma_{\mathbb{C}}$ which describes a subsystem. The subspace $\Gamma_{\textsc b}$ defined by
    \begin{align}
        \Gamma_{\textsc b} = \bigoplus_I \left(\Pi^+_I\Gamma_{\textsc a} \oplus \Pi^-_I \Gamma_{\textsc a}\right)
    \end{align}
    is the unique minimal $J$-invariant subspace that contains $\Gamma_{\textsc a}$. In addition, $\Gamma_{\textsc b}$ is a symplectic subspace. 
\end{theorem}
\begin{proof}
    From Lemma~\ref{lem:minimal_inv_subspace}, the unique minimal $J$-invariant subspace that contains $\Gamma_{\textsc a}$ is given by $\Gamma_{\textsc b}$, where we have used the fact that $J$ is diagonalizable and given by
    \begin{align}
        J=\sum_I\left(\ii \nu_I \Pi^+_I - \ii \nu_I \Pi^-_I \right). 
    \end{align}

    Since a symplectic subspace describing a subsystem is obtained by complexifying the corresponding real subspace of the classical phase space, $\Gamma_{\textsc a}=\Gamma_{\textsc a}^*$.
    From Eq.~\eqref{eq:projectors_basis_decomposition}, 
    \begin{align}
        \left(\Pi_I^{\pm}\gamma\right)^*=\Pi_I^{\mp} \gamma^*
    \end{align}
    holds for any vector $\gamma\in\Gamma_{\mathbb{C}}$. Therefore, by using $\Gamma_{\textsc a}=\Gamma_{\textsc a}^*$, we obtain 
    \begin{align}
        \left(\Pi_I^{\pm}(\Gamma_{\textsc a})\right)^*=\Pi_I^{\mp}\left(\Gamma_{\textsc a}\right).
    \end{align}
    Consequently, 
    \begin{align}
        \Gamma_{\textsc b}^*&= \bigoplus_I \left(\left(\Pi^+_I\Gamma_{\textsc a}\right)^* \oplus \left(\Pi^-_I \Gamma_{\textsc a}\right)^*\right)\\
        &=\bigoplus_I \left(\Pi_I^-\Gamma_{\textsc a}\oplus \Pi^+_I \Gamma_{\textsc a}\right)\\
        &=\Gamma_{\textsc b}.
    \end{align}
    Since $\Gamma_{\textsc b}$ is also $J$-invariant, Lemma~\ref{lem:J_inv_symplectic_subspace} implies that $\Gamma_{\textsc b}$ is a symplectic subspace. 
\end{proof}

Now, suppose that $\Gamma_{\textsc a}$ is a symplectic subspace of $\Gamma_{\mathbb{C}}$ which describes a correlated subsystem. In this case, $\Gamma_{\textsc a}$ is not $J$-invariant. Therefore, $\Gamma_{\textsc a}\subsetneq\Gamma_{\textsc b}$, or equivalently, the subspace
\begin{align}
    \Gamma_{{\textsc a}_{cp}} \coloneqq \Pi^\perp_{\textsc a}\left[ \bigoplus_I \left(\Pi^+_I\Gamma_{\textsc a} \oplus \Pi^-_I \Gamma_{\textsc a}\right)\right]
\end{align}
is not empty. Since $\Gamma_{\textsc a}$ and $\Gamma_{\textsc b}$ are symplectic subspaces, so is $\Gamma_{{\textsc a}_{cp}}$, proving the existence of the correlation partner. Its uniqueness follows from the uniqueness of $\Gamma_{\textsc b}$.

\section{Proof of existence of the entanglement partner}\label{app:proof_symplectic_subspace_ep}

To establish existence, it is sufficient to show that $\Gamma_{\text{\protect\textsc{A}}_{ep}}$ is a symplectic subspace, since this is equivalent to requiring that $\Gamma_{\text{\protect\textsc{A}}_{ep}}$ define a bona fide degree of freedom.

Since $(J^{T_{\textsc{A}}})^a_{\ b}=-\hbar \Omega^{ac}(\sigma^{T_{\textsc{A}}})_{cb}$, where $(\sigma^{T_{\textsc{A}}})_{cb}=(T_{\textsc{A}})^{d}_{\ c}(T_{\textsc{A}})^{e}_{\ b}\sigma_{de}$, we get
\begin{align}
        \Omega_{ac}(J^{T_{\textsc{A}}})^c_{\ b}=-\hbar(\sigma^{T_{\textsc{A}}})_{ab}.
\end{align} 
Using the symmetry of $(\sigma^{T_{\textsc{A}}})_{ab}$, this equation can be re-written as 
\begin{align}
        \Omega_{cb}(J^{T_{\textsc{A}}})^c_{\ a}=\hbar(\sigma^{T_{\textsc{A}}})_{ab}.
\end{align} 
Therefore, for arbitrary vectors $u, v \in\Gamma_{\mathbb{C}}$, we have
\begin{align}
        \sigma(T_{\textsc{A}}u,T_{\textsc{A}}v)&=u^av^b(\sigma^{T_{\textsc{A}}})_{ab}\nonumber\\
        &=\hbar^{-1}\Omega(J^{T_{\textsc{A}}}u,v)\nonumber\\
        &=\ii \langle (J^{T_{\textsc{A}}}u)^*,v\rangle \nonumber \\
        &=\ii \langle J^{T_{\textsc{A}}}(u^*),v\rangle, \label{eq:transpose_inner_prod}
\end{align}
where we have used the fact that $J^{T_{\textsc{A}}}$ is real in the last line. 

We denote by $\{e_I\}_{I=1}^N$ the positive-norm eigenvectors of $J^{T_{\textsc{A}}}$ such that
\begin{align}
    J^{T_{\textsc{A}}}e_I^{T_{\textsc{A}}}=\ii \nu_I^{T_{\textsc{A}}} e_I^{T_{\textsc{A}}},\quad \nu_I^{T_{\textsc{A}}}>0\label{eq:symp_e_vect_transposed}
\end{align}
and
\begin{align}
    \langle e_I^{T_{\textsc{A}}},e_J^{T_{\textsc{A}}}\rangle =-\langle e_I^{T_{\textsc{A}}*},e_J^{T_{\textsc{A}}*}\rangle =\delta_{IJ},\quad \langle e_I^{T_{\textsc{A}}},e_J^{T_{\textsc{A}}*}\rangle=0.
\end{align}
Without loss of generality, we may assume that the eigenvalues are ordered nondecreasingly.
When $J$ is non-PPT$_\textsc{A}$, some of the symplectic eigenvalues are smaller than one. We therefore assume that
\begin{align}
    0<\nu_1^{T_{\textsc{A}}}\leq \cdots \leq \nu_m^{T_{\textsc{A}}} <1 \leq \nu_{m+1}^{T_{\textsc{A}}}\leq \cdots \leq \nu_N^{T_{\textsc{A}}}\label{eq:ordering_symp_ev_transposed}
\end{align}
for some $m$ ($1\leq m\leq N$). Denoting by
\begin{align}
    E_+^{T_\textsc{A}}\coloneqq \mathrm{span}\{e_I^{T_{\textsc{A}}}\}_{I=1}^m,\label{eq:definition_Eplus}
\end{align}
the subspace spanned by eigenvectors of $J^{T_\textsc{A}}$ with symplectic eigenvalues smaller than one is expressed as
\begin{align}
    E^{T_\textsc{A}}=E_+^{T_\textsc{A}}\oplus E_+^{T_\textsc{A}*}.
\end{align}
Propositions~\ref{prop:EPpartner} and \ref{prop:entanglement_partner_general} state 
\begin{align}
    \Gamma_{{\textsc a}_{\mathrm{ep}}}=\Pi_{\textsc a}^\perp E^{T_{\textsc a}}
\end{align}
is the entanglement partner of $A$. 

In order to show that entanglement partner exists, i.e., that $ \Gamma_{{\textsc a}_{ep}}$ is a symplectic subspace, we first prove several lemmas.
\begin{lemma}
    For any nonzero
    \begin{align}
        x=\sum_{I=1}^mc_I e_I^{T_{\textsc{A}}} \in E_{+}^{T_\textsc{A}}
    \end{align}
    we have
    \begin{align}
        \langle\Pi_{\textsc{A}}^\perp x,\Pi_{\textsc{A}}^\perp x\rangle >0.
    \end{align}
    In particular, this implies $\Pi_{\textsc{A}}^\perp|_{E_{+}^{T_\textsc{A}}}$ is injective, and therefore,
    \begin{align}
        \dim F=m,\quad F\coloneqq \Pi_{\textsc{A}}^\perp E_{+}^{T_\textsc{A}}.
    \end{align}
\end{lemma}
\begin{proof}
    For nonzero $x=\sum_{I=1}^mc_I e_I$, we define
    \begin{align}
        p\coloneqq \langle x,x\rangle =\sum_I |c_I|^2>0.
    \end{align}
    From Eq.~\eqref{eq:transpose_inner_prod}, we have
    \begin{align}
        s\coloneqq \sigma (T_{\textsc{A}} x^*,T_{\textsc{A}}x)=\ii \langle J^{T_{\textsc{A}}}x,x\rangle=\sum_{I}\nu_I|c_I|^2. 
    \end{align}

    Since 
    \begin{align}
        T_{\textsc{A}}(\Omega_{\textsc{A}}\oplus \Omega_{\bar{\textsc{A}}})T_{\textsc{A}}=(-\Omega_{\textsc{A}})\oplus \Omega_{\bar{\textsc{A}}},
    \end{align}
    we obtain
    \begin{align}
        \langle T_{\textsc{A}}x , T_{\textsc{A}} x\rangle =-\langle \Pi_{\textsc{A}} x ,\Pi_{\textsc{A}} x\rangle+\langle \Pi_{\textsc{A}}^\perp x ,\Pi_{\textsc{A}}^\perp x\rangle.
    \end{align}
    By using
    \begin{align}
        \langle x, x \rangle = \langle \Pi_{\textsc{A}} x ,\Pi_{\textsc{A}} x\rangle+\langle \Pi_{\textsc{A}}^\perp x ,\Pi_{\textsc{A}}^\perp x\rangle,
    \end{align}
    we get
    \begin{align}
        \langle T_{\textsc{A}}x , T_{\textsc{A}} x\rangle=-p+2\langle \Pi_{\textsc{A}}^\perp x ,\Pi_{\textsc{A}}^\perp x\rangle.
    \end{align}
    Applying the uncertainty relation $\sigma(\gamma^*,\gamma)+\langle \gamma,\gamma\rangle\geq 0$ for $\gamma\coloneqq T_{\textsc{A}}x$, we get
    \begin{align}
        0\leq s- p +2\langle \Pi_{\textsc{A}}^\perp x ,\Pi_{\textsc{A}}^\perp x\rangle,
    \end{align}
    implying that
    \begin{align}
        \langle \Pi_{\textsc{A}}^\perp x ,\Pi_{\textsc{A}}^\perp x\rangle\geq \frac{1}{2}(p-s)=\frac{1}{2}\sum_{I}(1-\nu_I)|c_I|^2>0,
    \end{align}
    where we have used $\nu_I<1$. 
\end{proof}

\begin{lemma}\label{lem:Gamma_A_ep}
    $\Gamma_{\textsc{A}_{\mathrm{ep}}}=\Pi_{\textsc{A}}^\perp E^{T_{\textsc{A}}}=F\oplus F^*$ and therefore, 
    \begin{align}
        \dim \Gamma_{\textsc{A}_{\mathrm{ep}}}= 2m.
    \end{align}
\end{lemma}
\begin{proof}
    Since $\Pi_{\textsc{A}}^\perp$ is real, we have 
    \begin{align}
        \Pi_{\textsc{A}}^\perp\left((E_+^{T_\textsc{A}})^*\right)=\left(\Pi_{\textsc{A}}^\perp E_+^{T_\textsc{A}}\right)^*=F^*.
    \end{align}
    Therefore,
    \begin{align}
        \Gamma_{\textsc{A}_{ep}}=\Pi_{\textsc{A}}^\perp\left(E^{T_\textsc{A}}\right)=\Pi_{\textsc{A}}^\perp\left(E_+^{T_\textsc{A}}\oplus (E_+^{T_\textsc{A}})^*\right)=F+F^*.
    \end{align}

    Suppose that there exists a nonzero element in $ F\cap F^*$, which we denote by $z$. One one hand, since $z\in F$, $\langle z,z\rangle >0$. On the other hand, since $z\in F^*$, and therefore, there exists nonzero $f\in F$ such that $z=f^*$. Thus,
    \begin{align}
        \langle z,z\rangle = \langle f^*,f^*\rangle =-\langle f,f\rangle <0,
    \end{align}
    which contradicts with $\langle z,z\rangle >0$. Consequently, $F\cap F^*=\{0\}$, and therefore, 
    \begin{align}
        \Gamma_{\textsc{A}_{ep}}=F\oplus F^*= \Pi_{\textsc{A}}^\perp\left(E_+^{T_\textsc{A}}\right)\oplus \Pi_{\textsc{A}}^\perp\left(E_+^{T_\textsc{A}*}\right).
    \end{align}
\end{proof}

\begin{lemma}\label{lem:symp_subspace_lemma}
    Let $F$ be a subspace that is positive definite with respect to $\langle \cdot,\cdot \rangle$. Then 
    \begin{align}
        W\coloneqq F\oplus F^*
    \end{align}
    is a symplectic subspace. 
\end{lemma}

\begin{proof}
    Suppose that $z\in W$ satisfies
    \begin{align}
        \langle z, w \rangle=0\quad \forall w \in W.\label{eq:orthogonal_z_w}
    \end{align}
    We prove that this condition implies $z=0$. 
    Since $W=F\oplus F^*$, there uniquely exist $f,g\in F$ such that $z= f+g^*$. We define
    \begin{align}
        A\coloneqq \langle f, f\rangle \geq 0,\quad B\coloneqq \langle g,g\rangle \geq 0. 
    \end{align}
    
    Equation~\eqref{eq:orthogonal_z_w} with $w=f$ implies
    \begin{align}
        0=\langle z, w \rangle =\langle f, f \rangle + \langle g^*,f\rangle,
    \end{align}
    i.e.,
    \begin{align}
         \langle g^*,f \rangle =-A.\label{eq:A}
    \end{align}
    On the other hand, Eq.~\eqref{eq:orthogonal_z_w} with $w=g^*$ implies
    \begin{align}
        0=\langle f, g^* \rangle + \langle g^*,g^*\rangle,
    \end{align}
    i.e.,
    \begin{align}
        B=\langle f, g^* \rangle,
    \end{align}
    where we have used $\langle g^*,g^*\rangle=-\langle g,g\rangle$. 
    Since $B$ is real, this equation also implies
    \begin{align}
        B=B^*=\langle f, g^* \rangle^*=\langle g^*,f\rangle.\label{eq:B}
    \end{align}
    From Eqs.~\eqref{eq:A} and \eqref{eq:B}, we get
    \begin{align}
        A=-B.
    \end{align}
    From $A\geq 0$ and $B\geq 0$, this implies $A=B=0$. Since $F$ is positive definite with respect to $\langle\cdot,\cdot\rangle$, we get
    \begin{align}
        z=0. 
    \end{align}
    Therefore, the restriction of $\langle \cdot,\cdot\rangle$ to $W$ is nondegenerate. 

    It automatically follows that restriction of $\Omega$ to $W$ is nondegenerate. Indeed, since
    \begin{align}
        \langle u,v\rangle =\frac{1}{\ii\hbar}\Omega (u^*,v), 
    \end{align}
    if $z^*\in W$ satisfies
    \begin{align}
        \Omega(z^*,w)=0\quad \forall w\in W,
    \end{align}
    then
    \begin{align}
        \langle z,w\rangle=0\quad \forall w\in W.
    \end{align}
    As we have shown above, this condition implies $z=0$, which, in turn, means $z^*=0$. 
\end{proof}

\begin{proof}[Proof of the existence of entanglement partner]
    From Lemma~\ref{lem:Gamma_A_ep}, 
    \begin{align}
        \Gamma_{\textsc{A}_{\mathrm{ep}}}=\Pi_{\textsc{A}}^\perp E^{T_{\textsc{A}}}=F\oplus F^*,
    \end{align}
    where $F\coloneqq  \Pi_{\textsc{A}}^\perp E_{+}^{T_\textsc{A}}$. 
    Applying Lemma~\ref{lem:symp_subspace_lemma} to $F$, we find $\Gamma_{\textsc{A}_{\mathrm{ep}}}$ is a $2m$-dimensional symplectic subspace, which defines an $m$-mode subsystem. Note that since $\dim \Gamma_{\textsc{A}_{\mathrm{ep}}}= 2m$, and $m\leq N_{\textsc{A}}$ as shown in Appendix~\ref{numbdim}, the dimension of $\Gamma_{\textsc{A}_{\mathrm{ep}}}$ is at most $2N_{\textsc{A}}$. 

    We now decompose the total phase space as
    \begin{align}
        \Gamma_{\mathbb{C}}=\Gamma_{\textsc{A}}\oplus \Gamma_{\textsc{A}_{\mathrm{ep}}}\oplus \Gamma_{R}=\Gamma_{\textsc{A}}\oplus\Gamma_{\bar{\textsc{A}}}
    \end{align}
    where $\Gamma_R$ is a $2(N-(N_{\textsc{A}}+m))$-dimensional subspace symplectic orthogonal to $\Gamma_{\textsc{A}}\oplus \Gamma_{\textsc{A}_{\mathrm{ep}}}$, and $\Gamma_{\bar{\textsc{A}}}=\Gamma_{\textsc{A}_{\mathrm{ep}}}\oplus \Gamma_{R}$. 
    
    From Eqs.~\eqref{eq:symp_e_vect_transposed} and \eqref{eq:ordering_symp_ev_transposed}, the negativity between $\textsc{A}$ and its complement system, denoted by $E_{\mathcal{N}}(A:\bar{A})$ is given by
    \begin{align}
        E_{\mathcal{N}}(A:\bar{A})=-\sum_{I=1}^m\log_2 \nu_I^{T_{\textsc{A}}}.
    \end{align}
    On the other hand, since $e_I\in \Gamma_{\textsc{A}}\oplus \Gamma_{\textsc{A}_{\mathrm{ep}}}$, $e_I$ is also an eigenvector of the partially transposed complex structure $J_{\textsc{A}\textsc{A}_{\mathrm{ep}}}^{T_{\textsc{A}}}$ with symplectic eigenvalue $\nu_I^{T_{\textsc{A}}}$. Indeed, by using the projector $\Pi_{\textsc{A}\textsc{A}_{\mathrm{ep}}}$ onto $\Gamma_{\textsc{A}\textsc{A}_{\mathrm{ep}}}$, we have
    \begin{align}
        J_{\textsc{A}\textsc{A}_{\mathrm{ep}}}^{T_{\textsc{A}}}e_I^{T_{\textsc{A}}}&=\Pi_{\textsc{A}\textsc{A}_{\mathrm{ep}}}J^{T_{\textsc{A}}}\Pi_{\textsc{A}\textsc{A}_{\mathrm{ep}}}e_I^{T_{\textsc{A}}}\\
        &=\Pi_{\textsc{A}\textsc{A}_{\mathrm{ep}}}\ii\nu_I e_I^{T_{\textsc{A}}}=\ii\nu_I e_I^{T_{\textsc{A}}}.
    \end{align}
    where in the first line we have used $J_{\textsc{A}\textsc{A}_{\mathrm{ep}}}^{T_{\textsc{A}}}=\Pi_{\textsc{A}\textsc{A}_{\mathrm{ep}}}J^{T_{\textsc{A}}}\Pi_{\textsc{A}\textsc{A}_{\mathrm{ep}}}$, which follows from the fact that $\Pi_{\textsc{A}\textsc{A}_{\mathrm{ep}}}$ commutes with $T_{\textsc{A}}$. Therefore,
    \begin{align}
        E_{\mathcal{N}}(A:A_{\mathrm{ep}})\geq -\sum_{I=1}^m\log_2 \nu_I^{T_{\textsc{A}}}=E_{\mathcal{N}}(A:\bar{A})
    \end{align}
    However, since the logarithmic negativity is monotonic under LOCC, in particular under partial trace, we also have
    \begin{align}
        E_{\mathcal{N}}(A:A_{\mathrm{ep}})\leq E_{\mathcal{N}}(A:\bar{A}),
    \end{align}
    implying that
    \begin{align}
        E_{\mathcal{N}}(A:A_{\mathrm{ep}})=E_{\mathcal{N}}(A:\bar{A}).
    \end{align}
    That is, $A_{\mathrm{ep}}$ captures all the logarithmic negativity entanglement between $A$ and its complement. 
\end{proof}

\section{Proof of the uniqueness of entanglement partner}\label{app:UniquenessProof_EntanglementPartner}
In this subsection, we prove that for a subsystem $A$ and its complement $\bar{A}$ in an $N$-mode subsystem, if a subsystem $B$ of $\bar{A}$ reproduces the full logarithmic negativity between $A$ and $\bar{A}$, then $B$ must necessarily contain the entanglement partner $A_{\mathrm{ep}}$. In other words, $A_{\mathrm{ep}}$ is the unique minimal subsystem of $\bar{A}$ that reproduces all the logarithmic negativity between $A$ and its complement.

In this section, we adopt the notation on the eigenvectors of partially transposed complex structures in the previous section, namely, Eqs.~\eqref{eq:symp_e_vect_transposed}--\eqref{eq:definition_Eplus}. 

\begin{lemma}\label{lem:sesquilinear_form_Q}
    Define a sesquilinear form $Q:\Gamma_{\mathbb{C}}\times \Gamma_{\mathbb{C}}\to \mathbb{C}$ by
    \begin{align}
        Q(u,v)\coloneqq \langle u,-\ii J^{T_{\textsc a}} v\rangle,\quad u,v\in\Gamma_{\mathbb{C}}. 
    \end{align}
    Fix a symplectic eigenvalue $\nu_I^{T_{\textsc a}}$ of $J^{T_{\textsc a}}$. 
    If a vector $f\in \Gamma_{\mathbb{C}}$ satisfies $ \langle f,f\rangle =1$ and 
    \begin{align}
        \langle f , e_L^{T_{\textsc a}}\rangle =0 \quad \forall L\text{ s.t. }\nu_L^{T_{\textsc a}}<\nu_I^{T_{\textsc a}},\label{eq:vector_cutoff}
    \end{align}
    then
    \begin{align}
        Q(f,f)\geq \nu_{I}^{T_{\textsc{a}}}.
    \end{align}
    Moreover, equality holds if and only if $f$ belongs to the positive-norm eigenspace of $J^{T_{\textsc{A}}}$ with symplectic eigenvalue $\nu_{I}^{T_{\textsc{a}}}$. 
\end{lemma}
\begin{proof}
    From Eq.~\eqref{eq:vector_cutoff}, the vector $f$ can be expanded as
    \begin{align}
        f=\sum_{L;\,\nu_L\geq \nu_I}\alpha_L e_L^{T_{\textsc a}}+\sum_{K=1}^N\beta_Ke_K^{T_{\textsc a}*}
    \end{align}
    with $\alpha_I,\beta_I\in\mathbb{C}$. Since $f$ is normalized, i.e., $\langle f,f\rangle=1$, the coefficients satisfy
    \begin{align}
        \sum_{L;\,\nu_L\geq \nu_I}|\alpha_L|^2-\sum_{K=1}^N|\beta_K|^2=1.\label{eq:Bog_normalization}
    \end{align}
    Since
    \begin{align}
        -\ii J^{T_{\textsc a}}(\cdot)= \sum_{K=1}^N\left(\nu_K^{T_{\textsc a}}e_K^{T_{\textsc a}}\langle e_K^{T_{\textsc a}},\cdot\rangle+\nu_K^{T_{\textsc a}}(e_K^{T_{\textsc a}})^*\langle (e_K^{T_{\textsc a}})^*,\cdot\rangle  \right),
    \end{align}
    we obtain
    \begin{align}
        Q(f,f)
        &=\sum_{L;\,\nu_L\geq \nu_I}\nu_L^{T_{\textsc a}}|\alpha_L|^2+\sum_{K=1}^N\nu_K^{T_{\textsc a}}|\beta_K|^2 .
    \end{align}
    By using Eq.~\eqref{eq:Bog_normalization}, we further get
    \begin{align}
        &Q(f,f)-\nu_I^{T_{\textsc a}}\nonumber \\
        &=\sum_{L;\,\nu_L\geq \nu_I}(\nu_L^{T_{\textsc a}}-\nu_I^{T_{\textsc a}})|\alpha_L|^2+\sum_{K=1}^N(\nu_K^{T_{\textsc a}}+\nu_I^{T_{\textsc a}})|\beta_K|^2  
    \end{align}
    Since each term in the right-hand side is non-negative, we find
    \begin{align}
        Q(f,f)\geq\nu_I^{T_{\textsc a}} 
    \end{align} 
    and equality holds if and only if $\beta_K=0$ for all $K$ and $\alpha_L=0$ for all $L$ such that $\nu_L \neq \nu_I$. 
\end{proof}

\begin{lemma}\label{lem:ordering_ev_subsystem}
    Let $B$ be a subsystem of $\bar{A}$. We denote by $J_{AB}^{T_{\textsc a}}=\Pi_{AB}J^{T_{\textsc a}}\Pi_{AB}$ the partially transposed complex structure of the reduced state on $AB$. We denote the symplectic eigenvalues of $J_{AB}^{T_{\textsc a}}$, ordered nondecreasingly, by
    \begin{align}
    0<\mu_1^{T_{\textsc{A}}}\leq \cdots \leq \mu_{N_{AB}}^{T_{\textsc{A}}},
    \end{align}
    where $N_{AB}$ denotes the number of modes of the system $AB$. 
    Then,
    \begin{align}
        \mu_I^{T_{\textsc{A}}}\geq \nu_I^{T_{\textsc{A}}}\quad \forall I=1,...,N_{AB}.
    \end{align}
\end{lemma}
\begin{proof}
    We denote by $\{f_I\}_{I=1}^N$ the eigenvectors of $J_{AB}^{T_{\textsc a}}$ such that
    \begin{align}
        J_{AB}^{T_{\textsc a}}f_I =\ii \mu_I^{T_{\textsc{A}}}f_I ,\quad \langle f_I,f_J\rangle=\delta_{IJ}.
    \end{align}

    Fix $I\in \{1,\cdots,k\}$, and define 
    \begin{align}
        V_I\coloneqq \mathrm{span}\{f_1,\cdots,f_I\},
    \end{align}
    which is positive definite with respect to $\langle\cdot,\cdot\rangle$. We consider a set of $I-1$ conditions for a vector $f\in V_I$ 
    \begin{align}
        \langle f, e_1\rangle=\cdots = \langle f, e_{I-1}\rangle =0.
    \end{align}
    Since $\dim V_I= I$, there exists a solution $f\in V_I$ satisfying these $I-1$ conditions. After normalizing the solution, we can assume $\langle f,f\rangle =1$. Applying Lemma~\ref{lem:sesquilinear_form_Q} to this vector $f$, we have
    \begin{align}
        Q(f,f)\geq  \nu_{I}^{T_{\textsc{a}}}.
    \end{align}

    Since $f\in V_I$ and normalized, it can be expanded as
    \begin{align}
        f=\sum_{K=1}^Ic_Kf_K,\quad \sum_{K=1}^I|c_K|^2=1.
    \end{align}
    On the other hand, since $f\in \Gamma_A\oplus \Gamma_B$, we obtain
    \begin{align}
        Q(f,f)&=\langle f,-\ii J^{T_{\textsc a}} f\rangle\\
        &=\langle\Pi_{AB} f,-\ii J^{T_{\textsc a}} \Pi_{AB}f\rangle\\
        &=\langle f,-\ii J_{AB}^{T_{\textsc a}}f\rangle\\
        &=\sum_{K=1}^I\mu_K^{T_{\textsc{A}}}|c_K|^2\\
        &\leq \mu_I^{T_{\textsc{A}}}\sum_{K=1}^I|c_K|^2= \mu_I^{T_{\textsc{A}}},
    \end{align}
    where in the last line, we used $\mu_K^{T_{\textsc{A}}}\leq \mu_I^{T_{\textsc{A}}}$ for $K\leq I$. Therefore,
    \begin{align}
        \mu_I^{T_{\textsc{A}}}\geq  Q(f,f)\geq  \nu_{I}^{T_{\textsc{a}}}. 
    \end{align}
\end{proof}

By using this lemma, we establish the following key observation.
\begin{lemma}\label{lem:negativity_same_ev}
    Let $B$ be a subsystem of $\bar{A}$. If 
    \begin{align}
        E_{\mathcal{N}}(A:B)=E_{\mathcal{N}}(A:\bar{A}),
    \end{align}
    then
    \begin{align}
        m\leq N_{AB}
    \end{align}
    and
    \begin{align}
        \mu_I^{T_{\textsc{A}}}=\nu_{I}^{T_{\textsc{a}}}
    \end{align}
    for all $I=1,\cdots,m\leq N_{AB}$. 
\end{lemma}
\begin{proof}
    From Lemma~\ref{lem:ordering_ev_subsystem}, $ \mu_I^{T_{\textsc{A}}}\geq \nu_{I}^{T_{\textsc{a}}}$ for all $I=1,...,N_{AB}$. Since $-\log_2(x)$ is decreasing on $(0,\infty)$, this implies
    \begin{align}
        -\log_2\mu_I^{T_{\textsc{A}}}\leq -\log_2 \nu_{I}^{T_{\textsc{a}}}
    \end{align}
    and equality holds if and only if $\mu_I^{T_{\textsc{A}}}=\nu_{I}^{T_{\textsc{a}}}$. Therefore,
    \begin{align}
        E_{\mathcal{N}}(A:\bar{A})&=\sum_{I=1}^N\max\{ 0,-\log_2\nu_I^{T_{\textsc{A}}}\}\\
        &\geq \sum_{I=1}^{N_{AB}}\max\{ 0,-\log_2\nu_I^{T_{\textsc{A}}}\}\label{eq:N_to_N_AB}\\
        &\geq \sum_{I=1}^{N_{AB}}\max\{ 0,-\log_2\mu_I^{T_{\textsc{A}}}\}\label{eq:mu_nu}\\
        &= E_{\mathcal{N}}(A:B).
    \end{align}
    If $E_{\mathcal{N}}(A:B)=E_{\mathcal{N}}(A:\bar{A})$, then both equalities must hold. Equality in Eq.~\eqref{eq:N_to_N_AB} holds if and only if $\nu_I^{T_{\textsc{A}}}\geq 1$ for all $I=N_{AB}+1,...,N$, or equivalently,
    \begin{align}
        m\leq N_{AB}.
    \end{align}
    Equality in Eq.~\eqref{eq:mu_nu} holds if and only if
    \begin{align}
       \mu_I^{T_{\textsc{A}}}=\nu_{I}^{T_{\textsc{a}}}\quad (I=1,\cdots,m). 
    \end{align}
\end{proof}

\begin{lemma}\label{lem:en_subspace}
    Let $B$ be a subsystem of $\bar{A}$. If 
    \begin{align}
        E_{\mathcal{N}}(A:B)=E_{\mathcal{N}}(A:\bar{A}),
    \end{align}
    then
    \begin{align}
        E^{T_\textsc{A}}\subset \Gamma_A\oplus\Gamma_B.
    \end{align}
\end{lemma}
\begin{proof}
    Let 
    \begin{align}
       0< \xi_1<\cdots<\xi_k<1
    \end{align}
    be the distinct values among $0<\nu_1^{T_\textsc{A}}\leq \cdots\leq \nu_m^{T_\textsc{A}}<1$. We denote by $V_i$ the positive-norm eigenspace of $J^{T_\textsc{A}}$ with symplectic eigenvalue $\xi_i$. Similarly, we denote by $V_{AB,i}$ the positive-norm eigenspace of $J^{T_\textsc{A}}_{AB}$ with symplectic eigenvalue $\xi_i$. By Lemma~\ref{lem:negativity_same_ev}, the multiplicity of $\xi_l$ in $J^{T_\textsc{A}}_{AB}$ is the same as the multiplicity in $J^{T_\textsc{A}}$ for any $l=1,...,k$, i.e.,
    \begin{align}
        \dim V_{i}=\dim V_{AB,i}\label{eq:equal_dim_ev}
    \end{align}
    for all $i=1,\cdots,k$. 

    We prove
    \begin{align}
        V_{i}=V_{AB,i} \quad (i=1,...,k).\label{eq:ind}
    \end{align}
    by induction. For any unit vector $f\in V_{AB,1}\subset \Gamma_{A}\oplus\Gamma_B$, we have
    \begin{align}
        Q(f,f)=\langle f,-\ii J^{T_{\textsc a}} f\rangle=\langle f,-\ii J_{AB}^{T_{\textsc a}} f\rangle=\xi_1.
    \end{align}
    By Lemma~\ref{lem:sesquilinear_form_Q}, we get $f\in V_{\xi_1}$. From Eq.~\eqref{eq:equal_dim_ev}, this implies
    \begin{align}
        V_{1}=V_{AB,1}.
    \end{align}
    Suppose 
    \begin{align}
        V_{i}=V_{AB,i} \quad (i=1,...,l-1).\label{eq:m1}
    \end{align}
    Take any unit vector $f\in V_{AB,l}$. Since $f$ is symplectically orthogonal to any vector in $V_{AB,i}$ with $i\neq l$, Eq.~\eqref{eq:m1} implies that it is also orthogonal to any vector in $\bigoplus_{i=1}^{l-1}V_{i}$. Since
    \begin{align}
        Q(f,f)=\xi_l,
    \end{align}
    by Lemma~\ref{lem:sesquilinear_form_Q}, $f\in V_{\xi_l}$. Therefore, from Eq.~\eqref{eq:equal_dim_ev}, we obtain
    \begin{align}
        V_{l}=V_{AB,l}.
    \end{align}

    From the definition in Eq.~\eqref{eq:definition_Eplus}, we have
    \begin{align}
        E_+^{T_\textsc{A}}\coloneqq \mathrm{span}\{e_I^{T_{\textsc{A}}}\}_{I=1}^m=\bigoplus_{i=1}^kV_i.
    \end{align}
    Therefore, Eq.~\eqref{eq:ind} implies
    \begin{align}
         E_+^{T_\textsc{A}}=\bigoplus_{i=1}^kV_{AB,i}\subset \Gamma_A\oplus\Gamma_B.
    \end{align}
    Since $\Gamma_A\oplus\Gamma_B$ is closed under complex conjugation, we also get
    \begin{align}
         E_+^{T_\textsc{A}*}\subset \Gamma_A\oplus\Gamma_B,
    \end{align}
    and hence
    \begin{align}
        E^{T_\textsc{A}}=E_+^{T_\textsc{A}}\oplus E_+^{T_\textsc{A}*}\subset \Gamma_A\oplus\Gamma_B. 
    \end{align}
\end{proof}

As a consequence, we prove the uniqueness of the entanglement partner.
\begin{theorem}
    Let $B$ be a subsystem of $\bar{A}$. If 
    \begin{align}
        E_{\mathcal{N}}(A:B)=E_{\mathcal{N}}(A:\bar{A}),
    \end{align}
    then
    \begin{align}
        \Gamma_{A_{ep}}\subset \Gamma_B.
    \end{align}
\end{theorem}
\begin{proof}
    From Lemma~\ref{lem:en_subspace},         
    \begin{align}
        E^{T_\textsc{A}}\subset \Gamma_A\oplus\Gamma_B. 
    \end{align}
    Therefore, 
    \begin{align}
        \Gamma_{{\textsc a}_{ep}}=\Pi_{\textsc a}^\perp E^{T_\textsc{A}} \subset \Pi_{\textsc a}^\perp (\Gamma_{\textsc a}\oplus\Gamma_B)=\Gamma_B,
    \end{align}
    where we have used $\Gamma_B\subset \Gamma_{\bar{A}}$ in the rightmost equality. 
\end{proof}
This theorem proves that $A_{\mathrm{ep}}$ is the unique minimal subsystem of $\bar{A}$ that reproduces the full logarithmic negativity between $A$ and $\bar{A}$. In particular, if $B$ has the same number of modes as $A_{\mathrm{ep}}$ and satisfies $ E_{\mathcal{N}}(A:B)=E_{\mathcal{N}}(A:\bar{A})$, then $B=A_{\mathrm{ep}}$.

%\bibliography{references.bib,ref_ky}

\end{document}